\documentclass[12pt,draftcls,onecolumn]{IEEEtran}
\usepackage{mathptmx} 
\usepackage{latexsym,bm}
\usepackage{amsmath}
\usepackage{amsfonts}
 \usepackage{amssymb,amsmath}
  \usepackage{amsthm}
 \usepackage{graphicx}
\usepackage{epstopdf}
\usepackage{epsfig}       
 \usepackage{amsfonts}
 \usepackage{mathrsfs}
\usepackage{indentfirst}
\usepackage{graphicx}      
\usepackage{epsfig} 
\usepackage{epstopdf}
\usepackage{times}
\usepackage{bbm}
\usepackage[numbers,sort&compress]{natbib}
\usepackage{cleveref}
\usepackage{color}
\usepackage[justification=centering]{caption}
\usepackage{subfigure}

\allowdisplaybreaks
\renewcommand{\thesection}{\arabic{section}}
 \newtheorem{theorem}{Theorem}[section]

\newtheorem{lemma}{Lemma}[section]
\newtheorem{remark}{Remark}[section]

\newtheorem{assumption}{Assumption}[section]
\newtheorem{definition}{Definition}[section]
\ifCLASSINFOpdf
\else
\fi
\begin{document}

\def\ba{\begin{array}}
\def\ea{\end{array}}
\def\ban{\begin{eqnarray*}}
\def\ean{\end{eqnarray*}}
\def\bd{\begin{description}}
\def\ed{\end{description}}
\def\be{\begin{equation}}
\def\ee{\end{equation}}
\def\bna{\begin{eqnarray}}
\def\ena{\end{eqnarray}}
\allowdisplaybreaks
\title{Graphon Particle Systems, Part II: Dynamics of Distributed  Stochastic Continuum Optimization}
%
%
%

\author{Yan~Chen,
        Tao~Li,~\IEEEmembership{Senior Member,~IEEE}, and Xiaofeng Zong,~\IEEEmembership{Member,~IEEE}
\thanks{This work was funded by the National Natural Science Foundation of
China under Grants No. 62261136550 and No. 62522319.  This paper was presented in part at the 62nd IEEE Conference on Decision and Control, Marina Bay Sands, Singapore, Dec. 13-15, 2023. Corresponding author: Tao Li.}
\thanks{Yan Chen is   with the State Key Laboratory of Mathematical Sciences, Academy of Mathematics and Systems Science, Chinese Academy of Sciences, Beijing 100190, China (e-mail: yanchen2026@amss.ac.cn).}
\thanks{Tao Li is with the Key Laboratory of Management, Decision and Information Systems, Institute of Systems Science, Academy of Mathematics and Systems Science, Chinese Academy of Sciences,  Beijing 100190, China, and also with School of
Mathematical Sciences, University of Chinese Academy of Sciences, Beijing 100149, China (email: litao@amss.ac.cn).}
\thanks{Xiaofeng Zong is with the School of Artificial Intelligence and Automation, China University of Geosciences, Wuhan 430074, China, and Hubei Key Laboratory of Advanced Control and Intelligent Automation for Complex Systems, Wuhan 430074, China. (e-mail: zongxf@cug.edu.cn).}
}

%
%

\markboth{Journal of \LaTeX\ Class Files, June~2024}%
{Shell \MakeLowercase{\textit{et al.}}: Bare Demo of IEEEtran.cls for IEEE Journals}
%



\maketitle

\begin{abstract}
We study the distributed optimization problem over a graphon with a continuum of nodes, which is regarded as the limit of the distributed networked
optimization as the number of nodes goes to infinity. Each node has a private local cost function. The global cost function, which
all nodes cooperatively minimize, is the integral of the local cost functions on the node set. We propose
 stochastic gradient descent and  gradient tracking algorithms over the graphon.
 We establish a general lemma for the upper bound estimation related to a class of time-varying differential inequalities with negative linear terms, based upon which, we prove that for both kinds
  of algorithms, the second moments of the nodes' states are uniformly bounded.  Especially, for the stochastic gradient tracking algorithm, we transform the convergence analysis into the asymptotic property of coupled  nonlinear differential inequalities with time-varying coefficients and  develop a decoupling method. For both kinds of algorithms, we  show that  by choosing the time-varying algorithm gains properly, all nodes' states achieve $\mathcal{L}^{\infty}$-consensus for a connected graphon. Furthermore, if the local cost functions are strongly convex, then all nodes' states converge to the minimizer of the global cost function and the   auxiliary states in the stochastic gradient tracking algorithm converge to the gradient value  of the global cost function at the minimizer uniformly in mean square.

\end{abstract}


%
\IEEEpeerreviewmaketitle

\section{Introduction}
\hspace*{12pt}
\label{sec:introduction}
\IEEEPARstart{I}{n} a distributed optimization problem over a network, all nodes cooperatively
optimize a global cost function which is the sum of   local cost functions, and each node only knows its own local cost function. Distributed optimization involving information exchange among nodes over a large-scale network can be found applications in distributed machine learning (\cite{B. McMahan}), multi-agent target tracking (\cite{D. Li}), distributed resource allocation (\cite{R. Madan}), and so on. The dimensions of these algorithms explode as the number of nodes increases, and it is of interest to investigate the limiting case as the number of nodes tends to infinity. In fact, games and optimal control problems with a continuum of individuals have been studied intensively in the field called mean field games, which was pioneered independently by Huang,   Caines and Malham\'{e} (\cite{M. Huang}) and Lasry and Lions (\cite{J. M. Lasry06}), respectively. They attempt to understand the behaviors of the limiting systems of the dynamic games with a large number of individuals. In the past decades, there has been an increasing intention in mean field games and their applications (\cite{Lachapelle A}-\cite{Cardaliaguet P}).

Motivated by the distributed optimization over large-scale networks and the developing theory of mean-field control and games, we investigate the limiting model of the distributed optimization problem as the number of nodes tends to infinity, that is, the distributed optimization problem over a graphon with a continuum of nodes. The problem is formulated as follows. Let $[0,1]$ be the set of a continuum of nodes, each element of which corresponds to a node. The connecting structure among nodes is given by the graphon $A$, which is a symmetric measurable function from $[0,1]\times [0,1]$ to $[0,1]$ (\cite{L. Lovasz}). 
Any node $p \in [0,1]$ has a private local cost function $V(p,x):[0,1] \times \mathbb{R}^{n}\to \mathbb{R}$, which is strongly convex and continuously differentiable w.r.t. $x\in \mathbb{R}^{n}$ and  is integrable w.r.t. $p \in [0,1]$.  The objective of all nodes is to cooperatively solve the optimization problem
\begin{align}
\min \limits_{x \in \mathbb{R}^{n}}V(x)\triangleq\int_{[0,1]}V(p,x)dp, \label{globalgoal}
\end{align}
 where $x\in\mathbb{R}^n$ is the optimization variable, $V(x)$ is the global cost function to be optimized, and the private local cost function  $V(p,\cdot)$ is only known to node $p$. One hopes to find the unique minimizer of $V(x)$ denoted by $x^{*}$.

  In real-world scenarios, optimization problems are frequently encountered in uncertain environments. The randomness may arise from mini-batch sampling in deep learning (\cite{2012Chen}) or from measurement noise in distributed tracking tasks (\cite{Li2019}). Consequently, people can only depend on  the noisy approximations of gradients instead of exact ones.
Besides, in the distributed optimization over the network with finite nodes,  all nodes interact through the underlying network. The interactions among nodes depend on their labels and so are heterogeneous. In the graphon mean field theory, the concept of graph limit is introduced into the mean field theory, which provides a powerful tool for modeling the heterogeneous interactions among a large number of individuals (\cite{ERHAN.BAYRAKTAR}-\cite{Gao2}). Representing the heterogeneous interactions among nodes in terms of the coupled mean field terms based on the graphon, we  follow the discrete-time distributed stochastic gradient descent (D-SGD) algorithm in \cite{CFFFL2025} for finite nodes and   propose the following continuous-time D-SGD algorithm for the problem (\ref{globalgoal}).  For any node $ p \in [0,1]$,
\begin{align}
dx_p(t)=&\alpha_{1}(t)\int_{\mathbb{R}^n\times [0,1]}A(p,q)(x-x_p(t))\mu_{t}(dx,dq)dt -\alpha_{2}(t)\nabla_{x} V(p,x_p(t))dt\notag\\
&-\alpha_{2}(t)\Sigma_{1} dw_p(t),\label{Mckeanvlasov0}
\end{align}
where $x_p(t)\in \mathbb{R}^n$ is the state of node $p$ at time $t$, representing its local estimate of $x^{*}$;  $\nabla_{x} V(p,x_p(t))  \\ \in\mathbb{R}^n$ is the gradient value of the local cost function at   $x_{p}(t)$; $\int_{\mathbb{R}^n \times [0,1]}A(p,q)(x-x_p(t))\mu_{t}(dx,dq)$ is the coupled mean field term based on the graphon $A$.
Let $(\Omega, \mathcal{F}, \mathcal{P})$ be a complete probability space with a  family of non-decreasing $\sigma$-algebras $\{\mathcal{F}_{t},  t \geqslant 0\}\subseteq \mathcal{F}$. For any $t \geqslant 0$, $\mu_{t}(dx,dq)$ is the distribution on $\mathbb{R}^n\times [0,1]$ and satisfies the following conditions. (i) The marginal distribution $\mu_{t}(\cdot,dq)$  is always the uniform distribution on
$[0,1]$, that is, $ \mu_{t}(\cdot,dq)=dq, \ \forall \ t \geqslant 0$. (ii) Given $q \in [0,1]$, the conditional distribution $\mu_{t}(dx|q)$ is the distribution of $x_q(t)$. Here, $\{(w_{p}(t),\mathcal{F}_{t}),  t \geqslant 0,   p\in [0,1]\}$ is a family of independent $n$-dimensional standard Brownian motions   (see Remark 1.1 in  \cite{Y. Chen})  and the initial states $\{x_{p}(0),\  p\in [0,1]\}$ are adapted to $\mathcal{F}_{0}$ and independent of $\{ w_{p}(t),   t \geqslant 0,   p\in [0,1]\}$. The terms $\alpha_{1}(t)$ and
$\alpha_{2}(t)$ are time-varying algorithm gains and $\Sigma_{1} \in \mathbb{R}^{n\times n}.$

We also propose the following  distributed stochastic gradient tracking (D-SGT)  algorithm
\begin{equation}\label{MCKEANTRACKING}
\left\{\begin{array}{l}
dz_{p}(t)=\beta_{3}(t)\int_{[0,1]\times \mathbb{R}^{n}}A(p,q)(z-z_{p}(t))\mu_{t,q}(dz)dqdt -\beta_{1}(t) y_{p}(t)dt\\
\ \ \ \ \ \ \ \ \ -\beta_{1}(t)\beta_{3}(t)\int_{[0,1]\times \mathbb{R}^{n}}A(p,q)(y-y_{p}(t))
\nu_{t,q}(dy)dqdt,\\
dy_{p}(t)=\beta_{3}(t)\int_{[0,1]\times \mathbb{R}^{n}}A(p,q)(y-y_{p}(t))\nu_{t,q}
(dy)dqdt+\beta_{2}(t) H\left(V\left(p,z_{p}(t)\right)\right)dz_{p}(t) \\
\ \ \ \ \ \ \ \ \ \ \
+\beta_{2}(t)\eta_{p}(t)dt +\beta_{2}^{'}(t) \nabla_{x} V\left(p,z_{p}(t)\right)dt,\\
\end{array}\right.
\end{equation}
 $ \forall \ p \in [0,1]$, where  $z_p(t)\in \mathbb{R}^n$ is the state of node $p$ at time $t$, representing its local estimate of $x^{*}$;  $y_{p}(t) \in \mathbb{R}^n$  is the auxiliary state of node $p$ at time $t$,  tracking
the average
$\nabla_{x}\big(\int_{[0,1]}V(p,z_{p}(t))dp\big)$ and satisfying that $E\left[y_{p}(0)\right]=E\left[\nabla_{x} V(p,z_p(0))\right]$;
$\nabla_{x} V\left(p,z_p(t)\right)\in\mathbb{R}^n$  is the gradient value of the local cost function  at   $z_{p}(t)$; $H\left(V\left(p,z_{p}(t)\right)\right)$ is the Hessian matrix of the local cost function  at  $z_{p}(t)$;
$\mu_{t,q}(dz)$ and $\nu_{t,q}(dy)$ are the distributions of $z_{q}(t)$ and $y_{q}(t)$;
$\int_{[0,1]\times \mathbb{R}^{n}}A(p,q)(z-z_{p}(t))\mu_{t,q}(dz)dq$ and $\int_{[0,1]\times \mathbb{R}^{n}}A(p,q)(y-y_{p}(t))
\nu_{t,q}(dy)dq$ are the coupled mean field terms of the states and the auxiliary states based on the graphon $A$. Here, $\left\{(\eta_{p}(t), \mathcal{F}_{t}), t \geqslant 0,  p\in [0,1]\right\}$ is a family of independent $n$-dimensional continuous  stochastic processes,  and the initial states $\{z_{p}(0), p \in [0,1]\}$ and  auxiliary states  $\{y_{p}(0), p \in [0,1]\}$ are adapted to $\mathcal{F}_{0}$ and independent of $ \{\eta_{p}(t),\ t \geqslant 0,\\  p\in [0,1] \}$.   The terms $\beta_{1}(t)$,
$\beta_{2}(t)$ and $\beta_{3}(t)$ are time-varying algorithm gains and $\beta_{2}^{'}(t)$ is the derivative of  $\beta_{2}(t)$ w.r.t. $t$.

Both  systems (\ref{Mckeanvlasov0}) and
(\ref{MCKEANTRACKING}) belong to the following graphon particle system
\begin{align}
&dz_{p}(t)\notag\\
=& \Big[c_{1}(t)\textstyle{\int_{[0,1]\times  \mathbb{R}^{m}}} A(p,q)(z-z_{p}(t))\mu_{t,q}(dz)dq+c_{2}(t)
\textstyle{\int_{[0,1]\times  \mathbb{R}^{m}}} A(p,q)\bar{f}(q,p,z_{p}(t),z,t)\mu_{t,q}(dz)dq\notag\\
&+c_{4}(t)\xi_{p}(t)
+c_{3}(t)g(p,z_{p}(t),t) \Big]dt+c_{5}(t)\Sigma dw_{p}(t),\label{combinitwoalgorithm}
\end{align}
$ \forall \ p \in [0,1]$, where $\bar{f}(p,q,z_{p}(t),z,t)=f(q,z,t)-f(p,z_{p}(t), t)$, $f(p,z,t): [0,1] \times \mathbb{R}^{m}\times [0,\infty) \to \mathbb{R}^{m}$  and  $g(p,z,t): [0,1] \times \mathbb{R}^{m}\times [0,\infty) \to \mathbb{R}^{m}$ are the functions satisfying appropriate
 conditions; $\left\{(\xi_{p}(t), \mathcal{F}_{t}),  t \geqslant 0,   p\in [0,1]\right\}$ is a family of independent  $m$-dimensional continuous stochastic processes; the processes $\{w_{p}(t), t \geqslant 0, p \in [0,1]\}$ and $\{\xi_{p}(t),   t \geqslant 0, p \in [0,1]\}$ are mutually independent;  the initial states $\{z_{p}(0), p \in [0,1]\}$  are adapted to $\mathcal{F}_{0}$ and independent of $\left\{\xi_{p}(t),\ t \geqslant 0, \ p\in [0,1]\right\}$ and $\{w_{p}(t), t \geqslant 0, p \in [0,1]\}$; $c_{i}(t), \ i=1, \ \ldots,\ 5$ are the time-varying coefficients, $\Sigma \in \mathbb{R}^{m \times m}$ and $m$ is a positive integer.

Up to now, most  existing works
 (\cite{ERHAN.BAYRAKTAR}-\cite{Bet G20}) focused on the existence and uniqueness of
 the solutions for different graphon particle systems and the
 convergence of finite particle systems to graphon particle
 systems. Only few works (\cite{Bayraktar22}-\cite{ERHAN.BAYRAKTAR22}) are concerned with the asymptotic properties of the graphon particle systems.  Bayraktar and Wu (\cite{Bayraktar22}) showed that the distribution of each node's state converges to a limiting distribution as time goes to infinity.  They also provided an exponential  concentration bound  for the Wasserstein distance between the empirical distribution  and the integral of the limiting  distributions on the node set  in \cite{ERHAN.BAYRAKTAR22}.
 Note that all aforementioned works on the graphon particle systems only prove the existence of limiting distributions but do not characterize what these limiting distributions specifically are, particularly, they do not reveal the relation between the limiting distributions and system dynamics. However, for many practical problems, people are more interested in how the limiting distribution is related  to the system dynamics. In particular, for the problem (\ref{globalgoal}) and the algorithms  (\ref{Mckeanvlasov0}) and (\ref{MCKEANTRACKING}), people expect to figure out whether the states  $\{x_{p}(t),  t \geqslant 0,  p \in [0,1]\}$  and $\{z_{p}(t),  t \geqslant 0,  p \in [0,1]\}$ converge to the minimizer of the global cost function under some proper assumptions.

Motivated by the above, we investigate the asymptotic properties of  the  graphon particle systems (\ref{Mckeanvlasov0}) and  (\ref{MCKEANTRACKING}).
We prove that if the graphon is connected and the local cost functions are strongly convex, then by properly choosing algorithm gains, both the states $\{x_{p}(t), t \geqslant 0, p \in [0,1]\}$ in (\ref{Mckeanvlasov0}) and $\{z_{p}(t), t \geqslant 0, p \in [0,1]\}$ in (\ref{MCKEANTRACKING}) converge to the minimizer of the global cost function in mean square. The main contributions are listed as follows.
\begin{itemize}
\item We prove that the $\mathcal{L}^{2}$-consensus implies  $\mathcal{L}^{\infty}$-consensus for the system (\ref{combinitwoalgorithm}) if the integral of the second moments of all nodes' states on the node set is uniformly bounded. The introducing of time-varying algorithm gains removes the requirement on the strong convexity constant of the local cost functions in  (\ref{Mckeanvlasov0}) and  (\ref{MCKEANTRACKING}), which is introduced in \cite{Bayraktar22} for time-invariant graphon particle systems. This leads to a time-varying general system (\ref{combinitwoalgorithm}) and  poses difficulties in establishing the relationship between the $\mathcal{L}^{2}$-consensus  $\lim\limits_{t \to \infty} \int_{[0,1]}\|E[z_{p}(t)]-\int_{[0,1]}E[z_{q}(t)]dq\|^{2}dp=0$  and  the $\mathcal{L}^{\infty}$-consensus $\lim\limits_{t \to \infty}\sup_{p \in [0,1]} \|E[z_{p}(t)]-$ $\int_{[0,1]}E[z_{q}(t)]dq\|^{2}\\ =0$. To this end, we give a key lemma to estimate the  upper  bounds of a class of  functions  satisfying   time-varying differential inequalities with negative linear terms, so as to obtain  the relationship between the $\mathcal{L}^{2}$-consensus  and $\mathcal{L}^{\infty}$-consensus.
  \item We obtain the $\mathcal{L}^{2}$-consensus for the D-SGD algorithm   (\ref{Mckeanvlasov0}) under the connected graphon by choosing the algorithm gains properly.  It is also proved that if the local cost functions are strongly convex, then $\int_{[0,1]}E\big[\|x_{p}(t)\|^{2}\big]dp$ is uniformly bounded, and then  the $\mathcal{L}^{\infty}$-consensus is also achieved. This in turn derives that all nodes' states converge to the minimizer of the global cost function uniformly in mean square.  Besides, we quatify how the convergence rate  of $\mathcal{L}^{2}$-consensus relates to the parameters of the system dynamics (\ref{Mckeanvlasov0}), especially the algebraic connectivity of the graphon.
  \item  For the  D-SGT algorithm (\ref{MCKEANTRACKING}), we prove that if the local cost functions are  strongly convex, then the nodes' states converge  to the minimizer of the global cost function  and the auxiliary states converge to the gradient value of the global cost function at the minimizer  uniformly in mean square, respectively. Note that the convergence analysis for the double-variable system (\ref{MCKEANTRACKING})  is more  challenging. Since the states and the auxiliary states are coupled by the time-varying algorithm gains, the analysis method for the system (\ref{Mckeanvlasov0}) is  no longer applicable. We firstly develop a decoupling method for the asymptotic properties of a class of coupled nonlinear differential  inequalities. Then, we obtain the $\mathcal{L}^{2}$-consensus of the states and the transformed auxiliary states under the connected graphon and the strongly convex local cost functions. Finally, the corresponding optimization is solved by comparison theorem and the relationship between the $\mathcal{L}^{2}$-consensus and  $\mathcal{L}^{\infty}$-consensus for the general system (\ref{combinitwoalgorithm}).

\end{itemize}

 The rest of the paper is organized as follows. In Section II, the definition of the graphon and its property, and some assumptions are presented. Section III gives the main results, containing the relationship between the $\mathcal{L}^{2}$-consensus and $\mathcal{L}^{\infty}$-consensus for  the  system (\ref{combinitwoalgorithm}), the convergence of the D-SGD algorithm (\ref{Mckeanvlasov0}), and  the convergence of the D-SGT algorithm (\ref{MCKEANTRACKING}). In Section IV, the simulation examples are given. In Section V, the conclusions and future works are given.

\textbf{Notation}: Denote the $n$-dimensional Euclidean space by $\mathbb{R}^{n}$ and the Euclidean norm by $\left\| \cdot \right\|$. For a given matrix $A \in \mathbb{R}^{n\times n}$, $\operatorname{Tr}(A)$ denotes its trace. For a given  vector $x \in \mathbb{R}^{n}$, $x^{\mathsf{T}}$ denotes its transpose.
Denote $L^2( [0,1], \mathbb{R}^n)=\{f:  [0,1]\rightarrow  \mathbb{R}^{n}, \ f \ \text{is measurable}, \ \int_{[0,1]}  \|f(x)\|^{2}dx
 < \infty\}$. Denote the set of all bounded linear operators from
 $L^2\left( [0,1], \mathbb{R}^n\right)$ to $L^2\left( [0,1],\ \mathbb{R}^n\right)$ by $\mathcal{L}\left(L^2\left( [0,1],\ \mathbb{R}^n\right)\right)$. Denote  the inner product on $L^2\left([0,1], \ \mathbb{R}^n\right)$ by
$\langle \cdot, \cdot\rangle_{L^2\left([0,1], \mathbb{R}^n\right)}$, that is, for any given $f, g \in L^2\left([0,1],\ \mathbb{R}^n\right)$, $\langle f, g\rangle_{L^2\left([0,1], \ \mathbb{R}^n\right)} \triangleq\int_{[0,1]}f^{\mathsf{T}}(x)g(x)dx$. For a given function $f: F \to \mathbb{R}$, $\operatorname{supp}(f)=\{x \in F: f(x) \neq 0\}$ denotes the support set of $f$. For a given random vector $X \in \mathbb{R}^{n}$,
 denote  its  mathematical expectation  and   distribution   by $E[X]$
 and $\mathcal{L}(X)$, respectively.
Denote the set of  probability measures on
$\mathbb{R}^{n}$ by $\mathscr{P}(\mathbb{R}^{n})$.
Denote the set of probability measures on  $\mathcal{C}_{T}^{n}$ by  $\mathscr{P}(\mathcal{C}_{T}^{n})$.
For a given measurable space $(F, \mathscr{G})$ and $x \in F$, where $\mathscr{G}$ is a $\sigma$-algebra on $F$, the Dirac measure $\delta_x$ at $x$ is defined by $\delta_x(A)=1$ if $x \in A$  and $\delta_x(A)=0$ otherwise, $\forall \ A \in \mathscr{G}$.

\section{Preliminaries}
This work is the companion paper of  \cite{Y. Chen}, in which we have proved the existence and uniqueness of the solution to the system (\ref{combinitwoalgorithm}) and the law of large numbers.  Moreover, some preliminaries about the graphon theory and T-SGD algorithm were reported. So in this paper, we only introduce some necessary information about the graphon. One can refer to \cite{ M2014, CM2019,Benoit Bonnet} for more information.

\vskip 1.5mm
For a given graphon $W$, the Graphon-Laplacian $\mathbb{L}_{W} \in \mathcal{L}\left(L^2\left( [0,1],   \mathbb{R}^n\right)\right)$ generated by $W$ is given by, for any  $\ z \in L^{2}([0,1],  \mathbb{R}^{n})$, $(\mathbb{L}_{W} z)(p)=\int_{[0,1]}W(p,q) (z(p)-z(q))dq, \ \forall \ p \in [0,1].$
For a graphon $W$, the algebraic connectivity of $W$ is defined by
%
\begin{align}
\lambda_2(\mathbb{L}_{W})=\inf _{z \in \mathscr{C}^{\perp}} \frac{\langle\mathbb{L}_{W} z, z\rangle_{L^2\left([0,1], \ \mathbb{R}^n\right)}}{\langle z, z\rangle_{L^2\left([0,1], \ \mathbb{R}^n\right)}^2} \geqslant 0,\label{Algebricconnect}
\end{align}
where $\mathscr{C}^{\perp}=\{z \in L^2\left([0,1], \ \mathbb{R}^n\right): \int_{[0,1]}z(p)dp=0\}$.
By Proposition 4.9 in \cite{Benoit Bonnet}, for the graphon $W$, the algebraic connectivity can also be written as
\begin{align}\label{Algebricconnectibalance}
\lambda_2(\mathbb{L}_{W})=&\inf _{z \notin \mathscr{C}} \frac{\int_{[0,1]\times [0,1]}W(p,q)z^{\mathsf{T}}(p)(z(p)-z(q))
dqdp}{\int_{[0,1]}\|z(p)-\int_{[0,1]}z(q)dq\|^{2}dp},
\end{align}
where $\mathscr{C}=\{z \in L^2([0,1], \ \mathbb{R}^n): z(\cdot) \ \text{is constant} \text{ over}\ [0,1]\}$.

\vskip 1.5mm
\begin{definition}\label{stronglyconnecte}(\cite{Benoit Bonnet})
For a graphon $W$, if the following conditions hold, then the graphon $W$ is said to be connected.
\begin{itemize}
\item[(i)] For any $p \in [0,1]$ and $q \in [0,1] \backslash\{p\}$, there exists an integer $m \geqslant 1$ and a finite sequence $\left(l_k\right)_{1 \leqslant k \leqslant m} \subset [0,1]$ satisfying that $p=l_1, \ q=l_m$ and $l_{k+1} \in \operatorname{supp}\left(W\left(l_k, \cdot\right)\right)$, $\forall \ k \in\{1, \ldots, m-1\}$.
\item[(ii)] $\inf \limits_{p \in [0,1]} \int_{[0,1]} W(p,q) \mathrm{d}q>0$.
\end{itemize}

\end{definition}
\vskip 1.5mm

The following lemma shows the connection between the algebraic connectivity  and the connectivity of a graphon.
\vskip 1.5mm

\begin{lemma}\label{STEWARTconnected}(\cite{Benoit Bonnet})
The graphon $W$ is connected in the sense of Definition \ref{stronglyconnecte}  if and only if $\lambda_2(\mathbb{L}_{W})>0$.
\end{lemma}

\vskip 1.5mm
We make the following assumptions on the graphon and the local cost functions in (\ref{globalgoal}).
\vskip 1.5mm

\begin{assumption}\label{assumption0}
\indent
\begin{itemize}
\item[(i)] Graphon $A$ is connected.

\item[(ii)]
There exists a constant $\kappa>0$, such that $\|\nabla_{x} V(p,x_{1})-\nabla_{x} V(p,x_{2})\| \leqslant \kappa \| x_{1}- x_{2}\| $, $\forall \ x_{1},\ x_{2} \\ \in \mathbb{R}^{n}, \ p \in [0,1]$. There exist constants $\sigma_{v}>0$  and $C_{v}>0$, such that $\|\nabla_{x} V(p,x)\| \leqslant \sigma_{v}\|x\|+C_{v}$,  $\forall \ x \in \mathbb{R}^{n}, \ p \in [0,1]$.
\item[(iii)] The local cost function $V(p,x)$ is uniformly strongly convex w.r.t. $x$, that is, there exists $\kappa_{2}>0$, such that $(x_{1}-x_{2})^{\mathsf{T}} (\nabla_{x} V(p,x_{1})-\nabla_{x} V(p,x_{2})) \geqslant \kappa_{2}\|x_{1}-x_{2}\|^{2}, \ \forall \ x_{1}, \ x_{2} \ \in \mathbb{R}^{n},\ p \in [0,1].$
\end{itemize}
\end{assumption}

\section{Main Results}

 In this section, the relationship between $\mathcal{L}^{2}$-consensus and $\mathcal{L}^{\infty}$- consensus,  convergence of D-SGD
algorithm, and convergence of D-SGT algorithm are investigated, respectively. To maintain continuity, we relegate the proofs of the lemmas and theorems to the appendix.

\subsection{Relationship Between $\mathcal{L}^{2}$-Consensus and $\mathcal{L}^{\infty}$-Consensus}
In this subsection, we prove that   the $\mathcal{L}^{2}$-consensus implies $\mathcal{L}^{\infty}$-consensus for the  system (\ref{combinitwoalgorithm}) under some conditions.


\begin{assumption}\label{assumption10}
There exists a nonnegative constant $\lambda_{1}$, such that $ \|f(p,z_{1},t)-f (p,z_{2},t  ) \|+ \|g(p,z_{1},t)-g (p,z_{2},t  ) \| \leqslant \lambda_{1}\|z_{1}-z_{2}\|, \ \forall \ z_{1},\ z_{2} \in \mathbb{R}^{m}, \ t \geqslant 0, \ p \in [0,1]$; there  exist   nonnegative constants $\lambda_{11}$ and $\lambda_{12}$ such that
$\|f(p,z,t)\|+ \|g(p,z,t)\| \leqslant \lambda_{11}\|z\|+\lambda_{12}$,  $\forall \ z \in \mathbb{R}^{m}, \  t \geqslant 0, \   p \in [0,1]$; there exist nonnegative constants $\lambda_{3}$ and $\lambda_{4}$, such that for any $\epsilon>0$,  there exists $\delta> 0$, such that if $\|t_{1}-t_{2}\| < \delta$, then $ \|f(p,z,t_{1})-f (p,z,t_{2} ) \|^{2} +  \|g(p,z,t_{1})-g (p,z,t_{2}) \|^{2} <\epsilon\left(\lambda_{3}\|z\|^{2}+\lambda_{4}\right),\ \forall \  t_{1}, \  t_{2} \geqslant 0,  \ z \in \mathbb{R}^{m},   p \in [0,1]$; $f(p,z,t)$ and $g(p,z,t)$ are measurable w.r.t. $p$, $\forall  \ z\in \mathbb{R}^{m}, \ t \geqslant 0$;
the map $[0,1] \ni p \mapsto \mu_{0,p}=\mathcal{L}(z_{p}(0)) \in \mathscr{P}(\mathbb{R}^{m})$ is  measurable and there exists a constant $\varsigma \geqslant 0$ such that $\sup_{p \in [0,1]} E\left[\|z_{p}(0)\|^{2}\right]  \leqslant \varsigma$.
\end{assumption}
\vskip 1.5mm
\begin{assumption}\label{assumption41}
 The map $[0,1] \ni p\mapsto \mathcal{L}(\xi_{p}(t))$ is measurable, $t\geqslant 0$;
  $E\left[\xi_{p}(t)\right]=0, \forall \  p \in [0,1], \ t \geqslant 0$; there exists  $r_{1} \geqslant 0$, such that $\sup_{t \geqslant 0, \ p \in [0,1]}E\left[\|\xi_{p}(t)\|^{2}\right]\leqslant r_{1}$; $\xi_{p}(\cdot)$ satisfies that, for any $\epsilon>0$, there exists $\delta >0$, such that if $|t_{1}-t_{2}|<\delta$, then $E\left[\|\xi_{p}(t_{1})-\xi_{p}(t_{2})\|^{2}\right]
<\epsilon, \forall \ t_{1},\ t_{2}\in [0,\infty), \ p \in [0,1]$.
\end{assumption}

\vskip 1.5mm
\begin{assumption}\label{assumption12}
The time-varying coefficients satisfy that
$c_{1}(t)>0$, $c_{2}(t)\geqslant 0$, $c_{3}(t)\geqslant 0$, $c_{4}(t)\geqslant 0$, $c_{5}(t)\geqslant 0$, $\forall \ t \geqslant 0$, $c_{i}(t), \ i=1,\ldots,5$ are continuous w.r.t. $t$,
 $\lim_{t \to \infty}\frac{c_{i}(t)}{c_{1}(t)}=0, i=2,\ldots,5$, $\int_{0}^{\infty}c^{2}_{5}(t)dt< \infty$, $\int_{0}^{\infty}c_{1}(t)dt=+\infty$ and $\lim_{t \to \infty}c_{1}(t)=0.$
\end{assumption}
%

\vskip 1.5mm

The following lemma illustrates that the variances of the nodes' states tend to zero.

\vskip 1.5mm
\begin{lemma}\label{jointvariance}
For the graphon particle system (\ref{combinitwoalgorithm}), if Assumption
\ref{assumption0} (i) and Assumptions \ref{assumption10}-\ref{assumption12} hold, then
\begin{align}
&\lim_{t \rightarrow \infty} \sup_{p \in [0,1]} E\big[ \|z_{p}(t)-E[z_{p}(t)] \|^{2}\big]=0.\label{variancevanish}
\end{align}
\end{lemma}

\vskip 1.5mm

%

To give the relation between the $\mathcal{L}^{2}$-consensus and $\mathcal{L}^{\infty}$-consensus, we also need the following lemma to show the time-varying upper bounds of a class of functions satisfying   time-varying differential inequalities with negative linear terms.

\vskip 1.5mm

\begin{lemma}\label{seconesstmate}
If $y(\cdot): [0, \infty) \rightarrow[0, \infty)$ satisfies
\begin{align}\label{comparisioninequality}
y'(t) \leqslant - a_1(t) y(t) + a_2(t) \sqrt{y(t)}+ a_3(t), \quad \forall \ t \geqslant 0,
\end{align}
where
$a_1(\cdot): [0, \infty) \rightarrow (0, \infty)$,\ $a_2(\cdot),\ a_3(\cdot): [0, \infty) \rightarrow [0, \infty)$, and $a_{i}(\cdot), \ i=1,2,3$ are continuous, then
\begin{align}
  y(t) \leqslant& \max \Bigg\{y(0),\bigg(\sup _{s\in [0,t]} \frac{a_{2}(s)}{2 a_{1}(s)}+\bigg(\sup _{s\in [0,t]}\frac{1}{4} \frac{a_{2}^{2}(s)}{a_{1}^{2}(s)} +\sup _{s\in [0,t]} \frac{a_{3}(s)}{a_{1}(s)}\bigg)^{\frac{1}{2}}\bigg)^{2} \Bigg\}, \quad \forall \ t \geqslant 0.\label{seconesstmateb}
\end{align}
\end{lemma}

\vskip 1.5mm

By Lemmas \ref{jointvariance}-\ref{seconesstmate}, we give the following theorem which shows that the $\mathcal{L}^{2}$-consensus implies  $\mathcal{L}^{\infty}$-consensus for the system (\ref{combinitwoalgorithm}) if the integral of the second moments of all nodes' states on the node set is uniformly bounded.

\vskip 1.5mm
\begin{theorem}\label{jointdecoupleconsensus}
For the graphon particle system (\ref{combinitwoalgorithm}), if Assumption \ref{assumption0} (i) and Assumptions \ref{assumption10}-\ref{assumption12} hold,  $\lim_{t \to \infty} \int_{[0,1]}\big\|E[z_{p}(t)] -\int_{[0,1]}E[z_{q}(t)]dq\big\|^{2}dp=0$ and $\sup_{t \geqslant 0}\int_{[0,1]}E\left[\|z_{p}(t)\|^{2}\right]dp<\infty$, then
 \begin{equation} \label{linftycon}
 \lim_{t \to \infty}\sup_{p \in [0,1]}\Big\|E[z_{p}(t)]-\int_{[0,1]}E[z_{q}(t)]dq\Big\|^{2}=0.
 \end{equation}
\end{theorem}

\subsection{Convergence of  D-SGD  Algorithm}

In this subsection, we  prove the convergence of the D-SGD algorithm (\ref{Mckeanvlasov0}).

\vskip 1.5mm

Denote $\mu_{t}(dx|q)=\mu_{t,q}(dx)$. Then $\mu_{t}(dx,dq)=\mu_{t,q}(dx)  dq$. Therefore, (\ref{Mckeanvlasov0}) can be written as
 \begin{align}\label{Mckeanvlasov1}
dx_p(t) =&\alpha_{1}(t)\int_{ [0,1]}A(p,q)\left( \int_{\mathbb{R}^n}(x-x_p(t))\mu_{t,q}(dx)\right)dqdt \notag\\
 &-\alpha_{2}(t)\nabla_{x} V(p,x_p(t))dt-\alpha_{2}(t)\Sigma_{1} dw_p(t).
 \end{align}

We give the following assumptions on the  algorithm (\ref{Mckeanvlasov0}) for the convergence analysis.
\vskip 1.5mm
\begin{assumption}\label{assumptiondescent02}
 The map $[0,1] \ni p \mapsto \mu_{0,p}=\mathcal{L}(x_{p}(0)) \in \mathscr{P}(\mathbb{R}^{n})$ is measurable and there exists $\zeta_{2} >0$ such that  $\sup_{p \in [0,1]}E\big[\|x_{p}(0)\|^{2}\big] \leqslant \zeta_{2}$.
\end{assumption}
\vskip 1.5mm

\begin{assumption}\label{assumptiondescent1}
The time-varying algorithm gains satisfy that $\alpha_{1}(t)>0$, $\alpha_{2}(t)>0$, $\forall \ t \geqslant 0$, $\alpha_{1}(t)$ and  $\alpha_{2}(t)$ are continuous w.r.t. $t$,   $\int_{0}^{\infty} \alpha_{2}(t) dt = \infty$,  $\int_{0}^{\infty}\alpha_{2}^{2}(t) dt < \infty$, $\lim_{t \rightarrow \infty} \frac{\alpha_{2}(t)}{\alpha_{1}(t)}=0$ and $\lim_{t \rightarrow \infty} \alpha_{1}(t)=0$.
\end{assumption}

\vskip 1.5mm
\begin{remark}
 Assumption \ref{assumption0} (i) guarantees that information can be adequately exchanged among the nodes, thereby enabling the finding of the minimizer of the  global cost function;  Assumption \ref{assumption0}  (ii)-(iii)   are commonly used in \cite{CFFFL2025}, \cite{Pu}-\cite{A. Nedic}  for the distributed optimization problems with finite nodes.   Assumption \ref{assumptiondescent02} is for the uniqueness and existence of the   solution to  (\ref{Mckeanvlasov0}). Assumption \ref{assumptiondescent1} is for the algorithm gains, which means that the vanishing rates of the algorithm gains should be properly selected  to ensure convergence.
Note that Assumption \ref{assumptiondescent1}   requires that $\alpha_{2}(t)$ decays faster than $\alpha_{1}(t)$, which  makes each node not stuck in the minimum of its own local cost function.  Similar assumptions on the algorithm gains have been used for discrete-time stochastic gradient descent algorithms over finite graphs in \cite{CFFFL2025}.\hfill  $\blacksquare$
\end{remark}

%


\vskip 1.5mm

The following lemma illustrates that all nodes' states achieve $\mathcal{L}^{\infty}$-consensus.
\vskip 1.5mm

 \begin{lemma}\label{consensusstochagra}
 For the problem (\ref{globalgoal}) and the algorithm (\ref{Mckeanvlasov1}), if Assumption \ref{assumption0} and Assumptions \ref{assumptiondescent02}-\ref{assumptiondescent1} hold, then there exists $K_{0} \geqslant 0$, such that
 \begin{align}
\sup_{t \geqslant 0,p\in [0,1]}E\big[\|x_{p}(t)\|^{2}\big]\leqslant&  K_{0},\label{qiwangjiftyj}\\
 \int_{[0,1]}\big\|Z_{p}(t)\big\|^{2}dp
 \leqslant & \Psi_{0}(0,t)\zeta+
\int_{0}^{t}8\Big( \sigma_{v}K_{0}+ C_{v} K_{0}^{\frac{1}{2}}\Big)
\alpha_{2}(s)\Psi_{0}(s,t)ds,\label{seppedofcons}\\
\lim_{t \rightarrow \infty} \sup\limits_{p \in [0,1]}\big\|Z_{p}(t)\big\|^{2}= & 0,\label{boundedsecmom}
 \end{align}
  where $Z_{p}(t)=E[x_{p}(t)]-\int_{[0,1]} E[x_{q}(t)]dq$, $\Psi_{0}(s,t)=e^{-2\lambda_2(\mathbb{L}_{A})\int_{s}^{t}\alpha_{1}(s^{\prime})ds^{\prime}}$ and $\lambda_{2}(\mathbb{L}_{A})$ is the algebraic connectivity of the graphon $A$ defined by (\ref{Algebricconnect}).
 \end{lemma}

\vskip 1.5mm
Then we  prove that the integral of the expectations of the states on the node set converges to the minimizer of the global cost function.
By Assumption
\ref{assumption0} (iii), we know that $V(x)$ is strongly convex w.r.t. $x$ and $\nabla_{x}V(p,x)$ is continuous w.r.t. $x$. Then,
 $\nabla_{x}V(x^{*})=\int_{[0,1]}\nabla_{x}V(p,x^*)dp=0$.
 \vskip 1.5mm
\begin{lemma}\label{meanoptimal}
For the problem (\ref{globalgoal}) and the algorithm (\ref{Mckeanvlasov1}), if Assumption \ref{assumption0} and Assumptions
\ref{assumptiondescent02}-\ref{assumptiondescent1} hold, then $\lim_{t \rightarrow \infty} \|\int_{[0,1]}E[x_{p}(t)]dp-x^{*}\|^{2}=0.$
\end{lemma}
\vskip 1.5mm

Finally, we show that the state of each node converges to the minimizer of the global cost function in mean square.
\vskip 1.5mm

\begin{theorem}\label{optimalstgra}
For the problem (\ref{globalgoal}) and the algorithm (\ref{Mckeanvlasov0}), if Assumption \ref{assumption0} and Assumptions \ref{assumptiondescent02}-\ref{assumptiondescent1} hold, then
\begin{align}
\lim_{t \to \infty}\sup_{p \in [0,1]}E\big[ \|x_{p}(t)-x^{*} \|^{2}\big]=0.\label{extoopt}
\end{align}
\end{theorem}
\vskip 1.5mm


\begin{remark}
    Bayraktar and Wu  (\cite{Bayraktar22}) assumed that the dissipativity of the drift term is  strictly twice greater than the Lipschitz constant of the interaction term. For the systems (\ref{Mckeanvlasov0}) and (\ref{MCKEANTRACKING}), this assumption is equivalent to the strong convexity constant of the local cost functions being greater than two, which is not reasonable for distributed optimization problems.
In Assumption \ref{assumption0} (iii), the local cost functions are only assumed to be strongly convex and there is no further requirement on the strong convexity constant.

 For the system (\ref{Mckeanvlasov0}), we introduce the   time-varying
algorithm gains to relax the  requirement. The introducing of time-varying
algorithm gains removes the requirement on the strong
convexity constant of the local cost functions, while it poses
difficulties in the uniform boundedness of the second moments
of all nodes' states, that is, the method for  the uniform boundedness  in  \cite{Bayraktar22} is not applicable. To this end, we develop  Lemma \ref{seconesstmate}  and choose the algorithm gains properly, and  finally prove that the second moments of all nodes' states are
uniformly bounded in Lemma \ref{consensusstochagra}.

  Besides, Bayraktar and Wu  (\cite{Bayraktar22}) proved the existence of the limiting distributions of the nodes' states, while we not only prove the existence of the limiting distributions but also reveal that the limiting distribution is right the Dirac measure at the minimizer of the global cost function. Besides, Bayraktar and Wu  (\cite{Bayraktar22}) proved that   all nodes' states converge in distribution, while we prove the convergence in mean square, which is stronger than convergence in distribution.\hfill  $\blacksquare$
\end{remark}

\vskip 1.5mm
\begin{remark}
The graphon particle system (\ref{Mckeanvlasov0}) is equivalent to the following system in  distribution. Given the initial state $x(0)=x_{P}(0)$,
\begin{align}
dx(t)=&\alpha_{1}(t)\int_{\mathbb{R}^n \times [0,1]}  A(P,q)(x-x(t))\mu_{t}(dx,dq)dt
 -\alpha_{2}(t)\nabla_{x} V(P,x(t))dt-\alpha_{2}(t)\Sigma_{1} dw(t),\label{jointdis}
\end{align}
where $P$ is   uniformly distributed on $[0,1]$ and  for any $t \geqslant 0$, $\mu_{t}(dx,dq)$ is the distribution on $\mathbb{R}^n\times [0,1]$ and satisfies the following conditions. (i) The marginal distribution $ \mu_{t}(\cdot,dq) $ is always the uniform distribution on $[0,1]$, that is, $\mu_{t}(\cdot,dq)=dq, \ \forall \ t \geqslant 0$. (ii) The marginal distribution $\mu_{t}(dx,\cdot)=\int_{[0,1]}\mu_{t}(dx|q)dq$ is the distribution of $x(t)$, where $\mu_{t}(dx|q)$ is the conditional distribution of $x(t)$ given $P=q$. Here, $\{w(t),\ t \geqslant 0\}$ is an $n$-dimensional standard Brownian motion. Notice that $\mu_{t}(dx|q)$ is also the distribution of $x_{q}(t)$ in (\ref{Mckeanvlasov0}). Therefore,
by Theorem \ref{optimalstgra} and  Lemma 4.7 in \cite{Kallenberg},
  $\mu_{t}(dx|q)$ in (\ref{jointdis}) weakly converges to $\delta_{x^{*}}(dx)$ uniformly. Then, the distribution $ \mu_{t}(dx,\cdot) $ weakly converges to $\delta_{x^{*}}(dx)$. \hfill  $\blacksquare$
\end{remark}

We will  study the spatio-temporal discretization  of the  algorithm \eqref{Mckeanvlasov0}, which will be used in the simulation of the system \eqref{Mckeanvlasov0}.

  Consider the   spatio-temporal approximation  of the  algorithm \eqref{Mckeanvlasov0}  as shown in \cite{Y. Chen}.
For any given positive integer $N$, define a step graphon $A^N$ as $A^N(p,q)=A(\frac{i}{N},\frac{j}{N})$, $p\in (\frac{i-1}{N},\frac{i}{N}]$, $q\in (\frac{j-1}{N},\frac{j}{N}]$, $i,j=1,2,...,N$.
Given a step-size $h$,
the system \eqref{Mckeanvlasov0} can be discretized into the following system.  For   $i =1,2,...,N$ and $k=0,1,2,...,$
\begin{align}\label{EQinuous}
   x_i^{N,h}((k+1)h)
= & x_i^{N,h}(kh) -  h \alpha_2(kh) \nabla_{x} V\left(\frac{i}{N},x_i^{N,h}(kh)\right)  \notag\\
&  + h \alpha_1(kh) \frac{1}{N}\sum_{j=1}^{N}A^N \Big(\frac{i}{N},\frac{j}{N} \Big)\left(x_j^{N,h}(kh )-x_i^{N,h}(kh )\right) - \alpha_2(kh)\Sigma_1  \Delta_k w_{\frac{i}{N}},
\end{align}
where $\Delta_kw_{\frac{i}{N}}= w_{\frac{i}{N}}((k+1)h)-w_{\frac{i}{N}}(kh)$ and $x_i^{N,h}(0)=x_{\frac{i}{N}}(0),\ i =1,2,...,N$.
%
 \vskip 2mm
 
 To show the connection between the system \eqref{EQinuous} and the
limiting state $x^*$ of the  system \eqref{Mckeanvlasov0}, the following extra assumptions are need.

 \vskip 2mm
 \begin{assumption}\label{VDLIP}
There exists $L_0>0$, such that $\|\nabla_{x}  V(p_1,x)-\nabla_{x}  V(p_2,x)\|\leq L_0 |p_1-p_2|\|x\|,\ \forall \ p_1,\\  p_2\in [0,1], \ x\in \mathbb{R}^n$. 
\end{assumption}

 \vskip 2mm
 \begin{assumption}\label{asssdd}
The time-varying algorithm gains satisfy that $\lim_{t \rightarrow \infty} \frac{\alpha_{1}^2(t)}{\alpha_{2}(t)}=0$,
 $\alpha_1(t),$ $|\dot\alpha_1(t)|$ and $|\dot\alpha_2(t)|$ are eventually non-increasing, and for any given $l$,
\[
 \alpha_{1}(t)\alpha_{1}(t-l)+|\dot\alpha_1(t-l)|+|\dot\alpha_2(t-l)|=o(\alpha_2(t)),\ t\to\infty.
\]
\end{assumption}

 \vskip 2mm
\begin{assumption}\label{continuousgrapho}
\begin{itemize}
  \item[(i)] The two-dimensional Lebesgue measure of the discontinuity set of  $A$ is zero.
  \item[(ii)] For any $p\in[0,1]$, the function $q\mapsto A(p,q)$
has finite discontinuity points and is Lipschitz continuous on each
continuity interval.
  \item[(iii)] The degree function
$
d_A(p)=\int_0^1 A(p,q)\,dq
$
is Lipschitz continuous on $[0,1]$.
\end{itemize}

\end{assumption}

  \vskip 2mm

The following theorem establishes a mean-square convergence estimate for the
states of the  discretized system \eqref{EQinuous} toward the
limiting state \(x^*\) of the  system \eqref{Mckeanvlasov0}.
The resulting error bound consists of a time-decaying term, a temporal
discretization term, and a network discretization term, which vanish as
$k \to\infty$, \(h\to0\), and \(N\to\infty\), respectively.

\begin{theorem}\label{hheNwucha}
For the  system  (\ref{EQinuous}), if Assumptions \ref{assumption0} and  \ref{assumptiondescent02}-\ref{continuousgrapho} hold,
then  there exist  $h_0$,  $T$, $M$, $M(h_0)$, $C_1,\ C_2,\ C_3,\ C_4>0$  such that, for any $h  \in (0, h_0)$,
\begin{align}
 \frac{1}{N}\sum_{i=1}^{N}
E\left[\left\|x_i^{N,h}(kh)-x^*\right\|^2\right]
\leq
\phi_1(kh)+\phi_2(h)+\phi_3(N),\ \forall \  k\geq  \left\lceil\frac{h_0+T}{h}\right\rceil, \label{lisantox}
\end{align}
where
 $
\phi_1(kh)
=e^{-\int_{h_0+T}^{kh}2\kappa_2\alpha_2(s) ds} \left(M+M(h_0)\right) + e^{-2\lambda_2(\mathbb{L}_{A}) \int_0^{kh}\alpha_1(s)ds}M
  +
\int_0^{kh}
\big[ \alpha_2(s)(M+ 2\sigma_v^2 M+ 2C_v^2 ) +2\alpha_2^2(s)\operatorname{Tr}\left(\Sigma_1\Sigma_1^\top\right)
\big]e^{- 2\lambda_2(\mathbb{L}_{A})\int_s^{kh}\alpha_1(r)dr}ds+ 2(\zeta_2+2\|x^*\|^2)
e^{-2\kappa_2 \int_{0}^{kh}\alpha_2(s)ds}
 +2 (M+\|x^*\|^2)^{\frac12}\\
\int_{0}^{kh}\alpha_2(s)
e^{-2\kappa_2 \int_{s}^{kh}\alpha_2(s^{\prime})ds^{\prime}}
\big[
e^{-\lambda_2(\mathbb L_A)\int_0^s\alpha_1(\tau)d\tau}M^{\frac12}
+
\big(
\int_0^s
e^{-2\lambda_2(\mathbb L_A)\int_\tau^s\alpha_1(r)dr}
\big[
\alpha_2(\tau)(M+2\sigma_v^2M+2C_v^2)
+
2\alpha_2^2(\tau)
\operatorname{Tr}\left(\Sigma_1\Sigma_1^\top\right)
\big]d\tau
\big)^{\frac12}
\big]ds,$
$
\phi_2(h)=   C_1 \sqrt h+C_2 h$,
$\phi_3(N)
=C_3\Big(\big|
\lambda_2(\mathbb L_A)
-
\dfrac{\lambda_2(L^N)}{N}
\big|+\big|
\lambda_2(\mathbb L_A)
-
\dfrac{\lambda_2(L^N)}{N}
\big|^{\frac{1}{2}}\Big)
+\frac{C_4}{N}$,  $\lambda_2(L^N)$ is the second smallest eigenvalue of the graph Laplacian matrix  $L^N$ corresponding to the weighted graph with adjacency matrix
$
 (A^N (\frac{i}{N},\frac{j}{N} ))_{1\leq i,j\leq N}.$ Moreover, 
 $\lim\limits_{k\to\infty}\phi_1(kh)=0$, $\lim\limits_{h\to0}\phi_2(h)=0$ and $\lim\limits_{N\to\infty}\phi_3(N)=0.$
\end{theorem}

\subsection{Convergence of D-SGT Algorithm}
In this subsection, we prove  the   convergence of the D-SGT algorithm  (\ref{MCKEANTRACKING}).
\vskip 1.5mm

We give some assumptions on the system (\ref{MCKEANTRACKING}).
\vskip 1.5mm

\begin{assumption}\label{assumption3}
 The time-varying algorithm gains satisfy that  $\beta_{1}(t)> 0$,  $\beta_{2}(t)> 0$,  $\beta_{3}(t)>0$, $\forall\ t \geqslant 0$,  $\beta_{2}(0)=1$, $\beta_{1}(t)$ and $\beta_{3}(t)$ are continuous w.r.t. $t$, $\beta_{2}(t)$ is differentiable w.r.t. $t$,
$\int_{0}^{\infty} \beta_{3}(t) dt = \infty$, $\int_{0}^{\infty} \beta_{1}(t) \beta_{2}(t) dt = \infty$, $\lim_{t \rightarrow \infty}\frac{\beta_{1}(t)}{\beta_{3}(t)}=0$,
$\lim_{t \rightarrow \infty}\frac{\beta_{2}(t)}{\beta_{1}(t)}=0$  and $\lim_{t \rightarrow \infty}\beta_{3}(t)=0$.
\end{assumption}

\vskip 1.5mm
\begin{assumption}\label{assumption2}
The map $[0,1] \ni p \mapsto \mathcal{L}(z_{p}(0), y_{p}(0)) \in \mathscr{P}(\mathbb{R}^{2n})$ is measurable and there exist $\zeta$ and $ \zeta_{0} >0$ such that $\sup_{p \in [0,1]}E\big[\|z_{p}(0)\|^{2}\big]\leqslant \zeta$ and  $\sup_{p \in [0,1]}E\big[\|y_{p}(0)\|^{2}\big] \leqslant \zeta_{0}$.
\end{assumption}
\vskip 1.5mm

\begin{assumption}\label{assumption5}
The map $[0,1] \ni p\mapsto \mathcal{L}(\eta_{p}(t))$ is measurable, $t\geqslant 0$; $E\left[\eta_{p}(t)\right]=0, \forall \ p \in [0,1],\ t \geqslant 0$; there exists $b_{1} \geqslant 0$ such that $\sup_{t \geqslant 0, \ p \in [0,1]}E\left[\|\eta_{p}(t)\|^{2}\right] \leqslant b_{1}$; for any $\epsilon>0$, there exists $\delta >0$, such that if $|t_{1}-t_{2}|<\delta$, then $E\big[ \|\eta_{p}(t_{1})-\eta_{p}(t_{2}) \|^{2}\big]<\epsilon, \forall \ t_{1}, \ t_{2} \in [0,\infty),\ p \in [0,1]$.
\end{assumption}
\vskip 1.5mm

Inspired by \cite{Bin}, by the  transformation $\widetilde{y}_{p}(t)=y_{p}(t)-\beta_{2}(t) \nabla_{x} V\left(p,z_{p}(t)\right)$, we have the following transformed graphon particle system
\begin{align}\label{disgremck}
\left\{\begin{array}{l}
dz_{p}(t)=\big(-\beta_{1}(t) \widetilde{y}_{p}(t)-\beta_{1}(t)\beta_{2}(t)\nabla_{x} V\left(p,z_{p}(t)\right)\big)dt\\
\ \ \ \ \ \ \ \ \ \ \ +\beta_{3}(t) \int_{[0,1]\times \mathbb{R}^{n}}A(p,q)(z-z_{p}(t))\mu_{t,q}(dz)dq dt\\
\ \ \ \ \ \ \ \ \ \ \ -\beta_{1}(t) \beta_{2}(t)\beta_{3}(t)\int_{[0,1]\times \mathbb{R}^{n}}A(p,q)\big(\nabla_{x} V(q,z)  -\nabla_{x} V\left(p,z_{p}(t)\right)\big)\mu_{t,q}(dz)dq dt \\
\ \ \ \ \ \ \ \ \ \ \ \ -\beta_{1}(t)\beta_{3}(t) \int_{[0,1]\times \mathbb{R}^{n}}A(p,q)\left(y-\widetilde{y}_{p}(t)\right)\widetilde{\nu}_{t,q}(dy)dqdt,\\
d\widetilde{y}_{p}(t)=\beta_{3}(t)\int_{[0,1]\times \mathbb{R}^{n}}A(p,q)\left(y-\widetilde{y}_{p}(t)\right)
\widetilde{\nu}_{t,q}(dy)dqdt+\beta_{2}(t)\eta_{p}(t)dt\\
\ \ \ \ \ \ \ \ \ \ \  +\beta_{2}(t)\beta_{3}(t)\int_{[0,1]\times \mathbb{R}^{n}}A(p,q)\big(\nabla_{x} V(q,z)   -\nabla_{x} V\left(p,z_{p}(t)\right)\big)\mu_{t,q}(dz)dqdt,
\end{array}\right.
\end{align}
where $\mu_{t,q}(dz)$ and $\widetilde{\nu}_{t,q}(dy)$ are the distributions of $z_{p}(t)$ and $\widetilde{y}_{p}(t)$. Here, $\widetilde{y}_{p}(t)$ is called the transformed auxiliary state.


\vskip 1.5mm

\vskip 1.5mm
\begin{remark}
Assumptions \ref{assumption2}-\ref{assumption5} and the continuity of the algorithm gains in Assumption \ref{assumption3} are imposed to guarantee the existence   and  uniqueness  of the  solution to  (\ref{disgremck}).
The limiting properties of the algorithm gains in Assumption \ref{assumption3} are used to establish  the convergence of the algorithm   (\ref{disgremck}).
The condition $\beta_2(0)=1$ in Assumption \ref{assumption3}, together with $E[y_p(0)]=E[\nabla_xV(p,z_p(0))]$, is imposed to ensure
$
\int_{[0,1]}E[\widetilde y_p(t)]dp=0,\ \forall\ t\geq0.
$
If $\beta_2(0)\neq1$, this zero-average property generally fails due to the arbitrariness of $z_p(0)$, and an additional bias term would need to be handled in the proof. A similar motivation for such an   initialization can be found in Remark 2.2 of \cite{Bin} on finite-node distributed stochastic gradient tracking algorithms.   \unskip\nobreak\hfill\mbox{$\blacksquare$}
\end{remark}

We transform the convergence analysis of the algorithm (\ref{disgremck}) into the asymptotic properties of a class of coupled differential inequalities with time-varying coefficients  and develop a decoupling method in the following lemma.
\vskip 1.5mm

\begin{lemma}\label{decotwotimescale}
If  $Y_{1}(\cdot), Y_{2}(\cdot): [0, \infty) \rightarrow[0, \infty)$ are differentiable and
\begin{align}
\frac{dY_{1}(t)}{dt} \leqslant & (-a_{1}(t)+a_{2}(t))Y_{1}(t)+a_{3}(t)Y_{2}(t)+a_{4}(t),\label{slowtime}\\
\frac{dY_{2}(t)}{dt} \leqslant & -b_{1}(t)Y_{2}(t)+b_{2}(t)Y_{2}^{\frac{1}{2}}(t)\big(Y_{1}^{\frac{1}{2}}(t)+Y_{3}(t)
\big)\label{fasttime}
\end{align}
hold, where the time-varying coefficients satisfy that
  $a_{1}(t)>0,$ $a_{i}(t)\geqslant 0,\ i=2,3,4,$ $\lim_{t \rightarrow \infty}\frac{a_{2}(t)}{a_{1}(t)}=0$, $\lim_{t \rightarrow \infty}\frac{a_{3}(t)}{a_{1}(t)}=0$, $\lim_{t \rightarrow \infty}\frac{a_{4}(t)}{a_{1}(t)}=0$, $\int_{0}^{\infty} a_{1}(t) dt = \infty$,
  $b_{1}(t)> 0$, $b_{2}(t)\geqslant 0$, $b_{1}(t)$ and $b_{2}(t)$ are continuous w.r.t. $t$,   $\int_{0}^{\infty} b_{1}(t) = \infty$, $\sup_{t \geqslant 0}\frac{b_{2}(t)}{b_{1}(t)}<\infty$, $\sup_{t\geq 0} Y_{1}(t)<\infty$ and
   $\lim_{t \rightarrow \infty}Y_{3}(t)=0$,
  then
  \setlength{\abovedisplayskip}{4pt}
\setlength{\belowdisplayskip}{4pt}
 \begin{align}
 \lim_{t \rightarrow \infty}Y_{1}(t)=&0, \label{0slowtime}\\
 \lim_{t \rightarrow \infty}Y_{2}(t)=& 0.\label{1slowtime}
 \end{align}
\end{lemma}

\vskip 1.5mm
\begin{remark}
 The main idea of decoupling inequalities in the above lemma lies in that  the time-varying coefficients
of (\ref{fasttime}) have same orders, which together with Lemma \ref{seconesstmate} shows that $Y_{2}(t)$ can be bounded by $Y_{1}(t)$.  Replacing the upper bound of $Y_{2}(t)$  into the inequality of $Y_{1}(t)$ and using the comparison theorem, we can show (\ref{0slowtime}) and then
 (\ref{1slowtime}) follows.\hfill  $\blacksquare$
\end{remark}

\vskip 1.5mm

By the above lemma and Theorem \ref{jointdecoupleconsensus},  we show that the states and  transformed auxiliary states achieve $\mathcal{L}^{\infty}$-consensus and the integral of the expectations of the states on the node set tends to the minimizer of the global cost function.

\vskip 1.5mm

\begin{lemma}\label{sigleconseussss}
For the problem (\ref{globalgoal}) and the D-SGT algorithm (\ref{disgremck}), if Assumption \ref{assumption0} and  Assumptions \ref{assumption3}-\ref{assumption5} hold, then
\begin{align}
& \sup\limits_{t\geqslant 0,p \in [0,1]}E\big[ \|z_{p}(t)\|^{2}\big]<\infty, \label{unibouofz}\\
& \sup\limits_{t\geqslant 0,p \in [0,1]}E\big[\|\widetilde{y}_{p}(t) \|^{2}\big]<\infty, \label{unibouofyw}\\
&\lim \limits_{t \to \infty}\sup\limits_{p \in [0,1]}\Big\|E[z_{p}(t)]-\int_{[0,1]}E[z_{q}(t)]dq\Big\|^{2}=0,\label{zpconsensus1}\\
&\lim \limits_{t \to \infty}\sup\limits_{p \in [0,1]}\| E[\widetilde{y}_{p}(t) ]\|^{2}=0,\label{zpconsensus}\\
&\lim \limits_{t \rightarrow \infty}\Big\|\int_{[0,1]}E[z_{p}(t)]dp-x^{*}\Big\|^{2}=0.\label{l2qiroop}
\end{align}
\end{lemma}
\vskip 1.5mm

The following theorem shows that  all nodes' states and    auxiliary states
converge to the minimizer of the global cost function and  the gradient value  of the global cost function at the minimizer uniformly in mean square, respectively.
\vskip 1.5mm

\begin{theorem}\label{therelast}
For the problem (\ref{globalgoal}) and the D-SGT algorithm (\ref{MCKEANTRACKING}), if Assumption \ref{assumption0} and  Assumptions \ref{assumption3}-\ref{assumption5} hold, then
\begin{align}
&\lim \limits_{t \to \infty}\sup\limits_{p \in [0,1]}E \Big[\|z_{p}(t)-x^{*} \|^{2} \Big]=0,\label{tenoptimal}\\
&\lim \limits_{t \to \infty}\sup\limits_{p \in [0,1]}E \bigg[\Big\|y_{p}(t)-\nabla_{x}\Big(\int_{[0,1]}  V(q,x^{*})dq\Big)\Big\|^{2}\bigg] =0. \label{tenoptimay}
\end{align}
\end{theorem}
\vskip 1.5mm

\section{Simulations}
\begin{figure}[htbp]
    \centering
  \hspace{-4mm}
    \begin{minipage}[c]{9cm}
        \includegraphics[width=9cm]{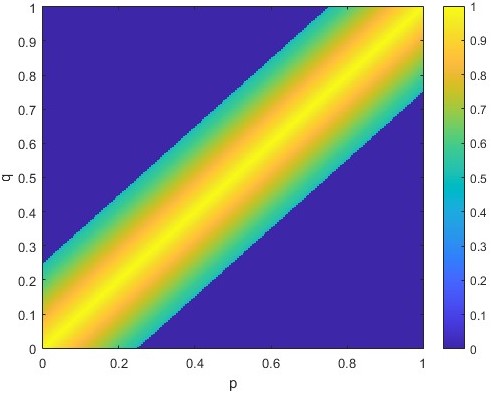}
        \caption{Graphon  $A$}\label{Fig0}
    \end{minipage}
\end{figure}

Consider the optimization problem \eqref{globalgoal} with the distributed stochastic gradient descent algorithm \eqref{Mckeanvlasov0}. We choose the   local cost functions as
 $V(p,x)=(x-x_0)^{T}R_p(x-x_0)+p\|x\|^2 + \sigma_p,$
and $R_p=\operatorname{diag}\{\frac{p}{2}+1, \frac{p}{2}+1\}$, $\sigma_p=p/2$,  where  $x=(x_1, x_2) \in \mathbb{R}^2$, $x_0=(x_{01},x_{02})$, $p \in [0,1]$.
Then, we know that $x^*=(\frac{5x_{01}}{7},\frac{5x_{02}}{7})$.
Graphon is given by $A(p,q)=(1-2|p-q|)I_{\{|p-q|\leqslant \frac{1}{4}\}},\ p,\ q \in [0,1]$ as shown in Fig.\ref{Fig0}, where $I_{\{|p-q| \leq 1 / 4\}}$ denotes the indicator function, which takes the value 1 if $|p-q| \leq 1 / 4$ and 0 otherwise. It can be verified that $A$ is connected following the method in \cite{Benoit Bonnet} and $A$ satisfies Assumption \ref{continuousgrapho}  by Lemma \ref{lamda2sklian1}.
The time-varying algorithm gains are  $\alpha_1(t) =1.5/(1 + t)^{0.6}$ and $\alpha_2(t)=1/(1 + t)^{0.85}$, and $\Sigma_1=\operatorname{diag}\{2, 2\}$.
The initial values $x_{i}^{N,h}(0)=(0,0.5),\ i=1,\cdots,N.$

\begin{figure}[htbp]
    \centering
    \begin{minipage}[c]{7.5cm}
        \includegraphics[width=7.5cm]{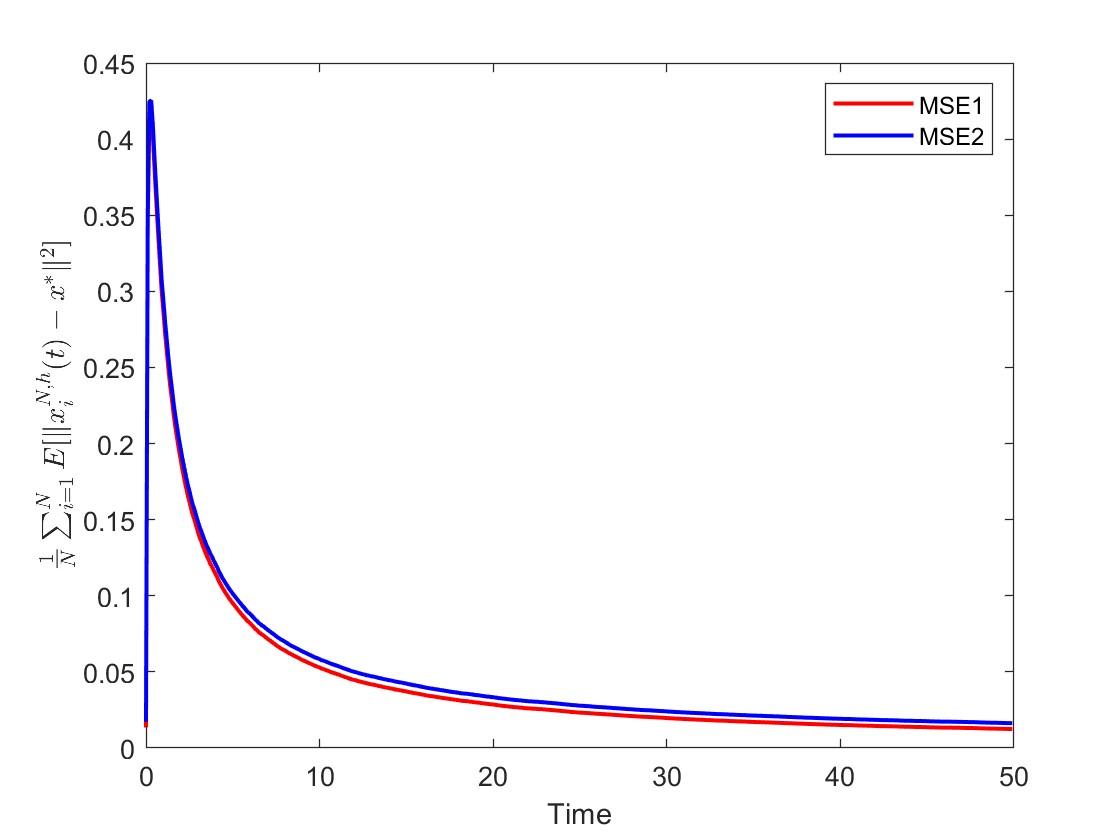}
    \end{minipage}
    \hspace{1mm}
    \begin{minipage}[c]{7.5cm}
        \includegraphics[width=7.5cm]{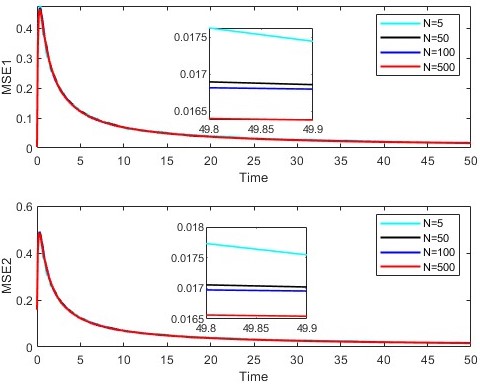}
    \end{minipage}
     \caption{Left: Mean square errors between  states and $x^{*}$, $N=500$, $\Delta t=0.1$; Right: Mean square errors between   states and $x^{*}$ for various  network sizes, $\Delta t=0.1$.  }\label{Fig1}
\end{figure}

\begin{figure}[htbp]
    \centering
    \begin{minipage}[c]{7.5cm}
        \includegraphics[width=7.5cm]{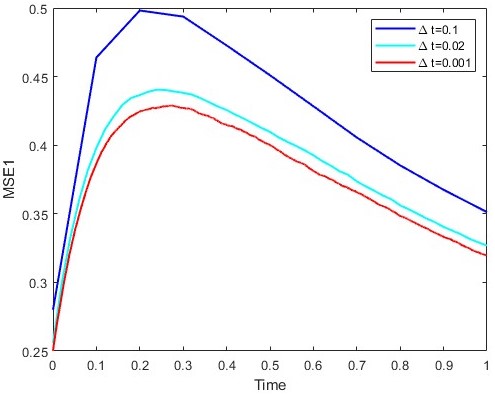}
    \end{minipage}
    \hspace{1.5mm}
    \begin{minipage}[c]{7.5cm}
        \includegraphics[width=7.5cm]{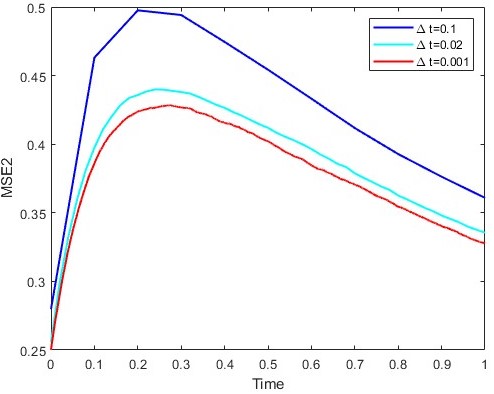}
    \end{minipage}
     \caption{Mean square errors between   states and $x^{*}$ for various step-sizes, $N=500$.}\label{Fig11}
\end{figure}

Then, we implement  (\ref{EQinuous}), which is the discretized system of  \eqref{Mckeanvlasov0}. Choose $x_0=(0.7,1.4)$.
 The left figure in Fig.\ref{Fig1} shows the decaying of the  mean square errors of the two components of the states relative to $x^{*}$, where the expectation is approximated by $500$ samples. It illustrates that all node states converge to $x^{*}$. Then, we choose $x_0=(0.07,0.14)$ and show the mean square errors between the states and $x^{*}$ under different network sizes $N$ in the right figure in  Fig.\ref{Fig1}, indicating that the errors decrease as the number of nodes increases.
  Finally, with $x_0=(0.7,1.4)$, Fig.\ref{Fig11} depicts the mean square errors of the two components of the states relative to $x^{*}$ with various step-sizes,  demonstrating that smaller step-sizes yield smaller mean square errors. The simulation results are consistent with the theoretical result in Theorem \ref{hheNwucha}.

\vskip -2mm

\section{Conclusions and Future Works}
We have proposed  the D-SGD and  D-SGT algorithms  over the graphon for solving the distributed optimization problem with a continuum of nodes.
 By establishing the lemma for the upper bound estimation related to a class of time-varying differential inequalities with negative linear terms,
we have proved the uniform boundedness of  the second moments of the nodes' states in both kinds of  algorithms. Besides,
we have proved that if the graphon is connected and the  time-varying algorithm gains are  chosen properly, then the   states in both  kinds of  algorithms  achieve  $\mathcal{L}^{\infty}$-consensus. Moreover, if the local cost functions are strongly convex, then the states in both  kinds of algorithms converge  to the minimizer of the global cost function  and the   auxiliary states in the D-SGT algorithm converge to the gradient value  of the global cost function at the minimizer uniformly in mean square.

Note that the analysis of the asymptotic properties of the graphon particle systems relies on the special linear interactions among nodes in the proposed two kinds of algorithms, while for many graphon particle systems,
such as   Kuramoto oscillator  (\cite{Gkogkas}),   neural mean-field   (\cite{Baladron}),  SIS epidemics   (\cite{Vizuete}) and so on, the interactions are nonlinear. This results in inapplicability of   the methods in this paper.
Combining graphon estimation techniques with distributed optimization over graphons is another promising direction for future research, particularly when the underlying graphon is unknown and must be learned from finite-node observations (\cite{Gao2015}-\cite{Zhang2024JMLR}).
 Besides, the graphon considered  here is static. In many practical scenarios, networks among nodes receive the feedback from nearby individuals and then make changes to better adapt to the world, such as adaptive Kuramoto-type network models in \cite{Gkogkas1}, which leads to a dynamic graphon.
The asymptotic properties of the graphon particle systems with dynamic graphons are still open so far, which is of major importance from an applied perspective but highly mathematically challenging. Moreover, explicit convergence rates are crucial for evaluating algorithm latency in practical settings, it is also worth analyzing these rates under specific selections of the algorithm gains.

\appendices


\renewcommand{\thesection}{\Roman{section}}
\renewcommand{\thesectiondis}{\Roman{section}}

\setcounter{equation}{0}
\renewcommand{\theequation}{A.\arabic{equation}}
\setcounter{lemma}{0}
\renewcommand{\thelemma}{A.\arabic{lemma}}
\section{}

\emph{Proof of Lemma \ref{jointvariance}:}
Noting that $\mu_{t,p}$ is the distribution of $z_p(t)$ in (\ref{combinitwoalgorithm}), the system (\ref{combinitwoalgorithm}) can be written as
\begin{align}
dz_{p}(t)
 =&\Big[c_{2}(t) \int_{[0,1]} A(p,q)\left(E\left[f(q,z_{q}(t),t)\right]-f(p,z_{p}(t),t)\right)dq\notag\\
 &
 +c_{1}(t)\int_{[0,1]} A(p,q)\left(E\left[z_{q}(t)\right]-z_{p}(t)
 \right)dq+c_{4}(t)\xi_{p}(t)
\notag\\
 &+c_{3}(t)g\left(p,z_{p}(t),t\right)\Big]dt+c_{5}(t)\Sigma dw_{p}(t).\label{excombinitwoalgorithm}
\end{align}
By Assumption \ref{assumption12} and Theorem 2.3.1  in \cite{Boksendal}, we have $E\big[\int_{0}^{t}c_{5}(s) \Sigma dw_{p}(s)\big]=0$.  Denote $$\hat{f}(q,p,z_{q}(t),z_{p}(t),t)=E[f(q,z_{q}(t),t)]- E[f(p,z_{p}(t),t)].$$ Then, by Assumption \ref{assumption41} and (\ref{excombinitwoalgorithm}), we have
\begin{align}
 dE[z_{p}(t)]  \notag=&c_{1}(t)\int_{[0,1]} A(p,q)
 \left(E[z_{q}(t)]-E[z_{p}(t)]
 \right)dqdt\notag\\
 & +c_{2}(t)\int_{[0,1]} A(p,q)\hat{f}(q,p,z_{q}(t),z_{p}(t),t)dqdt
 +c_{3}(t)E[g(p,z_{p}(t),t)]dt.\label{parameterexpecderijoint}
 \end{align}
Denote $S_{p}(t)=\|z_{p}(t)-E[z_{p}(t)]\|^{2}$.
By Theorem 2.1  in \cite{Y. Chen} and Assumption \ref{assumption12}, we have $ E\big[\int_{0}^{t}\| c_{5}(s) (z_{p}(s)-E[z_{p}(s)])^{\mathsf{T}} \Sigma  \|^{2}ds\big] \leqslant   E\big[\sup_{0\leqslant s \leqslant t}\|z_{p}(s) \|^{2}\big] \int_{0}^{t}c_{5}^{2}(s)ds   \| \Sigma \|^{2}  <  \infty$.  Then, by Theorem 2.3.1 in \cite{Boksendal}, we have $E[\int_{0}^{t}c_{5}(s)(z_{p}(s)-E[z_{p}(s)])^{\mathsf{T}}  \Sigma dw_{p}(s)] =0.$
By (\ref{parameterexpecderijoint}), $E\big[(z_{p}(t)
-E[z_{p}(t)]
)^{\mathsf{T}} g(p,E[z_{p}(t)],t)\big]   =0$, $E\big[ (z_{p}(t) -E[z_{p}(t)]
 )^{\mathsf{T}}  (f(p, E[z_{p}(t)],t)
-   E[f(p,z_{p}(t),t)]
 )\big]=0$,  Assumptions \ref{assumption10}-\ref{assumption41} and   It\^{o} formula, we have
\begin{align}
&\frac{dE\left[S_{p}(t)\right] }{dt}\notag\\
=&\int_{[0,1]}
A(p,q)dq\big(-2c_{1}(t)E\left[S_{p}(t)\right] +2c_{2}(t)E\big[(z_{p}(t)-E[z_{p}(t)])^{\mathsf{T}}\notag\\
& \times \big(E[f(p,z_{p}(t),t)]
-f(p,z_{p}(t),t)\big)\big]\big)\notag\\
&+2c_{3}(t)
E\left[(z_{p}(t)-E[z_{p}(t)])^{\mathsf{T}}g(p,z_{p}(t),t)\right]\notag\\
&+2c_{4}(t)
E\left[(z_{p}(t)-E[z_{p}(t)])^{\mathsf{T}}\xi_{p}(t)\right]+\operatorname{Tr}
\left(\Sigma^{\mathsf{T}}\Sigma \right)c_{5}^{2}(t)\notag\\
\leqslant & \left(-2c_{1}(t)\inf\limits_{p \in [0,1]}A_{p}+c_{4}(t) \right)E\left[S_{p}(t)\right] +2c_{2}(t)A_{p}E\big[\|z_{p}(t)  -E[z_{p}(t)]\|\|f(p,z_{p}(t),t) \notag\\
&  -f(p,E[z_{p}(t)],t)\|\big] +2c_{3}(t)
E\big[ \|z_{p}(t) -E[z_{p}(t)]
 \|   \|g(p,z_{p}(t),t)-g(p,E[z_{p}(t)],t) \|\big]\notag\\
&
 +c_{4}(t)r_{1}  +\operatorname{Tr}\left(\Sigma^{\mathsf{T}}\Sigma \right)c_{5}^{2}(t)\notag\\
\leqslant &\phi(t)E\left[S_{p}(t)\right] +c_{4}(t)r_{1}+\operatorname{Tr}(\Sigma^{\mathsf{T}}
\Sigma)c_{5}^{2}(t),\notag
\end{align}
where $A_{p}=\int_{[0,1]}
A(p,q)dq$ and $\phi(t)=-2c_{1}(t)\inf_{p \in [0,1]}A_{p}  +c_{4}(t)+2\lambda_{1}(c_{2}(t)+ c_{3}(t))
$. This together with the comparison theorem (\cite{Michel A. N.}) gives
$
E\left[S_{p}(t)\right]
 \leqslant  e^{\int_{0}^{t} \phi(s)ds}E[S_{p}(0)]+   \int_{0}^{t}\big(c_{4}(s)  r_{1}
 +\operatorname{Tr}(\Sigma^{\mathsf{T}}
\Sigma)
   c_{5}^{2}(s)\big)
e^{\int_{s}^{t} \phi(s^{\prime})ds^{\prime}}ds
.
$
By Assumption   \ref{assumption10}, we have $$\sup_{p\in[0,1]}E\big[\|S_{p}(0)\|^{2}\big]\leqslant \sup\limits_{p\in[0,1]} E\big[\|z_{p}(0)\|^{2} \big]     \leqslant \varsigma.$$ Then,  we have
\begin{align}
\sup\limits_{p\in[0,1]}E\left[S_{p}(t)\right]
 \leqslant& e^{\int_{0}^{t} \phi(s)ds} \varsigma+\int_{0}^{t}\big(c_{4}(s)r_{1}+\operatorname{Tr}(\Sigma^{\mathsf{T}}
\Sigma) c_{5}^{2}(s)\big)e^{\int_{s}^{t} \phi(s^{\prime})ds^{\prime}}ds.\label{doublecompare}
\end{align}
By Assumption \ref{assumption12}, we know that there exists $T \geqslant 0$, such that if $ t \geqslant T$, then $ 2\lambda_{1} \frac{c_{2}(t)+c_{3}(t)}{c_{1}(t)}
+\frac{c_{4}(t)}{c_{1}(t)}\leqslant \inf_{p \in [0,1]}A_{p},$  which together with $\int_{0}^{\infty}c_{1}(t)dt=\infty$, Assumption
\ref{assumption0} (i) and Definition \ref{stronglyconnecte} gives
\begin{align}
\int_{0}^{\infty} \phi(s)ds
= -\infty.\label{jointdivergence}
\end{align}
For the first term on the r.h.s. of (\ref{doublecompare}), by the above equality, we have
 \setlength{\abovedisplayskip}{-3pt}
\setlength{\belowdisplayskip}{3pt}
\begin{align}
\lim\limits_{t \to \infty}  e^{\int_{0}^{t} \phi(s)ds}\varsigma=0.\label{doublecomparefirst}
\end{align}
For the second term on the r.h.s. of (\ref{doublecompare}), by Assumption \ref{assumption0} (i), Assumption \ref{assumption12}, (\ref{jointdivergence}) and L'Hospital's rule, we have $\lim_{t \to \infty} \int_{0}^{t}(c_{4}(s)r_{1}+\operatorname{Tr}(\Sigma^{\mathsf{T}}
\Sigma)c_{5}^{2}(s))e^{\int_{s}^{t} \phi(s^{\prime})ds^{\prime}}ds=0,$
which together with (\ref{doublecompare}) and (\ref{doublecomparefirst}) gives (\ref{variancevanish}).
$\hfill\blacksquare$
\vskip 1.5mm

\emph{Proof of Lemma \ref{seconesstmate}:}
By (\ref{comparisioninequality}) and $a_1(t)>0, \ \forall \ t \geqslant 0$, we have $ y'(t) \leqslant -a_{1}(t) (\sqrt{y(t)}-\frac{a_{2}(t)}{2a_{1}(t)} )^{2}
+\frac{a_{2}^{2}(t)}{4a_{1}(t)}+a_{3}(t).$
Therefore,  we know that if $$\sqrt{y(t)} > \frac{a_{2}(t)}{2 a_{1}(t)}+\bigg(\frac{1}{4} \frac{a_{2}^{2}(t)}{a_{1}^{2}(t)}  + \frac{a_{3}(t)}{a_{1}(t)}\bigg)^{\frac{1}{2}},$$  then $y'(t)<0$ and $$y(t) \leqslant  \max \bigg\{y(0),\bigg(\sup _{s\in [0,t]} \frac{a_{2}(s)}{2 a_{1}(s)}  +\\ \bigg(\sup _{s\in [0,t]}\frac{1}{4} \frac{a_{2}^{2}(s)}{a_{1}^{2}(s)}  +\sup _{s\in [0,t]} \frac{a_{3}(s)}{a_{1}(s)}\bigg)^{\frac{1}{2}}\bigg)^{2} \bigg\}, \quad \forall \ t \geqslant 0,$$ which leads to (\ref{seconesstmateb}). $\hfill\blacksquare$
\vskip 1.5mm

\emph{Proof of Theorem \ref{jointdecoupleconsensus}:}
Denote $B_{p}(t)=E\left[\|z_{p}(t)\|^{2}\right]$ and $R_{p}(t)=\big\|E[z_{p}(t)]-\int_{[0,1]} E[z_{q}(t)]dq\big\|^{2}$.
By $\sup_{t \geqslant 0}\int_{[0,1]}$ $E\left[\|z_{p}(t)\|^{2}\right]dp<\infty$, we know that there exists $K_{1}\geqslant 0$, such that
\setlength{\abovedisplayskip}{4pt}
\setlength{\belowdisplayskip}{4pt}
\begin{align}
\sup_{t \geqslant 0}\int_{[0,1]}E\left[\|z_{p}(t)\|^{2}\right]dp\leqslant K_{1}.\label{boundofintegra}
\end{align}
By Theorem 2.1  in \cite{Y. Chen} and Assumption \ref{assumption12}, we have $E\big[\int_{0}^{t}\|2c_{5}(s)  z_{p}^{\mathsf{T}}(s)
 \Sigma\|^{2}ds\big] \leqslant  4 \|\Sigma\|^{2}\sup_{s\in [0,t]} \\ E\big[\|z_{p}(s)\|^{2}\big]      \int_{0}^{t}c^{2}_{5}(s)ds
  <   \infty.$ Define $S(X,Y)=E[X^\mathsf{T}]E[Y]-E[X^\mathsf{T}Y]$ for $X,\ Y\in \mathbb{R}^n$.
  Then, by Theorem 2.3.1 in \cite{Boksendal}, we have  $$ E \Big[\int_{0}^{t} 2c_{5}(s)  z_{p}^{\mathsf{T}}(s) \Sigma dw_{p}(s)  \Big]=0,$$  which together with Assumptions \ref{assumption10}-\ref{assumption41}, It\^{o} formula, (\ref{combinitwoalgorithm}), (\ref{boundofintegra}), H{\"o}lder inequality and Jensen inequality gives
\begin{align}
 \frac{dB_{p}(t)}{dt}
=&2 c_{1}(t) \int_{[0,1]} A(p,q)\big(E[z_{p}^\mathsf{T}(t)]
E[z_{q}(t)]-E[\|z_{p}(t)\|^2]\big) dq+\operatorname{Tr}(\Sigma^{\mathsf{T}}\Sigma)c_{5}^{2}(t)\notag\\
&+2c_{2}(t)\int_{[0,1]}A(p,q)\big(E[z^{\mathsf{T}}_{p}(t)]E[f(q,z_{q}(t),t)
]\notag\\
 &-E[z^{\mathsf{T}}_{p}(t)f(p,z_{p}(t),t)]\big)dq
  +
 2c_{4}(t) E[z^{\mathsf{T}}_{p}(t)\xi_{p}(t)]+2c_{3}(t)
 E[z^{\mathsf{T}}_{p}(t)g(p,z_{p}(t),t)]\notag\\
 \leqslant &\Big(2c_{2}(t)-2 \int_{[0,1]} A(p,q)dqc_{1}(t)+c_{3}(t)+c_{4}(t)\Big)B_{p}(t)\notag\\
 &
 +2c_{1}(t)\int_{[0,1]}
 \|E[z_{q}(t)]\|dq  \|E[z_{p}(t)]\|\notag\\
 &+c_{2}(t) \Big(\int_{[0,1]} \|E[f(q,z_{q}(t),t)]\|^{2}dq
 \notag\\
 &+E\big[\|f(p,z_{p}(t),t)\|^{2}\big]\Big)
+c_{4}(t)E\big[\|\xi_{p}(t)\|^{2}\big]
 \notag\\
 &+c_{3}(t)
 E\big[\|g(p,z_{p}(t),t)\|^{2}\big]+\operatorname{Tr}(\Sigma^{\mathsf{T}}\Sigma)c_{5}^{2}(t)\notag\\
 \leqslant &\Big(2c_{2}(t)-2\inf\limits_{p \in[0,1]}\int_{[0,1]} A(p,q)dqc_{1}(t)+c_{3}(t)+c_{4}(t)\Big) B_{p}(t)\notag\\
 &  +2c_{1}(t)
 \Big(\int_{[0,1]} \|E[z_{q}(t)]\|^{2}dq
 \Big)^{\frac{1}{2}}  E\big[\|z_{p}(t)\|\big]
 +c_{2}(t)\int_{[0,1]} \|E[f(q,z_{q}(t),t)]\|^{2}dq
 \notag\\
 &+c_{2}(t)E\big[\|f(p,z_{p}(t),t)\|^{2}\big]+c_{4}(t)r_{1}\notag\\
 &
  +c_{3}(t)
 E\big[\|g(p,z_{p}(t),t)\|^{2}\big]
  +\operatorname{Tr}(\Sigma^{\mathsf{T}}\Sigma)c_{5}^{2}(t)\notag\\
 \leqslant & h_{1}(t) B_{p}(t)
 +2c_{1}(t)K_{1}^{\frac{1}{2}}B_{p}^{\frac{1}{2}}(t)
 +2\lambda_{12}^{2}c_{2}(t) +2\lambda_{12}^{2}c_{3}(t)+c_{4}(t)r_{1}
 +\operatorname{Tr}\left(\Sigma^{\mathsf{T}}\Sigma \right)c_{5}^{2}(t)\notag\\
 &+c_{2}(t)\int_{[0,1]}
 E\big[\|f(q,z_{q}(t),t)\|^{2}\big]dq\notag\\
 \leqslant & h_{1}(t) B_{p}(t)
 +2c_{1}(t)K_{1}^{\frac{1}{2}}  B_{p}^{\frac{1}{2}}(t)+2\lambda_{12}^{2}c_{3}(t) +c_{4}(t)r_{1}+\operatorname{Tr}(\Sigma^{\mathsf{T}}\Sigma )c_{5}^{2}(t)\notag\\
 &+2c_{2}(t)\bigg(2\lambda^{2}_{12}+
 \lambda^{2}_{11}\int_{[0,1]} E\big[ \|z_{q}(t) \|^{2}\big]dq
  \bigg)
\notag\\
  \leqslant & h_{1}(t)B_{p}(t)
 +h_{2}(t)B_{p}^{\frac{1}{2}}(t)+h_{3}(t),\label{Boundedsenconp}
 \end{align}
where $h_{1}(t)=-2\inf_{p \in[0,1]}\int_{[0,1]}A(p,q)dqc_{1}(t)+(2+2\lambda_{11}^{2})c_{2}(t)+(1+2\lambda_{11}^{2})c_{3}(t)+c_{4}(t)
$, $h_{2}(t)=2c_{1}(t)K_{1}^{\frac{1}{2}}$ and $h_{3}(t)=(2\lambda^{2}_{11}K_{1}+4\lambda^{2}_{12})c_{2}(t)
+2\lambda_{12}^{2}c_{3}(t)+c_{4}(t)r_{1}
 +\operatorname{Tr}(\Sigma^{\mathsf{T}}\Sigma)c_{5}^{2}(t)$. By Assumption \ref{assumption12}, there exists $T_{2}\geqslant 0$, such that
$$
 \big((2+2\lambda_{11}^{2})c_{2}(t)
 +(1+2\lambda_{11}^{2})c_{3}(t)+c_{4}(t)
 \big)\frac{1}{c_{1}(t)}
 <2\inf_{p \in[0,1]}\int_{[0,1]}A(p,q)dq,\ \forall \ t \geqslant T_{2},
$$
 that is, $h_{1}(t)<0, \ \forall \ t \geqslant T_{2}$. By Theorem 2.1  in \cite{Y. Chen}, there exists $L_{0}\geqslant 0$, such that $$\sup_{p \in [0,1],  t \in [0,T_{2}]}B_{p}(t)\leqslant L_{0}.$$ For any $t \geqslant T_{2}$, by (\ref{Boundedsenconp}) and Lemma \ref{seconesstmate},  we have
\begin{align}
B_{p}(t) \leqslant&  \max \bigg\{B_{p}(T_{2}), \bigg( \sup _{s\in [T_{2},t] } \frac{h_{2}(s)}{-2h_{1}(s)}+ \Big(\sup _{s\in [T_{2},t]}\frac{1}{4} \frac{h_{2}^{2}(s)}{h_{1}^{2}(s)}  +\sup _{s\in [T_{2},t]} \frac{h_{3}(s)}{-h_{1}(s)} \Big)^{\frac{1}{2}} \bigg)^{2}\bigg\}.\label{bounawyzero}
\end{align}
By Assumption \ref{assumption12}, we have $$ \lim_{t \to \infty}\frac{h_{2}(t)}{-2h_{1}(t)}
=\Big(\inf_{p \in[0,1]} \int_{[0,1]}A(p,q)dq\Big)^{-1}\frac{K_{1}^{\frac{1}{2}}}{2},$$  $\lim_{t \to \infty}\frac{h_{3}(t)}{-h_{1}(t)}=0$, and $\lim_{t \to \infty}\frac{1}{4} \frac{h_{2}^{2}(t)}{h_{1}^{2}(t)}$ $=\frac{K_{1}}{4}  \big(\inf_{p \in[0,1]}   \int_{[0,1]} A(p,q) dq\big)^{-2}.$
Then, there exist non-negative constants $L_{1},$ $L_{2}$ and $L_{3}$, such that $\sup_{t \geqslant 0}\frac{h_{2}(t)}{-2h_{1}(t)}\leqslant L_{1}$, $\sup_{t \geqslant 0} \frac{1}{4} \frac{h_{2}^{2}(t)}{h_{1}^{2}(t)}  \leqslant L_{2}$ and $\sup_{t \geqslant 0} \frac{h_{3}(t)}{-h_{1}(t)}\\ \leqslant   L_{3}$. Denote $L=\max\{ L_{0},   (\sqrt{L_{2}+L_{3}} + L_{1} )^{2}\}$. Then, by (\ref{bounawyzero}), we have $\sup_{t \geqslant 0}B_{p}(t) \leqslant L$. Noting that $L$ is independent of $p$,  we have
\begin{align}
\sup\limits_{p \in [0,1], \ t \geqslant 0}E\big[ \|z_{p}(t) \|^{2}\big]\leqslant L.\label{boudeforjoint}
\end{align}
 By $\lim_{t \to \infty}\int_{[0,1]}R_{p}(t)dp=0$, we know that, for any
$\epsilon >0$, there exists $T_{\epsilon}>0$, such that if $t\geqslant T_{\epsilon}$, then $\int_{[0,1]}R_{p}(t)dp < \epsilon^{2}$. For any  $t\geqslant T_{\epsilon}$, denote $S_{\epsilon}^{t}=\left\{p\in [0,1]: R_{p}(t)>\epsilon\right\}$. By Theorem 2.1  in \cite{Y. Chen} and (5.3.1) in \cite{L. Ambrosio}, we know that $S_{\epsilon}^{t}$ is a measurable set of $[0,1]$. For any $t\geqslant T_{\epsilon}$, we have
$\epsilon m(S_{\epsilon}^{t})< \int_{S_{\epsilon}^{t}}R_{p}(t)dp\leqslant
 \int_{[0,1]}R_{p}(t)dp <
\epsilon^{2},$
 that is, $m(S_{\epsilon}^{t})<\epsilon$. Let $\bar{z}_p(t)=E[z_{p}(t)]-\int_{[0,1]} E[z_{q}(t)]dq$ and $G_p(t)=E[g(p,z_{p}(t),t)]-\int_{[0,1]}E[g(q,z_{q}
 (t),t)]dq$. Taking the  derivative of $R_{p}(t)$ on $t \geqslant T_{\epsilon}$ and combining (\ref{parameterexpecderijoint}) in   Lemma \ref{jointvariance} with the symmetry of the graphon $A$ give
 \begin{align}
 \frac{dR_{p}(t)}{dt}
 =&2c_{1}(t)\bigg(
\bar{z}_p^{\mathsf{T}}(t) \int_{[0,1]} A(p,q)
 \bar{z}_{q}(t)
 dq - \int_{[0,1]} A(p,q)dqR_{p}(t) \bigg)\notag\\
 &+2c_{2}(t)\bar{z}_{p}(t)^{\mathsf{T}}\Big(\int_{[0,1]} A(p,q)(E[f(q,z_{q}(t),t)]-E[f(p,z_{p}(t),t)]\big)dq\Big)\notag\\
 &+2c_{3}(t)\bar{z}_{p}(t)^{\mathsf{T}}G_p(t)\notag\\
=:&J_{1p}(t)+J_{2p}(t)+J_{3p}(t).\label{beforsingle}
\end{align}
By  $C_{r}$ inequality, H{\"o}lder inequality, Jensen inequality and (\ref{boudeforjoint}), we have
\begin{align}
&2c_{1}(t)
\bar{z}_p^{\mathsf{T}}(t) \int_{[0,1]} A(p,q)
 \bar{z}_{q}(t)
 dq\notag\\
 \leqslant & 2c_{1}(t)R_{p}^{\frac{1}{2}}(t)\bigg(\int_{[0,1]}R_{q}(t)dq\bigg)^{\frac{1}{2}}\notag\\
 \leqslant & 2c_{1}(t)R_{p}^{\frac{1}{2}}(t)\epsilon^{\frac{1}{2}}\bigg( \sup_{q \in [0,1]}R_{q}(t)+1\bigg)^{\frac{1}{2}}\notag\\
 \leqslant & 2c_{1}(t)R_{p}^{\frac{1}{2}}(t)\epsilon^{\frac{1}{2}}\bigg(2 \sup_{q \in [0,1]}\left\|E[z_{q}(t)]\right\|^{2} +2\bigg\|\int_{[0,1]} E[z_{q^{\prime}}(t)]dq^{\prime}\bigg\|^{2}+1\bigg)^{\frac{1}{2}}\notag\\
  \leqslant & 2c_{1}(t)R_{p}^{\frac{1}{2}}(t)\epsilon^{\frac{1}{2}}\Big(2 L+2\int_{[0,1]} E\big[\left\|z_{q^{\prime}}(t)\right\|^{2}\big]dq^{\prime}+1\Big)^{\frac{1}{2}}\notag\\
    \leqslant & 2c_{1}(t)R_{p}^{\frac{1}{2}}(t) \epsilon^{\frac{1}{2}}(4L + 1 )^{\frac{1}{2}}.\label{beforsingle1}
\end{align}
By $C_{r}$ inequality, H{\"o}lder inequality, Jensen inequality,  Assumption \ref{assumption10}  and (\ref{boudeforjoint}), we have
\begin{align}
 J_{2p}(t)
\leqslant & c_{2}(t)\Bigg(R_{p}(t)+ \bigg\|\int_{[0,1] }A(p,q)\big( E\left[f\left(q,z_{q}(t),t\right)\right]-E[f(p,z_{p}(t),t)]\big)dq\bigg\|^{2}\Bigg)\notag\\
\leqslant &c_{2}(t)\bigg(R_{p}(t)+2\bigg\|\int_{[0,1] } A(p,q) E\left[f\left(q,z_{q}(t),t\right)\right]dq\bigg\|^{2}+2\left\|E\left[f(p,z_{p}(t),t)\right]\right\|^{2}\bigg)\notag\\
\leqslant &c_{2}(t)\bigg(R_{p}(t)+2\int_{[0,1] }\left\| E\left[f\left(q,z_{q}(t),t\right)\right]\right\|^{2}dq+2E\big[ \| f(p,z_{p}(t),t) \|^{2}\big]\bigg)\notag\\
\leqslant &c_{2}(t)R_{p}(t)+4c_{2}(t)\bigg(\lambda_{11}^{2}\int_{[0,1] }  E\big[ \| z_{q}(t) \|^{2}\big]dq  +2\lambda_{12}^{2} + \lambda_{11}^{2} E\big[ \| z_{p}(t) \|^{2}\big]\bigg)\notag\\
\leqslant &c_{2}(t)\big(R_{p}(t)+8\lambda_{11}^{2}L
+8\lambda_{12}^{2}\big).\label{beforsingle2}
\end{align}
By $C_{r}$ inequality, H{\"o}lder inequality, Jensen inequality, Assumption \ref{assumption10} and (\ref{boudeforjoint}), we have
\begin{align}
J_{3p}(t)
\leqslant &  2c_{3}(t)\|\bar{z}_p(t)\|G_p(t)\|\notag\\
 \leqslant &  2\sqrt{2}c_{3}(t)R_{p}^{\frac{1}{2}}(t) \Bigg(\bigg\|\int_{[0,1]}E\left[g(q,z_{q}(t),t)\right]dq\bigg\|^{2}
 + \|E [g(p,z_{p}(t),t) ] \|^{2}\Bigg)^{\frac{1}{2}}\notag\\
 \leqslant &  2\sqrt{2}c_{3}(t)R_{p}^{\frac{1}{2}}(t) \bigg(\sup_{q \in [0,1]}E\big[ \|g(q,z_{q}(t),t) \|^{2}\big]
 +E\big[ \|g(p,z_{p}(t),t) \|^{2}\big]\bigg)^{\frac{1}{2}}\notag\\
 \leqslant &  4\sqrt{2}c_{3}(t)R_{p}^{\frac{1}{2}}(t)\bigg(\lambda_{11}^{2}\sup_{q \in [0,1]}E\big[ \|z_{q}(t) \|^{2}\big]+\lambda_{12}^{2}\bigg)^{\frac{1}{2}}\notag\\
 \leqslant &  4\sqrt{2} (\lambda^{2}_{11}L+\lambda^{2}_{12} )^{\frac{1}{2}}c_{3}(t)R_{p}^{\frac{1}{2}}(t).\label{22kak}
\end{align}
By (\ref{boudeforjoint}), Jensen inequality and H{\"o}lder inequality, we have  $R_{p}(t)
\leqslant 4L$, which gives $R_{p}^{\frac{1}{2}}(t) \leqslant 2L^{\frac{1}{2}}, \forall \ t >0,\ p \in [0,1]$. Then, by (\ref{beforsingle})-(\ref{22kak}), we have 
\begin{align}
 \frac{dR_{p}(t)}{dt}
 \leqslant & -2c_{1}(t)\inf_{p\in[0,1]}\int_{[0,1]}A(p,q)dq R_{p}(t) + 4L^{\frac{1}{2}}\epsilon^{\frac{1}{2}}  (4L+1 )^{\frac{1}{2}}c_{1}(t)\notag\\
 & + 4\big(L+2\lambda_{11}^{2}L
+2\lambda_{12}^{2} \big)c_{2}(t)+8\sqrt{2} L^{\frac{1}{2}} \left(\lambda^{2}_{11}L+\lambda^{2}_{12}\right)^{\frac{1}{2}}c_{3}(t),\ \forall \  t\geqslant T_{\epsilon}.\notag
\end{align}
This together with  the comparison theorem  (\cite{Michel A. N.}) gives
\begin{align}
&\sup_{p \in [0,1]}R_{p}(t)\notag\\
\leqslant & \psi_{1}(T_{\epsilon},t)\sup_{p \in [0,1]}R_{p}(T_{\epsilon})+\int_{T_{\epsilon}}^{t}
4L^{\frac{1}{2}}  \epsilon^{\frac{1}{2}}(4L +1 )^{\frac{1}{2}}
c_{1}(s)\psi_{1}(s,t)ds\notag\\
& +\int_{T_{\epsilon}}^{t}
\Big(4\big(L+2
\lambda_{11}^{2}L
+2\lambda_{12}^{2}\big)c_{2}(s)  +8\sqrt{2}L^{\frac{1}{2}} \big(\lambda_{11}^{2}L +\lambda_{12}^{2}\big)^{\frac{1}{2}}c_{3}(s)\Big)
\psi_{1}(s,t)ds,\label{boudedsigleagent}
\end{align}
where $\psi_{1}(s,t)=e^{-2\inf\limits_{p \in [0,1]}\int_{[0,1]}A(p,q)dq \int_{s}^{t}c_{1}(s^{\prime})ds^{\prime}}$.
 By Assumption
\ref{assumption0}, Assumption  \ref{assumption12} and L'Hospital's rule, we have $$\lim_{t \to \infty} \int_{0}^{t}c_{1}(s)
\psi_{1}(s,t)ds=\frac{1}{2}\bigg(\inf_{p \in [0,1]}  \int_{[0,1]}A(p,q)dq\bigg)^{-1}.$$
Then, we know that there exists $K_{3}\geqslant 0$, such that $\sup_{t \geqslant 0}\int_{0}^{t}
\psi_{1}(s,t) c_{1}(s)ds\leqslant  K_{3}$. Therefore, for the second term on the r.h.s. of  (\ref{boudedsigleagent}), we have $$4L^{\frac{1}{2}}\epsilon^{\frac{1}{2}}  (4L  +1)^{\frac{1}{2}} \int_{T_{\epsilon}}^{t}c_{1}(s)
 \psi_{1}(s,t)ds
 \leqslant 4L^{\frac{1}{2}} \epsilon^{\frac{1}{2}}  (4L+1 )^{\frac{1}{2}}K_{3}.$$
 By the arbitrariness of $\epsilon$, for any $\delta>0$, there exists $\widetilde{\epsilon}>0$,  such that
$4L^{\frac{1}{2}} \widetilde{\epsilon}^{\frac{1}{2}}\big(4L +1\big)^{\frac{1}{2}}K_{3}   < \frac{\delta}{3}.$ By $\lim_{t \to \infty}\int_{[0,1]}R_{p}(t)dp=0$, we know that there exists $T_{\widetilde{\epsilon}}>0$, such that if $t\geqslant T_{\widetilde{\epsilon}}$, then $\int_{[0,1]}R_{p}(t)dp < \widetilde{\epsilon}^{2}$. Then, for any $t \geqslant T_{\widetilde{\epsilon}}$,  (\ref{boudedsigleagent}) can be written as
\begin{align}
\sup_{p \in [0,1]}R_{p}(t)
\leqslant & \psi_{1}(T_{\widetilde{\epsilon}},t)\sup_{p \in [0,1]}R_{p}(T_{\epsilon})+\frac{\delta}{3}\notag\\
&+\int_{T_{\widetilde{\epsilon}}}^{t}
\psi_{1}(s,t)\Big(4\big(L+2\lambda_{12}^{2}
 +2\lambda_{11}^{2}L\big)c_{2}(s) +8\sqrt{2}L^{\frac{1}{2}} \big(\lambda^{2}_{11}L+\lambda^{2}_{12}\big)^{\frac{1}{2}}
c_{3}(s)\Big)ds.\label{boudedsigleagent1}
\end{align}
By  Assumption \ref{assumption0} (i), Assumption \ref{assumption12} and $\sup_{p \in [0,1],t \geqslant 0}  R_{p}^{\frac{1}{2}}(t) \leqslant 2L^{\frac{1}{2}}$, we have
$$\lim_{t \to \infty}\psi_{1}(T_{\widetilde{\epsilon}},t)
\sup_{p \in [0,1]}R_{p}(T_{\widetilde{\epsilon}})=0.$$
Therefore, for the first term on the r.h.s. of (\ref{boudedsigleagent1}), there exists $T_{1}>0$, such that if $t > T_{1}$, then
\begin{align}
\psi_{1}(T_{\widetilde{\epsilon}},t)\sup_{p \in [0,1]}R_{p}(T_{\widetilde{\epsilon}})< \frac{\delta}{3}.\label{boudedsigleagent11}
\end{align}
  By Assumption \ref{assumption0} (i), Assumption \ref{assumption12} and L'Hospital's rule, we have
 $\lim _{t \to \infty}\int_{T_{\widetilde{\epsilon}}}^{t}\big(  4\big(L+2\lambda_{11}^{2}L
+2\lambda_{12}^{2}\big)c_{2}(s)
 +8\sqrt{2}L^{\frac{1}{2}} \big(\lambda^{2}_{11}L
 +\lambda^{2}_{12}\big)^{\frac{1}{2}}c_{3}(s)\big)
\psi_{1}(s,t)ds=0.$
Therefore, for the third term on the r.h.s. of  (\ref{boudedsigleagent1}), there exists $T_{11}> 0$ such that if $t > T_{11}$, then
\begin{align}
\int_{T_{\widetilde{\epsilon}}}^{t}\Big(& 4\big(L+2\lambda_{11}^{2}L
+2\lambda_{12}^{2}\big)c_{2}(s) +8\sqrt{2}L^{\frac{1}{2}} \big(\lambda_{11}^{2}L+\lambda_{12}^{2}\big)^{\frac{1}{2}}c_{3}(s)\Big)
\psi_{1}(s,t)ds<\frac{\delta}{3}.\notag
\end{align}
Therefore, for any $\delta >0$, taking $T=\max\left\{T_{1}, T_{11}\right\}$ and by (\ref{boudedsigleagent1})-(\ref{boudedsigleagent11}) and the above inequality, we know that, if $t \geqslant T$, then $\sup_{p \in [0,1]}R_{p}(t) < \delta$, that is,
(\ref{linftycon}) holds.
$\hfill\blacksquare$

\setcounter{equation}{0}
\renewcommand{\theequation}{B.\arabic{equation}}
\setcounter{lemma}{0}
\renewcommand{\thelemma}{B.\arabic{lemma}}
\section{}

\emph{Proof of Lemma \ref{consensusstochagra}:}
By Theorem \ref{jointdecoupleconsensus}, it's sufficient to prove $$\lim_{t \to \infty}\int_{[0,1]}\bigg\|E[x_{p}(t)]-\int_{[0,1]}E[ x_{q}(t)]dq\bigg\|^{2}dp =0$$ and (\ref{qiwangjiftyj}) for (\ref{boundedsecmom}).
By $\mu_{t,q}=\mathcal{L}\left(x_{q}(t)\right)$ in (\ref{Mckeanvlasov1}), the  system (\ref{Mckeanvlasov1})  can be written as
\begin{align}
 dx_p(t) =&\alpha_{1}(t)\int_{[0,1]}A(p,q)\left(E[x_{q}(t)]-x_p(t)\right)dqdt\notag\\
&-\alpha_{2}(t)\nabla_{x} V(p,x_p(t))dt-\alpha_{2}(t)\Sigma_{1} dw_p(t).\label{exMckean1}
\end{align}
By Assumption \ref{assumptiondescent1} and Theorem 2.3.1 in \cite{Boksendal}, we have $E\big[\int_{0}^{t}\alpha_{2}(s)  \Sigma_{1} dw_{p}(s)\big]=0$. This together with (\ref{exMckean1}) gives
\begin{align}
  dE[x_{p}(t)]
 =&\alpha_{1}(t)\int_{[0,1]}A(p,q)\left(E[x_{q}
 (t)]-E[x_{p}(t)]\right)dqdt  -\alpha_{2}(t)E\left[\nabla_{x}V
 \left(p,x_{p}(t)\right)\right]dt .\label{parameterexpecderi}
 \end{align}
Denote   $Y(t)=\int_{[0,1]}E\big[ \|x_{p}(t) \|^{2}\big]dp$ and $R(t)=\int_{[0,1]}\|E[x_{p}(t)]-\int_{[0,1]} E[x_{q}(t)]dq\|^{2}dp$.  By Assumption \ref{assumptiondescent1},  Corollary 3.1  in \cite{Y. Chen}
  and Theorem 2.3.1 in \cite{Boksendal}, we have $E\big[\int_{0}^{t} 2\alpha_{2}(s) x_{p}^{\mathsf{T}}(s)  \Sigma_{1}  dw_{p}(s)\big]=0$. Then, by It\^{o} formula and (\ref{exMckean1}), we have
\begin{align}
 dY(t)
=&2 \alpha_{1}(t)\int_{[0,1]\times[0,1]} A(p,q)\big(E[x_{p}^{\mathsf{T}}(t)]E[x_{q}(t)]
-E\big[ \|x_{p}(t) \|^{2}\big]\big)dqdp dt\notag\\
&+\alpha_{2}^{2}(t)\operatorname{Tr}(\Sigma_{1}^{\mathsf{T}}
\Sigma_{1}) dt -2\alpha_{2}(t)\int_{[0,1]}  E\left[x_{p}^{\mathsf{T}}(t) \nabla_{x} V(p,x_{p}(t))\right]dp dt.\label{expectationsquare}
\end{align}
By Assumption \ref{assumption0} (iii), we know that $$(x_{p}(t)-x_{p}(0))^{\mathsf{T}}  (\nabla_{x} V(p,x_{p}(t))-\nabla_{x} V(p,x_{p}(0))) \geqslant \kappa_{2}\|x_{p}(t) -  x_{p}(0)\|^{2}.$$
 Let $\|\cdot\|_{E}=E\big[ \|\cdot \|^{2}\big]^{\frac{1}{2}}$. Then, by Cauchy-Schwarz inequality and H{\"o}lder inequality, we have
\begin{align}
&-2\alpha_{2}(t) E \big[ x_{p}^{\mathsf{T}}(t)\nabla_{x}V(p,x_{p}(t)) \big]\notag\\
\leqslant & 2\alpha_{2}(t)  \Big(- \kappa_{2} E\big[\|x_{p}(t)\|^{2}\big] +
 \|x_{p}(t) \|_E  \|\nabla_{x}V(p,x_{p}(0))\|_E\notag\\
 &+ \|x_{p}(0)\|_E ( \|\nabla_{x}V(p,x_{p}(t))\|_E
   +       \|\nabla_{x}V(p,x_{p}(0))\|_E) +2\kappa_{2} \|x_{p}(t)\|_E \|x_{p}(0) \|_E  \Big).\notag
\end{align}
By Assumption \ref{assumption0} (ii) and $C_{r}$ inequality, we have $$\|\nabla_{x}V(p,x_{p}(t))\|^{2}  \leqslant 3 \big(\kappa^{2}\|x_{p}(t)\|^{2} + \|\nabla_{x}V(p,x_{p}(0))\|^{2}+  \kappa^{2}\|x_{p}(0)\|^{2}\big).$$ This together with the above inequality, $C_{r}$  inequality, Assumption \ref{assumption0} (ii) and Assumption \ref{assumptiondescent02} gives
\begin{align}
&-2\alpha_{2}(t) E\big[ x_{p}^{\mathsf{T}}(t)\nabla_{x}V(p,x_{p}(t))\big]\notag\\
\leqslant &2\alpha_{2}(t) \|x_{p}(t)\|_E\|\nabla_{x}V(p,x_{p}(0))\|_E
-2\kappa_{2}\alpha_{2}(t) E\|x_{p}(t)\|^{2}\notag\\
&
+2\alpha_{2}(t) E\big[ 3\kappa^{2} \left\|x_{p}(t)\right\|^{2}+3\left\|\nabla_{x}V(p,x_{p}(0))\right\|^{2}\notag\\
&+3\kappa^{2}\left \|x_{p}(0)\right\|^{2}\big]^{\frac{1}{2}}\|x_{p}(0)\|_E +4\alpha_{2}(t)\kappa_{2}\|x_{p}(t)\|_E  \|x_{p}(0) \|_E \notag\\
  &+2\alpha_{2}(t) \|x_{p}(0)\|_E \|\nabla_{x}V(p,x_{p}(0))\|_E  \notag\\
\leqslant
& -2\kappa_{2}\alpha_{2}(t) E\big[\|x_{p}(t)\|^{2}\big]+\Big(2\big(2\sigma_{v}^{2}\zeta_{2}
+2C_{v}^{2}\big)^{\frac{1}{2}}+
2\sqrt{3}\kappa\zeta_{2}^{\frac{1}{2}}  +4\kappa_{2}
\zeta_{2}^{\frac{1}{2}} \Big)\alpha_{2}(t)
\|x_{p}(t)\|_E\notag\\
& +\Big(2(\sqrt{3}+1) \zeta_{2}^{\frac{1}{2}} \big(2\sigma_{v}^{2}\zeta_{2}+2C_{v}^{2}\big)^{\frac{1}{2}}  +2\sqrt{3}\zeta_{2} \kappa \Big)\alpha_{2}(t).\label{neijivhed}
\end{align}
By Assumption \ref{assumption0} (i) and (\ref{Algebricconnectibalance}), we have
\begin{align}
&2 \alpha_{1}(t)\int_{[0,1]\times [0,1]} A(p,q)\big(E[x_{p}^{\mathsf{T}}(t)]
E[x_{q}(t)]-E\big[\|x_{p}(t)\|^{2}\big]\big)dqdp\notag\\
\leqslant & -2\alpha_{1}(t)\lambda_{2}(\mathbb{L}_{A})R(t)
\leqslant  0.\label{alinter}
\end{align}
This together with (\ref{expectationsquare})-(\ref{neijivhed})  gives
\begin{align}
 \frac{dY(t)}{dt}
 \leqslant  -l_{1}(t)Y(t)+l_{2}(t)Y^{\frac{1}{2}}(t) +l_{3}(t),\notag
\end{align}
where $l_{1}(t)=2\kappa_{2}\alpha_{2}(t) $, $l_{2}(t)=2\big((2\sigma_{v}^{2}\zeta_{2}+2C_{v})^{\frac{1}{2}}+
 \sqrt{3}\kappa \zeta_{2}^{\frac{1}{2}}+2\kappa_{2}\zeta_{2}^{\frac{1}{2}}\big)\alpha_{2}(t)$ and $l_{3}(t)=2\big(\left(\sqrt{3}+1\right) \\ \zeta_{2}^{\frac{1}{2}} (2\sigma_{v}^{2}\zeta_{2}+2C_{v}^{2})^{\frac{1}{2}}+ \sqrt{3}\zeta_{2} \kappa\big)\alpha_{2}(t)+\alpha_{2}^{2}(t)\operatorname{Tr}
\left(\Sigma_{1}^{\mathsf{T}}\Sigma_{1}\right)$.
By Assumption \ref{assumption0} (iii), Assumption \ref{assumptiondescent1} and Lemma \ref{seconesstmate}, we have
\begin{align}
Y(t)\leqslant &   \max \Bigg\{Y(0),\Bigg(\sup _{0\leqslant s \leqslant t} \frac{l_{2}(s)}{2 l_{1}(s)}+\bigg(\sup _{0\leqslant s \leqslant t}\frac{1}{4} \frac{l_{2}^{2}(s)}{l_{1}^{2}(s)} +\sup _{0 \leqslant s \leqslant t} \frac{l_{3}(s)}{l_{1}(s)}\bigg)^{\frac{1}{2}}\Bigg)^{2} \Bigg\}.\label{boundedsquarey}
 \end{align}
By Assumption \ref{assumptiondescent02}, we get $Y(0)= \int_{[0,1]}E\left[\|x_{p}(0)\|^{2}\right]\leqslant \sup_{p\in [0,1]} E\left[\|x_{p}(0)\|^{2}\right] \leqslant \zeta_{2}.$ By Assumption
\ref{assumptiondescent1}, we have
 $$ \lim\limits_{t \rightarrow \infty}\frac{l_{2}(t)}{2 l_{1}(t)}=\frac{1}{2\kappa_{2}}\Big(\big(2\sigma_{v}^{2}\zeta_{2}+C_{v}^{2}\big)^{\frac{1}{2}}+
(\sqrt{3}\kappa+2\kappa_{2})\zeta_{2}^{\frac{1}{2}}\Big),$$  $$ \lim_{t \rightarrow \infty}\frac{1}{4} \frac{l_{2}^{2}(t)}{l_{1}^{2}(t)}=\frac{1}{4\kappa_{2}^{2} }\Big( \big(2\sigma_{v}^{2}\zeta_{2}+2C_{v}^{2}\big)^{\frac{1}{2}}+
 (\sqrt{3}\kappa +2\kappa_{2})\zeta_{2}^{\frac{1}{2}}\Big)^{2}
$$  and $$\lim_{t \rightarrow \infty}  \frac{l_{3}(t)}{l_{1}(t)}=\frac{\zeta_{2}^{\frac{1}{2}}}{\kappa_{2}}\big( (\sqrt{3}+1)\big(2\sigma_{v}^{2}\zeta_{2}+2C_{v}^{2}\big)^{\frac{1}{2}}+\sqrt{3}\kappa
\zeta_{2}^{\frac{1}{2}}\big).$$
By the above three equalities, there exist non-negative constants
$M_{1}, \ M_{2}$ and $M_{3}$, such that $\sup_{t\geqslant 0}\frac{l_{2}(t)}{2 l_{1}(t)}\leqslant M_{1}$, $\sup_{t\geqslant 0}\frac{1}{4} \frac{l_{2}^{2}(t)}{l_{1}^{2}(t)}\leqslant M_{2}$ and $\sup_{t\geqslant 0}\frac{l_{3}(t)}{l_{1}(t)}\leqslant M_{3}$. Denote $K=\max\big\{\zeta_{2}, (\sqrt{M_{2}+M_{3}} \\ +M_{1})^{2}\big\}.$ Then, by (\ref{boundedsquarey}), we have $\sup_{t \geqslant 0 }Y(t) \leqslant K$. Then, similar to the proof of (\ref{boudeforjoint}) in Theorem \ref{jointdecoupleconsensus} and by Assumption \ref{assumption0} (i), we have  (\ref{qiwangjiftyj}).

Combining (\ref{parameterexpecderi}) and the symmetry of the graphon $A$ gives 
 \begin{align}
 \frac{d\big(\int_{[0,1]}E[x_{p}(t)]dp\big)}{dt}=-\alpha_{2}(t)
\int_{[0,1]}E\left[\nabla_{x}V
 \left(p,x_{p}(t)\right)\right]dp.\notag
 \end{align}
This together with (\ref{parameterexpecderi})  gives
\begin{align}
&\frac{dR(t)}{dt}\notag \\
=& 2\alpha_{1}(t)\int_{ [0,1]} E[x^{\mathsf{T}}_{p}(t)]\int_{[0,1]} A(p,q) (E[x_{q}
 (t)]-E[x_{p}(t)] )dqdp
  \notag \\
  &+2\alpha_{2}(t)\int_{ [0,1]} \bar{x}_{p}(t)^{\mathsf{T}} \bigg(\int_{[0,1]}E[\nabla_{x}V(q,x_{q}(t))]dq
  -E[\nabla_{x}V(p,x_{p}(t))]\bigg)dp=\bar{J}_{1}(t)+\bar{J}_{2}(t),\label{threetermR}
\end{align}
where $ \bar{x}_{p}(t)=E[x_{p}(t)]-\int_{[0,1]}E[ x_{q}(t)]dq$. Combining Assumption \ref{assumption0} (ii), $C_{r}$ inequality, Jensen inequality, H{\"o}lder inequality and (\ref{qiwangjiftyj}) gives
\begin{align}
&J_2(t)\notag\\
\leqslant & 2\alpha_{2}(t)\int_{ [0,1]} \|\bar{x}_{p}(t)\|\bigg\|\int_{[0,1]}E\left[\nabla_{x}V(q,x_{q}(t))\right]dq
 -E\left[\nabla_{x}V(p,x_{p}(t))\right]\bigg\|dp\notag\\
\leqslant
& 4\sqrt{2}\alpha_{2}(t)R^{\frac{1}{2}}(t)
 \bigg(\int_{[0,1]}
\big\|E\left[\nabla_{x}V(p,x_{p}(t))\right]\big\|^{2}dp
\bigg)^{\frac{1}{2}} \notag\\
\leqslant &4\sqrt{2}\alpha_{2}(t)R^{\frac{1}{2}}(t) \bigg(\int_{[0,1]}E\big[\|\nabla_{x}V(p,x_{p}(t))\|^{2}
\big]dp\bigg)^{\frac{1}{2}}
\notag\\
\leqslant &8\alpha_{2}(t)R^{\frac{1}{2}}(t) \bigg( \sigma_{v}^{2}\int_{[0,1]}E\big[\|x_{p}(t)
\|^{2}\big]dp+ C_{v}^{2}\bigg)^{\frac{1}{2}}
\notag\\
\leqslant &8\alpha_{2}(t)R^{\frac{1}{2}}(t)\big(
\sigma_{v}K_{0}^{\frac{1}{2}}+ C_{v}\big).\notag
\end{align}
Then, by  (\ref{Algebricconnectibalance}), (\ref{threetermR}) and the above inequality, we have
\begin{align}
\frac{dR(t)}{dt} \leqslant  8\alpha_{2}(t)R^{\frac{1}{2}}(t)(\sigma_{v} K_{0}^{\frac{1}{2}} + C_{v})-2\alpha_{1}(t)\lambda_2(\mathbb{L}_{A})R(t).\label{nearcompac}
\end{align}
By (\ref{qiwangjiftyj}) and Jensen  inequality, we get  $R(t)
\leqslant \int_{[0,1]}\|E[x_{p}(t)]\|^{2}dp \leqslant \int_{[0,1]}E\left[\|x_{p}(t)\|^{2}\right]dp \leqslant K_{0},$
and then $R^{\frac{1}{2}}(t)  \leqslant K_{0}^{\frac{1}{2}}$. This together with (\ref{nearcompac}) gives
\begin{align}
\frac{dR(t)}{dt} \leqslant -2\alpha_{1}(t)\lambda_2(\mathbb{L}_{A})R(t)+ 8\alpha_{2}(t)(\sigma_{v} K_{0}+ C_{v} K_{0}^{\frac{1}{2}}),\notag
\end{align}
which together with the comparison theorem (\cite{Michel A. N.}) gives
\begin{align}
R(t) \leqslant   \Psi_{0}(0,t)R(0)  +
\int_{0}^{t}\Big(\big(\sigma_{v}K_{0}+ C_{v} K_{0}^{\frac{1}{2}}\big) 8
\alpha_{2}(s)\Psi_{0}(s,t)
\Big)ds.\label{estimateconlast}
\end{align}
This together with  $R(0) \leqslant \zeta_{2} <\infty$ gives
 (\ref{seppedofcons}). Then,
by  (\ref{estimateconlast}), Assumption \ref{assumptiondescent1} and L'Hospital's rule, we have
\begin{align}
 \lim_{t \rightarrow \infty}  \bigg[\Psi_{0}(0,t)R(0)   +
\int_{0}^{t}\Big(\big(\sigma_{v}K_{0}+ C_{v} K_{0}^{\frac{1}{2}}\big) 8
\alpha_{2}(s)\Psi_{0}(s,t)
\Big)ds\bigg]=0.\notag
\end{align}
This together with (\ref{qiwangjiftyj}) and Theorem \ref{jointdecoupleconsensus} leads to (\ref{boundedsecmom}).
$\hfill\blacksquare$
\vskip 1.5mm

To prove Lemma \ref{meanoptimal}, we need the following lemma,  the proof of which is directly from Lemma \ref{jointvariance}.
\vskip 1.5mm
\begin{lemma}\label{variancediminish}
For the problem (\ref{globalgoal}) and the algorithm (\ref{Mckeanvlasov1}), if Assumption \ref{assumption0} and Assumptions \ref{assumptiondescent02}-\ref{assumptiondescent1} hold, then $\lim_{t \rightarrow \infty} \sup_{p \in [0,1]} E\big[\|x_{p}(t)-E[x_{p}(t)]\|^{2}\big]=0.$
\end{lemma}

\vskip 1mm

\emph{Proof of Lemma \ref{meanoptimal}:}
Denote $L(t)=\|\int_{[0,1]}E[x_{p}(t)]dp
  -x^{*}\|^{2}$, $L_{1}(t)=\int_{[0,1]}E[x_{p}(t)]dp$,  $L_{2}(t) = \sup\limits_{p \in [0,1]} E\big[\|x_{p}(t)
-E[x_{p}(t)]\|^{2}\big]$ and $L_{3}(t)= \sup\limits_{p \in [0,1]}\|E[x_{p}(t)]
-L_{1}(t)\|^{2}$. Noting that
 $$\nabla_{x}V(x^{*})=\int_{[0,1]}\nabla_{x}V(p,x^*)dp=0$$
 and by the symmetry of the graphon $A$ and (\ref{parameterexpecderi}) in Lemma 2.2, we have 
\begin{align}
 \frac{dL(t)}{dt}
=& 2\alpha_{2}(t)\left(x^{*}-L_{1}(t)\right)^{\mathsf{T}}
\Big(\int_{[0,1]} \widetilde{V}(p,t)  +\nabla_{x}V\left(p,L_{1}(t)\right)
-\nabla_{x}V(p,x^*)\big)dp\Big),\notag
\end{align}
where $\widetilde{V}(p,t)=E\left[\nabla_{x}V(p,x_{p}(t))\right] -\nabla_{x}V\left(p,L_{1}(t)\right)$.
 This together with Assumption \ref{assumption0}, H{\"o}lder inequality and Jensen inequality gives
\begin{align}
& \frac{dL(t)}{dt}\notag\\
\leqslant &  2\alpha_{2}(t)\bigg(L^{\frac{1}{2}}(t)
\bigg\|\int_{[0,1]} \widetilde{V}(p,t)dp\bigg\|- \kappa_{2} L(t)\bigg)\notag\\
\leqslant &2\alpha_{2}(t)\bigg( -\kappa_{2}L(t)+  \Big(\int_{[0,1]}
\|\widetilde{V}(p,t)\|^{2}dp
\Big)^{\frac{1}{2}}L^{\frac{1}{2}}(t) \bigg)\notag\\
\leqslant & 2\alpha_{2}(t)\Bigg(
\sqrt{2}L^{\frac{1}{2}}(t)\bigg(\int_{[0,1]}
\|E\left[\nabla_{x}V(p,x_{p}(t))\right]
 -\nabla_{x}V(p,E[x_{p}(t)])\|^{2}dp\bigg)^{\frac{1}{2}}- \kappa_{2} L(t)
\notag\\
&+ \sqrt{2} L^{\frac{1}{2}}(t)\bigg(\int_{[0,1]} \|\nabla_{x}
V(p,E[x_{p}(t)])
-\nabla_{x}V(p,L_{1}(t))
\|^{2}dp\bigg)^{\frac{1}{2}}\Bigg)\notag\\
\leqslant & 2\alpha_{2}(t)\Bigg( \sqrt{2}\kappa L^{\frac{1}{2}}(t)\Bigg(\sqrt{\int_{[0,1]} E\big[\|x_{p}(t)
-E[x_{p}(t)]\|^{2}\big]dp}
\notag\\
&+   \bigg(\int_{[0,1]} \|E[x_{p}(t)]
-L_{1}(t)\|^{2}dp\bigg)^{\frac{1}{2}}\Bigg)- \kappa_{2} L(t)\Bigg)\notag\\
\leqslant & 2\alpha_{2}(t)\big( -\kappa_{2}L(t)+ \sqrt{2}\kappa L^{\frac{1}{2}}(t)  \big(L^{\frac{1}{2}}_{2}(t)  +  L^{\frac{1}{2}}_{3}(t)\big)\big).\label{nearcomfopt}
\end{align}
By Lemma \ref{consensusstochagra}, $C_{r}$ inequality, H{\"o}lder inequality and Jensen inequality, we have $$L(t)\leqslant  2\big(\int_{[0,1]}E\big[\|x_{p}(t)\|^{2}\big]dp+\|x^{*}\|^{2}\big)
\leqslant  2(K_{0}+ \|x^{*}\|^{2})=:2\bar{C}
$$
 and $L^{\frac{1}{2}}(t)\leqslant  \sqrt{2\bar{C}}$, which together with (\ref{nearcomfopt}) gives
\begin{align}
 \frac{dL(t)}{dt}  \leqslant &
-2\kappa_{2}\alpha_{2}(t)L(t)+4\kappa\alpha_{2}(t) \sqrt{\bar{C}} \big(L^{\frac{1}{2}}_{2}(t) + L^{\frac{1}{2}}_{3}(t)\big).\notag
\end{align}
This together with the comparison theorem (\cite{Michel A. N.}) leads to
\begin{align}
L(t)
\leqslant& \psi_{2}(0,t)L(0)+4\kappa
\sqrt{\bar{C}}\int_{0}^{t}
\alpha_{2}(s)  L^{\frac{1}{2}}_{2}(s)
\psi_{2}(s,t)ds\notag\\
& +4\kappa
\sqrt{\bar{C}} \int_{0}^{t}
\alpha_{2}(s) L^{\frac{1}{2}}_{3}(s)
\psi_{2}(s,t)ds,\label{lastgraop}
\end{align}
where $\psi_{2}(s,t)=e^{-2\kappa_{2}\int_{s}^{t}\alpha_{2}(s^{\prime})ds^{\prime}}$.
For the first term on the r.h.s. of  (\ref{lastgraop}), by Assumption \ref{assumptiondescent02}, $C_{r}$ inequality and H{\"o}lder inequality, we have
 $$ L(0)=
\bigg\|\int_{[0,1]}E[x_{p}(0)]dp -x^{*}\bigg\|^{2}  \leqslant  2\bigg(\sup_{p \in [0,1]} \|E[x_{p}(0)] \|^{2} + \|x^{*} \|^{2}\bigg) \leqslant 2(\zeta_{2}+ \|x^{*} \|^{2}),$$
which together with Assumption
\ref{assumption0} (iii) and Assumption  \ref{assumptiondescent1} gives
\begin{align}
\lim_{t \rightarrow \infty}\psi_{2}(0,t)L(0)=0.\label{firstteofV}
\end{align}
For the second term on the r.h.s. of  (\ref{lastgraop}), by Assumption \ref{assumption0} (iii), Assumption \ref{assumptiondescent1},
Lemma \ref{variancediminish} and L'Hospital's rule, we have
\begin{align}
 \lim_{t \rightarrow \infty}4\kappa
\sqrt{\bar{C}}\int_{0}^{t}
\alpha_{2}(s)\psi_{2}(s,t) L^{\frac{1}{2}}_{2}(s) ds
=0.\label{seconteofV}
\end{align}
For the third term on the r.h.s. of (\ref{lastgraop}), by Assumption \ref{assumption0} (iii), Assumption \ref{assumptiondescent1},
 Lemma \ref{consensusstochagra} and L'Hospital's rule, we have
$$ \lim_{t \rightarrow \infty} 4\kappa
 \big(K_{0}+\|x^{*}\|^{2} \big)^{\frac{1}{2}}\int_{0}^{t}
\alpha_{2}(s)\psi_{2}(s,t) L^{\frac{1}{2}}_{3}(s) ds=0.$$
This together with (\ref{lastgraop})-(\ref{seconteofV})  gives $\lim\limits_{t \rightarrow \infty} L(t)=0. $ $\hfill\blacksquare$

\vskip  1mm
\emph{Proof of Theorem \ref{optimalstgra}:}
By $C_{r}$ inequality, we have
\begin{eqnarray*}
   &&\sup_{p \in [0,1]}E\big[ \|x_{p}(t)-x^{*} \|^{2}\big]\cr
&\leqslant&   3 \sup_{p \in [0,1]} E\big[ \|x_{p}(t)-E[x_{p}(t)] \|^{2}\big] +3\sup _{p \in [0,1]}\Big\|E[x_{p}(t)]
-\int_{[0,1]} E[x_{q}(t)]dq\Big\|^{2}\cr
&&+3 \Big\|\int_{[0,1]}E[x_{q}(t)]dq-x^{*}\Big\|^{2}.
\end{eqnarray*}
This together with Lemmas \ref{consensusstochagra}-\ref{meanoptimal} and Lemma \ref{variancediminish}  gives (\ref{extoopt}).   $\hfill\blacksquare$

\setcounter{equation}{0}
\renewcommand{\theequation}{C.\arabic{equation}}
\setcounter{lemma}{0}
\renewcommand{\thelemma}{C.\arabic{lemma}}
\section{}

\emph{Proof of Lemma \ref{decotwotimescale}:}
 By $\lim_{t \to \infty} {Y}_{3}(t)=0,$ we know that there exists  $N_{1} \geqslant 0$, such that
 \begin{align}
 \sup_{t\geqslant 0} Y_{3}(t)\leqslant N_{1}.\label{boundofY3}
 \end{align}
 Then, by (\ref{fasttime}), we have
 \begin{align}
 \frac{dY_{2}(t)}{dt} \leqslant  -b_{1}(t)Y_{2}(t)+b_{2}(t)Y_{2}^{\frac{1}{2}}(t)
 \big(Y_{1}^{\frac{1}{2}}(t)+N_{1}^{\frac{1}{2}}\big).\label{dairuboutoy2}
 \end{align}
By  $ \sup_{t \geqslant 0}\frac{b_{2}(t)}{b_{1}(t)}<\infty$ and $\sup_{t\geqslant 0}Y_{1}(t) <\infty$, we know that there exist  $N_{2},\ N_{3} \geqslant 0$, such that $$\sup_{t\geqslant}\frac{b_{2}(t)}{b_{1}(t)}\leqslant N_{2}$$ and $\sup_{t\geqslant 0}Y_{1}(t)\leqslant N_{3}$.
  Then, by $b_{1}(t)>0,$ (\ref{dairuboutoy2}), Lemma \ref{seconesstmate} and $C_{r}$ inequality, we have
 \begin{align}
Y_{2}(t)
\leqslant  & Y_{2}(0)+\sup_{0\leqslant s \leqslant t} \left(\frac{b_{2}(s)}{b_{1}(s)}
  \left(Y_{1}^{\frac{1}{2}}(s)+  N_{1}
\right)\right)^{2}\notag\\
\leqslant &Y_{2}(0)+
2 N_{2}^{2}(N_{3} +N_{1}^2),\   \forall \ t \  \geqslant 0.\label{boundofy2}
 \end{align}
This together with (\ref{slowtime}) gives
\begin{align}
\frac{dY_{1}(t)}{dt} \leqslant & \big(-a_{1}(t)+a_{2}(t)\big)Y_{1}(t) +(Y_{2}(0)+
2 N_{2}^{2}(N_{3} +N_{1}^2))a_{3}(t) +a_{4}(t).\label{twotime11}
\end{align}
By $a_{1}(t)>0,$ $a_{2}(t),\ a_{3}(t) \geqslant 0$ and
$\lim_{t\to\infty} \frac{a_{2}(t)}{a_{1}(t)}=0$, we know that there exists $T\geqslant 0$, such that if $t \geqslant T$, then $-a_{1}(t)+a_{2}(t)<0.$ Then, by (\ref{twotime11}) and  comparison theorem (\cite{Michel A. N.}), we have
\begin{align}
Y_{1}(t)\leqslant &\int_{T}^{t} \psi_{3}(s,t) ((Y_{2}(0)+
2 N_{2}^{2}(N_{3} +N_{1}))a_{3}(s)  +a_{4}(s))ds
 +\psi_{3}(T,t)Y_{1}(T),\ \forall \ t \geqslant T,\label{twotimese0}
\end{align}
where $\psi_{3}(s,t)=e^{\int_{s}^{t}(-a_{1}(s^{\prime})+a_{2}(s^{\prime}))ds^{\prime}}$.
By $\int_{0}^{\infty} a_{1}(t)dt=\infty$ and $\lim_{t\to\infty} \frac{a_{2}(t)}{a_{1}(t)}=0$, we have
\begin{align}
\int_{T}^{\infty}(-a_{1}(s)+a_{2}(s))ds=-\infty.\label{suma1infty}
\end{align}
For the first term on the  r.h.s. of (\ref{twotimese0}), by (\ref{suma1infty}), $\lim_{t \to \infty}\frac{a_{2}(t)}{a_{1}(t)}=0$, $\lim_{t \to \infty}\frac{a_{3}(t)}{a_{1}(t)}=0$,   $\lim_{t \to \infty}\frac{a_{4}(t)}{a_{1}(t)}=0$ and  L'Hospital's rule, we have
\begin{align}
\lim_{t \to \infty}\int_{T}^{t} &\psi_{3}(s,t) ((Y_{2}(0)+
2 N_{2}^{2}(N_{3} +N_{1}^2)) a_{3}(s) +a_{4}(s))ds=0.\label{twotimese1}
\end{align}
Noting that $Y_{1}(t)$ is continuous w.r.t. $t$, then we have $Y_{1}(T)<\infty$.
Then, for the second term on the r.h.s. of (\ref{twotimese0}), by (\ref{suma1infty}), we have
$
\lim_{t \to \infty}\psi_{3}(T,t)Y_{1}(T)=0.$
 This together with  (\ref{twotimese0}) and (\ref{twotimese1}) gives (\ref{0slowtime}).
 By (\ref{fasttime}), (\ref{boundofy2}) and comparison theorem (\cite{Michel A. N.}), we have
 \begin{align}
   Y_{2}(t)
\leqslant& e^{-\int_{0}^{t}b_{1}(s)ds }Y_{2}(0)+\int_{0}^{t} e^{-\int_{s}^{t}b_{1}(s^{\prime})ds^{\prime} }b_{2}(s) (Y_{2}(0)\notag\\
 &+
2 N_{2}^{2}(N_{3} +N_{1}^2) )^{\frac{1}{2}}
 \big(Y_{1}^{\frac{1}{2}}(s)+Y_{3}(s)\big)ds.\label{twotimese001}
 \end{align}
For the first term on the  r.h.s. of (\ref{twotimese001}), by $\int_{0}^{\infty}b_{1}(t)dt=\infty$, we have
\begin{align}
 \lim_{t \to \infty} e^{-\int_{0}^{t}b_{1}(s)ds }Y_{2}(0)=0.\label{twotimese0011}
\end{align}
By  (\ref{0slowtime}), we have $\lim_{t \to \infty}  Y_{1}^{\frac{1}{2}}(t) =0$. Then, for the second term on the r.h.s. of (\ref {twotimese001}), by $\lim_{t \to \infty}Y_{3}(t)=0$, $\sup_{t\geq 0}\frac{b_{2}(t)}{b_{1}(t)}<\infty$ and   L'Hospital's rule, we have $$\lim_{t \to \infty} \int_{0}^{t} e^{-\int_{s}^{t}b_{1}(s^{\prime})ds^{\prime} }b_{2}(s)\left(Y_{2}(0)+
2 N_{2}^{2}(N_{3} +N_{1})\right)^{\frac{1}{2}}
 (Y_{1}^{\frac{1}{2}}(s)+Y_{3}(s) )ds=0.$$
This together with (\ref{twotimese001}) and (\ref{twotimese0011}) gives (\ref{1slowtime}). $\hfill\blacksquare$
\vskip 1.5mm

Before we prove  Lemma \ref{sigleconseussss}, we need the following lemma whose proof is directly from  Lemma \ref{jointvariance}.

\vskip 1.5mm
\begin{lemma}\label{variancetenzero}
For the problem (\ref{globalgoal}) and the  algorithm (\ref{disgremck}), if Assumption \ref{assumption0} and  Assumptions \ref{assumption3}-\ref{assumption5} hold, then
\begin{align}
&\lim_{t \rightarrow \infty} \sup_{p \in [0,1]} E\big[ \|z_{p}(t)-E [z_{p}(t) ] \|^{2}\big]=0, \notag\\
&\lim_{t \rightarrow \infty} \sup_{p \in [0,1]} E\big[ \|\widetilde{y}_{p}(t)-E [\widetilde{y}_{p}(t) ] \|^{2}\big]=0.\notag
\end{align}
\end{lemma}
\vskip 1.5mm

\emph{Proof of Lemma \ref{sigleconseussss}:}
Denote $$ \mathcal{R}(t)=\int_{[0,1]}\bigg\|E[z_{p}(t)]-\int_{[0,1]}
E[z_{q}(t)]dq\bigg\|^{2}dp,$$ $\widetilde{\mathcal{R}}(t)=\int_{[0,1]}\|E[\widetilde{y}_{p}(t)]\|^{2}dp,$ $\mathcal{S}(t) = \|x^{*} -\int_{[0,1]}E[z_{p}(t)]dp \|^{2},$ $R_{1}(t)=\int_{[0,1]}E\big[z_{p}(t)\big]dp,$ $B(t)=\int_{[0,1]}E\big[\|z_{p}(t) -E [z_{p}(t) ]\|^{2}\big]dp$ and $\widetilde{\mathcal{Y}}(t)=\mathcal{R}(t)+\widetilde{\mathcal{R}}(t).$

At first, we prove
\begin{align}\label{ybounded}
 \sup_{t\geqslant 0} \widetilde{\mathcal{Y}}(t)
    <\infty.
\end{align}
 Noting that $\mu_{t,q}$ and $\widetilde{\nu}_{t,q}$ are the distributions of $z_{p}(t)$  and $\widetilde{y}_{p}(t)$, respectively and by Assumption
 \ref{assumption5},  we have
 \begin{align}
 &\frac{d E [z_{p}(t) ]}{dt}\notag\\
 =
&\beta_{3}(t)\int_{[0,1]}
A(p,q)\left(E\left[z_{q}(t)\right]-E\left[z_{p}(t)\right]\right)dq\notag\\
&-\beta_{1}(t)\beta_{3}(t)\int_{[0,1]}A(p,q)
\Big(E\left[\widetilde{y}_{q}(t)\right]-E\left[\widetilde{y}_{p}(t)\right]\notag\\
&+\beta_{2}(t)\big( E[\nabla_{x} V(q,z_{q}(t))]-E[\nabla_{x} V\left(p,z_{p}(t)\right)]\big) \Big)dq\notag\\
&-\beta_{1}(t)\left(E\left[\widetilde{y}_{p}(t)\right]
+\beta_{2}(t)E\left[\nabla_{x} V\left(p,z_{p}(t)\right)\right]\right) \label{deriaofezz}
 \end{align}
 and
 \begin{align}
 \frac{d E\left[\widetilde{y}_{p}(t)\right]}{dt}=& \beta_{3}(t)\int_{[0,1]}  A(p,q)\big(E\left[\widetilde{y}_{q}(t)\right]
-E\left[\widetilde{y}_{p}(t)\right]\big)dq\notag\\
 &  +\beta_{2}(t)\beta_{3}(t)\int_{[0,1]}A(p,q)\big(E[\nabla_{x} V(q,z_{q}(t))]
 -E[\nabla_{x} V\left(p,z_{p}(t)\right)]\big)dq.\label{deriaofezy}
 \end{align}
By  (\ref{MCKEANTRACKING}) and Assumption \ref{assumption3}, we have $E\left[\widetilde{y}_{p}(0)\right]=E\left[y_{p}(0)\right]-\beta_{2}(0)E\left[\nabla_{x} V(p,z_{p}(0))\right]=0$, which together with the above equality and   the symmetry of the graphon $A$ gives
\begin{align}
\int_{[0,1]}E\left[\widetilde{y}_{p}(t)\right]dp
=0.\label{meanyiszero}
\end{align}
Then, by (\ref{deriaofezz}), we have
\begin{align}
&\frac{d\mathcal{R}(t)}{dt}\notag\\
=&-2\beta_{1}(t)\int_{[0,1]}\big(E[z_{p}(t)]
-R_{1}(t)\big)^{\mathsf{T}}
E[\widetilde{y}_{p}(t)]dp\notag\\
&  -2\beta_{1}(t) \beta_{2}(t)\int_{[0,1]}\big(E[z_{p}(t)]
-R_{1}(t)\big)^{\mathsf{T}}
E[\nabla_{x} V\left(p,z_{p}(t)\right)]dp \notag\\
&  +2\beta_{3}(t)\int_{[0,1]}E [z^{\mathsf{T}}_{p}(t) ]
\bigg(\int_{[0,1]}A(p,q)(E\left[z_{q}(t)\right]-E\left[z_{p}(t)\right])dq\bigg)dp
\notag\\
& -2\beta_{1}(t)\beta_{3}(t)\int_{[0,1]}E [z^{\mathsf{T}}_{p}(t) ]
\bigg(\int_{[0,1]}A(p,q)\big(E\left[\widetilde{y}_{q}(t)\right]-
E\left[\widetilde{y}_{p}(t)\right]\big)dq\bigg)dp
\notag\\
&-2\beta_{1}(t)\beta_{2}(t) \beta_{3}(t)\int_{[0,1]}E[z^{\mathsf{T}}_{p}(t)]  \bigg( \int_{[0,1] }A(p,q) \big( E\left[\nabla_{x} V\left(q,z_{q}(t)\right)\right]\notag\\
&-E[\nabla_{x} V\left(p,z_{p}(t)\right)] \big)dq\bigg)dp\notag\\
&=:D_1(t)+D_2(t)+D_3(t)+D_4(t)+D_5(t).\label{secondederiR}
\end{align}
By $C_{r}$ inequality, we have
\begin{align}
 D_1(t)\leqslant  \beta_{1}(t)(\mathcal{R}(t)+ \widetilde{\mathcal{R}}(t)). \label{firstderiRineq}
\end{align}
By Assumption  \ref{assumption0} (ii),  $C_{r}$ inequality, Cauchy-Shwarz inequality  and Jensen inequality, we have, for any $\tau_{1}>0$,
\begin{align}
&D_2(t)\notag\\
\leqslant &\beta_{1}(t)\beta_{2}(t)\bigg( \tau_{1}\int_{[0,1]}\left\|E\left[z_{p}(t)\right]
-R_{1}(t)\right\|^2dp + \frac{1}{\tau_{1}}  \int_{[0,1]}
\big\|E\left[\nabla_{x} V\left(p,z_{p}(t)\right)\right]\big\|^2dp\bigg)\notag\\
\leqslant & \beta_{1}(t)\beta_{2}(t)\bigg( \tau_{1} \mathcal{R}(t)+ \frac{2}{\tau_{1}}   \int_{[0,1]}
\big\|\nabla_{x} V(p,x^{*})\big\|^2dp \notag\\
& + \frac{2}{\tau_{1}}  \int_{[0,1]} \big\|E\left[\nabla_{x} V\left(p,z_{p}(t)\right)-\nabla_{x} V(p,x^{*}) \right]\big\|^2dp\bigg)\notag\\
\leqslant & \beta_{1}(t)\beta_{2}(t)\bigg( \tau_{1} \mathcal{R}(t)+ \frac{2}{\tau_{1}}  \int_{[0,1]}
\big\|\nabla_{x} V(p,x^{*})\big\|^2dp \notag\\
& +\frac{ 2}{\tau_{1}}  \int_{[0,1]} E\left[\left\|\nabla_{x} V\left(p,z_{p}(t)\right)-\nabla_{x} V(p,x^{*})\right\|^2\right]dp\bigg)\notag\\
\leqslant & \tau_{1}\beta_{1}(t)\beta_{2}(t)\mathcal{R}(t)+ \frac{4}{\tau_{1}} \beta_{1}(t)\beta_{2}(t) \left(\sigma_{v}^{2}\|x^{*}\|^{2}
+C_{v}^{2}\right) \notag\\
& +
\frac{4\kappa^2 \beta_{1}(t)\beta_{2}(t)}{\tau_{1}}\bigg(  \int_{[0,1]} E\big[\left\|x^{*} - E[z_{p}(t)]\right\|^2\big]dp  +
 B(t) \bigg)\notag\\
\leqslant & \beta_{1}(t)\beta_{2}(t)\bigg(\Big(\tau_{1}+
\frac{ 8}{\tau_{1}}\kappa^2  \Big)\mathcal{R}(t)+ \frac{4}{\tau_{1}}   \left(\sigma_{v}^{2}\|x^{*}\|^{2}
+C_{v}^{2}\right)    + \frac{8}{\tau_{1}}\kappa^2  \mathcal{S}(t)+
\frac{ 4}{\tau_{1}} \kappa^2 B(t)\bigg).
\label{secondderiRineq}
\end{align}
  By (\ref{Algebricconnectibalance}), we have
\begin{align}
D_3(t)\leqslant -2\beta_{3}(t)\lambda_2\left(\mathbb{L}_{A}\right)\mathcal{R}(t).\label{thiredderiRineq}
\end{align}
By the symmetry of graphon $A$, H{\"o}lder inequality and  $C_{r}$ inequality, we have
\begin{align}
&D_4(t)\notag\\
=& -2\beta_{1}(t)\beta_{3}(t)\int_{[0,1]} \Big(E\left[z_{p}(t)\right]-\int_{[0,1]}  E\left[z_{q}(t)\right]dq\Big)^{\mathsf{T}} \notag\\
&\times \bigg(
\int_{[0,1]}A(p,q)\big(E\big[\widetilde{y}_{q}(t)\big] -E\big[\widetilde{y}_{p}(t)\big]\big)dq\bigg)dp\notag\\
\leqslant&  \beta_{1}(t)\beta_{3}(t)\bigg(\mathcal{R}(t)+  \int_{[0,1]}\bigg\|
\int_{[0,1]}A(p,q)\big(E\big[\widetilde{y}_{q}(t)\big]  -E\big[\widetilde{y}_{p}(t)\big]\big)dq\bigg\|^2 dp\bigg)\notag\\
\leqslant&  \beta_{1}(t)\beta_{3}(t)\mathcal{R}(t)+2 \beta_{1}(t)\beta_{3}(t)\int_{[0,1]}
\int_{[0,1]}A(p,q) \big(\left\|E\left[\widetilde{y}_{q}(t)\right] \right\|^2+\left\|E\left[\widetilde{y}_{p}(t)\right]\right\|^2 \big)dqdp
\notag\\
\leqslant&
\beta_{1}(t)\beta_{3}(t)\mathcal{R}(t)+4 \beta_{1}(t)\beta_{3}(t)\widetilde{\mathcal{R}}(t).
\label{fourthderiRineq}
\end{align}
By Assumption \ref{assumption0} (ii), $C_{r}$ inequality and Jensen inequality, we have
\begin{align}
&D_5(t)\notag\\
\leqslant &
\bar{\beta}(t)\int_{[0,1]\times[0,1]}
\Big(\left\|E\left[\nabla_{x} V\left(q,z_{q}(t)\right)\right]\right\|\left\|E\left[z_{p}(t)\right]\right\|+
\left\|E\left[\nabla_{x} V\left(p,z_{p}(t)\right)\right]\right\| \left\|E\left[z_{p}(t)\right]\right\|\Big)dqdp\notag\\
\leqslant
 & \bar{\beta}(t)\bigg(\int_{[0,1]}
 \left\|E\left[z_{p}(t)\right]\right\|^{2}dp
 + \int_{[0,1]}\left\|E\left[\nabla_{x} V\left(p,z_{p}(t)\right)\right]\right\|^{2}dp\bigg)\notag\\
\leqslant & \bar{\beta}(t)\bigg(2 \int_{[0,1]}  \|\nabla_{x} V\left(p,E\left[z_{p}(t)\right]\right) \|^{2} dp  +\int_{[0,1]}\left\|E\left[z_{p}(t)\right]\right\|^{2}dp
+2 \int_{[0,1]}E\big[ \|\nabla_{x} V\left(p,z_{p}(t)\right) \notag\\
&  -\nabla_{x} V\left(p,E\left[z_{p}(t)\right]\right) \|^{2}\big]dp\bigg)\notag\\
\leqslant & \bar{\beta}(t)\bigg(\int_{[0,1]}\left\|E\left[z_{p}(t)\right]\right\|^{2}dp +2\kappa^{2} B(t) +2  \int_{[0,1]} \|\nabla_{x} V\left(p,E\left[z_{p}(t)\right]\right)\|^{2} dp\bigg)\notag\\
\leqslant &
\bar{\beta}(t)\bigg(\left(1+4\sigma_{v}^{2}\right)\int_{[0,1]} \left\|E\left[z_{p}(t)\right]\right\|^{2}dp
 +2\kappa^{2}B(t)
+4C_{v}^{2}\bigg)\notag\\
\leqslant &\bar{\beta}(t)\big(
3\left(1+4\sigma_{v}^{2}\right) (
\mathcal{R}(t)+  \mathcal{S}(t)+
 \|x^{*}\|^{2}) +4C_{v}^{2} +2\kappa^{2} B(t)\big),
 \label{fifthderiRineq}
\end{align}
where $\bar{\beta}(t)=2\beta_{1}(t)\beta_{2}(t)\beta_{3}(t)$.
Combining (\ref{secondederiR})-(\ref{fifthderiRineq}) gives
\begin{align}
&\frac{d\mathcal{R}(t)}{dt}\notag\\
\leqslant & \Big(\beta_{1}(t)+\left(\tau_{1}+
\frac{8}{\tau_{1}}\kappa^2  \right)\beta_{1}(t)\beta_{2}(t)
-2\lambda_2(\mathbb{L}_{A})\beta_{3}(t)\notag\\
&+\beta_{1}(t) \beta_{3}(t)+3\left(2+8\sigma_{v}^{2}\right)
\beta_{1}(t)\beta_{2}(t)\beta_{3}(t)\Big) \mathcal{R}(t) \notag\\
& +\Big(\beta_{1}(t) +4\beta_{1}(t)\beta_{3}(t)\Big)\widetilde{\mathcal{R}}(t)
+\Big(\frac{8}{\tau_{1}}\kappa^2
\beta_{2}(t)+3(2+8
\sigma_{v}^{2}) \beta_{2}(t)\beta_{3}(t)\Big)\beta_{1}(t) \mathcal{S}(t) \notag\\
&
+4\kappa^{2}\beta_{1}(t)\beta_{2}(t)\Big(\frac{1}{\tau_{1}}
 +\beta_{3}(t)\Big)  B(t)
+\Big(\frac{4}{\tau_{1}}\sigma^{2}_{v}\beta_{2}(t)
+3(2+8\sigma_{v}^{2})
 \beta_{2}(t)\beta_{3}(t)\Big)\beta_{1}(t)
  \|x^{*}\|^{2}\notag\\
& +C^{2}_{v}\beta_{1}(t)\beta_{2}(t)\left(\frac{4}{\tau_{1}}+8 \beta_{3}(t)\right).\label{derivofR}
\end{align}
By (\ref{deriaofezy}), we have
\begin{align}
&\frac{d \widetilde{\mathcal{R}}(t)}{dt}\notag\\
=& 2\beta_{3}(t)\int_{[0,1]\times[0,1]} A(p,q)\big(E[\widetilde{y}_{p}^{\mathsf{T}}(t)]
E\left[\widetilde{y}_{q}(t)\right]-E\big[\widetilde{y}^{\mathsf{T}}_{p}(t)\big]
E\left[\widetilde{y}_{p}(t)\right]\big)dqdp\notag\\
&  +2\beta_{2}(t)
\beta_{3}(t)\int_{[0,1]\times[0,1]}
A(p,q) \big(E\big[\widetilde{y}^{\mathsf{T}}_{p}(t)\big]E\left[\nabla_{x} V\left(q,z_{q}(t)\right)\right]-E\big[\widetilde{y}^{\mathsf{T}}_{p}(t)\big]\notag\\
&\times
E\left[\nabla_{x} V\left(p,z_{p}(t)\right)\right]\big)dqdp.\label{derivaofwidR}
\end{align}
For the first term on the r.h.s. of the above equality, combining  (\ref{Algebricconnectibalance}) and (\ref{meanyiszero}) gives
\begin{align}
&2\beta_{3}(t)\int_{[0,1]\times[0,1]}A(p,q)\Big(E\big[\widetilde{y}_{p}^{\mathsf{T}}(t)\big]
E\big[\widetilde{y}_{q}(t)\big]-E\big[\widetilde{y}^{\mathsf{T}}_{p}(t)\big]
E\big[\widetilde{y}_{p}(t)\big]\Big)dqdp\notag\\
 \leqslant&
-2\beta_{3}(t)\lambda_2(\mathbb{L}_{A})\widetilde{\mathcal{R}}(t).\label{firsttidy}
\end{align}
For the second term on the r.h.s. of (\ref{derivaofwidR}),   by  Assumption  \ref{assumption0} (ii), $C_{r}$ inequality, Cauchy-Shwarz inequality and Jensen inequality, we have
\begin{align}
&2\beta_{2}(t)\beta_{3}(t)\int_{[0,1]\times[0,1]}A(p,q)\big(E\big[\widetilde{y}^{\mathsf{T}}_{p}(t)\big]E\left[\nabla_{x} V\left(q,z_{q}(t)\right)\right] -E\big[\widetilde{y}^{\mathsf{T}}_{p}(t)\big]
E\left[\nabla_{x} V\left(p,z_{p}(t)\right)\right]\big)dqdp\notag\\
\leqslant& 2\beta_{2}(t)\beta_{3}(t)\Bigg(\int_{[0,1]\times[0,1]}
\|E\big[\widetilde{y}_{p}(t)\big]\|
\big(\left\|E\left[\nabla_{x} V\left(q,z_{q}(t)\right)\right]\right\|  +
\left\|E\left[\nabla_{x} V\left(p,z_{p}(t)\right)\right]\right\|\big) dqdp\Bigg)\notag\\
\leqslant& 2\frac{\beta_{2}(t)\beta_{3}(t)}{\beta_{1}(t)}\int_{[0,1]}
\|E\big[\widetilde{y}_{p}(t)\big]\|^2dp +2\beta_{1}(t)\beta_{2}(t)\beta_{3}(t)\int_{[0,1]}
\left\|E\left[\nabla_{x} V\left(q,z_{q}(t)\right)\right]\right\|^2dq\notag\\
\leqslant& 2\frac{\beta_{2}(t)\beta_{3}(t)}{\beta_{1}(t)} \widetilde{\mathcal{R}}(t) +2\beta_{1}(t)\beta_{2}(t)\beta_{3}(t)\int_{[0,1]}
E\left[\left\|\nabla_{x} V\left(q,z_{q}(t)\right)\right\|^2\right]dq\notag\\
\leqslant& 2\frac{\beta_{2}(t)\beta_{3}(t)}{\beta_{1}(t)} \widetilde{\mathcal{R}}(t) +4\sigma_{v}^{2}\beta_{1}(t)\beta_{2}(t)\beta_{3}(t)  \int_{[0,1]}
E\left[\left\| z_{q}(t) \right\|^2\right]dq+4C_{v}^{2}\beta_{1}(t)\beta_{2}(t)\beta_{3}(t) \notag\\
\leqslant& 2\frac{\beta_{2}(t)\beta_{3}(t)}{\beta_{1}(t)} \widetilde{\mathcal{R}}(t) +16\sigma_{v}^{2}\beta_{1}(t)\beta_{2}(t)\beta_{3}(t) (B(t)+ \mathcal{R}(t)  +\mathcal{S}(t)+ \|x^*\|^2 )
+4C_{v}^{2}\beta_{1}(t)\beta_{2}(t)\beta_{3}(t). \notag
\end{align}
This together with  (\ref{derivaofwidR})-(\ref{firsttidy}) gives
\begin{align}
 & \frac{d \widetilde{\mathcal{R}}(t)}{dt}\notag\\
 \leqslant & \left(-2\lambda_2(\mathbb{L}_{A})
 + 2\frac{\beta_{2}(t)}{\beta_{1}(t)} \right)\beta_{3}(t)\widetilde{\mathcal{R}}(t)
+\beta_{1}(t)\beta_{2}(t)\beta_{3}(t)\notag\\
& \times \big(16\sigma_{v}^{2}\left(B(t)+ \mathcal{R}(t)+\mathcal{S}(t)+ \|x^*\|^2\right)+4C_{v}^{2}\big).\label{derivoftideR}
\end{align}
   Combining   (\ref{derivofR}) and (\ref{derivoftideR}) leads to
\begin{align}
 \frac{d\widetilde{\mathcal{Y}}(t)}{dt}
\leqslant   m_{1}(t) \widetilde{\mathcal{Y}}(t)+m_{2}(t)\mathcal{S}(t)+m_{3}(t),\label{ASSinequaliofwidy}
\end{align}
where $m_{1}(t)= -2\lambda_2(\mathbb{L}_{A})\beta_{3}(t)+\beta_{1}(t)+ (\tau_{1}+
\frac{8}{\tau_{1}}\kappa^2   )\beta_{1}(t)   \beta_{2}(t) +5\beta_{1}(t) \beta_{3}(t)+ 2\frac{\beta_{2}(t)}{\beta_{1}(t)}  \beta_{3}(t)  + (6+40\sigma_{v}^{2} )
\beta_{1}(t)\beta_{2}(t) \beta_{3}(t)$, $m_{2}(t)= \frac{8}{\tau_{1}}\kappa^2\beta_{1}(t)
\beta_{2}(t)+(6+40
\sigma_{v}^{2})\beta_{1}(t) \beta_{2}(t)\beta_{3}(t)$ and $m_{3}(t)=  \big(\frac{4}{\tau_{1}}\kappa^{2}\\ \beta_{1}(t)   \beta_{2}(t)
+(4\kappa^{2}+16\sigma_{v}^{2})\beta_{1}(t)\beta_{2}(t) \beta_{3}(t)\big)  B(t)
 +\big(\frac{4}{\tau_{1}}\sigma^{2}_{v}\beta_{2}(t)
   + (6+40\sigma_{v}^{2})
 \beta_{2}(t)\beta_{3}(t)\big)\beta_{1}(t)
\|x^{*}\|^{2} +C^{2}_{v}\beta_{1}(t)\beta_{2}(t)( \frac{4}{\tau_{1}}+12 \beta_{3}(t)).$
 By (\ref{deriaofezz})-(\ref{meanyiszero}),  Assumption  \ref{assumption0} (ii), $C_{r}$ inequality, H{\"o}lder inequality and Jensen inequality, we have
\begin{align}
&\frac{d\mathcal{S}(t)}{dt}\notag\\
=&-2\beta_{1}(t)\beta_{2}(t) (R_{1}(t)-x^{*})^{\mathsf{T}}\int_{[0,1]}\big(E\left[\nabla_{x} V\left(p,z_{p}(t)\right)\right] -\nabla_{x} V\left(p,R_{1}(t)\right)\big)dp\notag\\
&  -2\beta_{1}(t)\beta_{2}(t)\left(R_{1}(t)-x^{*}\right)^{\mathsf{T}}\int_{[0,1]}\left(\nabla_{x} V\left(p,R_{1}(t)\right)-\nabla_{x} V(p,x^{*})\right)dp\notag\\
\leqslant & -2\kappa_{2}\beta_{1}(t)\beta_{2}(t)\mathcal{S}(t)+ 2\beta_{1}(t)\beta_{2}(t)\mathcal{S}^{\frac{1}{2}}(t)\notag\\
&\times\Big(\int_{[0,1]}
E\big[\|\nabla_{x} V\left(p,z_{p}(t)\right)-\nabla_{x} V(p,R_{1}(t))\|^{2}\big]dp\Big)^{\frac{1}{2}}\notag\\
\leqslant & -2\kappa_{2}\beta_{1}(t)\beta_{2}(t)\mathcal{S}(t)+ 2\beta_{1}(t)\beta_{2}(t)\mathcal{S}^{\frac{1}{2}}(t) \bigg(\int_{[0,1]}
\kappa^{2}E\big[\|z_{p}(t)- R_{1}(t)\|^{2}\big]dp\bigg)^{\frac{1}{2}}\notag\\
\leqslant & 2\beta_{1}(t)\beta_{2}(t)\Big(-\kappa_{2}\mathcal{S}(t)+ \sqrt{2}\kappa \mathcal{S}^{\frac{1}{2}}(t)\big(B(t)+\mathcal{R}(t)\big)^{\frac{1}{2}}\Big)\notag\\
\leqslant & 2\beta_{1}(t)\beta_{2}(t)\big(-\kappa_{2}\mathcal{S}(t)+ \sqrt{2}\kappa \mathcal{S}^{\frac{1}{2}}(t)
\big(B^{\frac{1}{2}}(t)+\mathcal{R}^{\frac{1}{2}}(t)\big)\big)\notag\\
\leqslant & 2\beta_{1}(t)\beta_{2}(t)\big( -\kappa_{2}\mathcal{S}(t) +  \sqrt{2}\kappa \mathcal{S}^{\frac{1}{2}}(t)
\big(B^{\frac{1}{2}}(t)  +  \widetilde{\mathcal{Y}}^{\frac{1}{2}}(t)\big)\big).\label{deriavteofS}
\end{align}
By the above inequality, Cauchy-Shwarz inequality and $C_r$ inequality, we have, for any $\tau_{2}>0$,
$$
    \frac{d\mathcal{S}(t)}{dt}
\leqslant \beta_{1}(t)\beta_{2}(t)\Big(
 (-2\kappa_{2}+\tau_{2} )\mathcal{S}(t) +  \frac{ 4 \kappa^2}{\tau_{2}  }
 B(t) +  \frac{ 4 \kappa^2}{\tau_{2}  }
 \widetilde{\mathcal{Y}}(t) \Big).
$$
This together with (\ref{ASSinequaliofwidy}) leads to
\begin{align}
&d (\widetilde{\mathcal{Y}}(t)+\mathcal{S}(t) )\notag\\
\leqslant & \Big(m_{1}(t) + \frac{ 4 \kappa^2}{\tau_{2}  }\beta_{1}(t)\beta_{2}(t)  \Big)  \widetilde{\mathcal{Y}}(t)+ ( m_{2}(t)+ (-2\kappa_{2}+\tau_{2} )\beta_{1}(t)\beta_{2}(t)  ) \mathcal{S}(t)\notag\\
&     +  m_{3}(t)+\frac{ 4 \kappa^2}{\tau_{2}  } \beta_{1}(t)\beta_{2}(t)
 B(t).\label{heqilaibound}
\end{align}
Let $\tau_{1}=\frac{16\kappa^2}{\kappa_{2}}, \ \tau_{2}=\frac{\kappa_{2}}{2}$ and we have $m_{2}(t)+ (-2\kappa_{2}+\tau_{2} )\beta_{1}(t)\beta_{2}(t)=- \kappa_{2}  \beta_{1}(t)
\beta_{2}(t)+(6+40
\sigma_{v}^{2})\beta_{1}(t) \beta_{2}(t)\beta_{3}(t).$
This together with (\ref{heqilaibound}) gives
\begin{align}
&d\big(\widetilde{\mathcal{Y}}(t)+\mathcal{S}(t)\big)\notag\\
\leqslant & \Big(m_{1}(t) + \frac{ 4 \kappa^2}{\tau_{2}  }\beta_{1}(t)\beta_{2}(t)  \Big)  \widetilde{\mathcal{Y}}(t)+\big( - \kappa_{2}  \beta_{1}(t)
\beta_{2}(t)+(6+40
\sigma_{v}^{2})\beta_{1}(t) \beta_{2}(t)\beta_{3}(t)  \big) \mathcal{S}(t) \notag\\
&+  m_{3}(t)+\frac{ 4\kappa^2}{\tau_{2}  } \beta_{1}(t)\beta_{2}(t)
 B(t).\label{heqilaibound1}
\end{align}
By Assumption \ref{assumption3}, we have $\beta_{1}(t)+\big(\tau_{1}+
\frac{8}{\tau_{1}}\kappa^2  +\frac{ 4 \kappa^2}{\tau_{2}}  \big)\beta_{1}(t) \beta_{2}(t) +5\beta_{1}(t) \beta_{3}(t)+ 2\frac{\beta_{2}(t)}{\beta_{1}(t)}  \beta_{3}(t)  + (6+40\sigma_{v}^{2} )
\beta_{1}(t)\beta_{2}(t)  \beta_{3}(t) = o(\beta_{3}(t) ), \ t \to \infty$
and
$ (6+40
\sigma_{v}^{2})\beta_{1}(t) \beta_{2}(t)\beta_{3}(t)=o(\beta_{1}(t)\beta_{2}(t)), \ t \to \infty.$
Then, we know that there exists $T_{1}>0$, such that if $t\geqslant T_{1}$, then we have
$ (m_{1}(t) + \frac{4 \kappa^2}{\tau_{2}  }\beta_{1}(t)\beta_{2}(t)  )\\ \leqslant -\lambda_2(\mathbb{L}_{A})\beta_{3}(t) $
 and
 $- \kappa_{2}  \beta_{1}(t)
\beta_{2}(t)+(6+40
\sigma_{v}^{2})\beta_{1}(t) \beta_{2}(t)\beta_{3}(t) \leqslant - \frac{\kappa_{2} }{2} \beta_{1}(t)
\beta_{2}(t),$
which together with (\ref{heqilaibound1}) gives
\begin{align}
&d\big(\widetilde{\mathcal{Y}}(t)+\mathcal{S}(t)\big)\notag\\
\leqslant & -\lambda_2(\mathbb{L}_{A})\beta_{3}(t) \widetilde{\mathcal{Y}}(t)- \frac{\kappa_{2} }{2} \beta_{1}(t)
\beta_{2}(t) \mathcal{S}(t) +  m_{3}(t)+\frac{ 4\kappa^2}{\tau_{2}  } \beta_{1}(t)\beta_{2}(t)
 B(t), \ \forall \ t\geqslant T_{1}.\label{heqilaibound2}
\end{align}
By Assumption \ref{assumption3}, we have
$ \lim_{t\to \infty} \frac{ \frac{\kappa_{2} }{2} \beta_{1}(t)
\beta_{2}(t)}{\lambda_2(\mathbb{L}_{A})\beta_{3}(t) }=0.$
Then, we know that there exists $T_{2}>0$, such that if $t\geqslant T_{2}$, then
$  \frac{\kappa_{2} }{2} \beta_{1}(t)
\beta_{2}(t) \leqslant \lambda_2(\mathbb{L}_{A})\beta_{3}(t).$
This together with (\ref{heqilaibound2}) gives,  $$d\big(\widetilde{\mathcal{Y}}(t)+\mathcal{S}(t)\big)
\leqslant   - \frac{\kappa_{2} }{2} \beta_{1}(t)
\beta_{2}(t)  ( \widetilde{\mathcal{Y}}(t)+\mathcal{S}(t) ) +  m_{3}(t)+\frac{ 4 \kappa^2}{\tau_{2}  } \beta_{1}(t)\beta_{2}(t)
 B(t), \ \forall \ t\geqslant  \max\{T_{1}, T_{2}\}.$$
Denote $T=\max\{T_{1}, T_{2}\}$. By the above inequality, we have
\begin{align} \label{heqilaibound3}
 &\widetilde{\mathcal{Y}}(t)+\mathcal{S}(t)\notag\\
 \leqslant &  e^{\int_{T}^{t}-\frac{\kappa_{2} }{2} \beta_{1}(s)
\beta_{2}(s)ds }\left(\widetilde{\mathcal{Y}}(T)+\mathcal{S}(T)\right)+ \int_{T}^{t}\Big(   m_{3}(s) +\frac{ 2 \kappa^2}{\tau_{2}  } \beta_{1}(s)\beta_{2}(s)
 B(s) \Big)e^{\int_{s}^{t}-\frac{\kappa_{2} }{2} \beta_{1}(s^{\prime})
\beta_{2}(s^{\prime})ds^{\prime} }ds.
\end{align}
By Assumption \ref{assumption3},  we have
\begin{align}\label{heqilaibound4}
   \lim_{t\to \infty}  e^{\int_{T}^{t}-\frac{\kappa_{2} }{2} \beta_{1}(s)
\beta_{2}(s)ds }=0.
\end{align}
By the continuity of $\widetilde{\mathcal{Y}}(t)+\mathcal{S}(t)$ w.r.t $t$, we have $\widetilde{\mathcal{Y}}(T)+\mathcal{S}(T)<\infty.$ This together with (\ref{heqilaibound4}) gives that, for the first term on the r.h.s. of (\ref{heqilaibound3}), we have
\begin{align}\label{heqilaibound5}
   \lim_{t\to \infty}e^{\int_{T}^{t}-\frac{\kappa_{2} }{2} \beta_{1}(s)
\beta_{2}(s)ds }\big(\widetilde{\mathcal{Y}}(T)+\mathcal{S}(T)\big)=0.
\end{align}
By $B(t) \leqslant \sup_{p\in [0,1]}
E\big[\|z_{p}(t)  -E[z_{p}(t)]\|^{2}\big]$ and   Lemma
\ref{variancetenzero}, we have
\begin{align}
\lim_{t \to \infty}B(t)=0.\label{boundofB}
\end{align}
For the second term on the r.h.s. of (\ref{heqilaibound3}), by L'Hospital's rule,
Assumption \ref{assumption3} and  the above equality, we have $\lim\limits_{t\to \infty} \int_{T}^{t}e^{\int_{s}^{t}-\frac{\kappa_{2} }{2} \beta_{1}(s^{\prime})
\beta_{2}(s^{\prime})ds^{\prime} } (   m_{3}(s)+\frac{4 \kappa^2}{\tau_{2}  } \beta_{1}(s) \beta_{2}(s)
 B(s)  ) ds  =\frac{  \sigma^{2}_{v}
\|x^{*}\|^{2} +C^{2}_{v}  }{2\kappa^2}.$
This together with (\ref{heqilaibound3}) and (\ref{heqilaibound5}) gives
$  \lim\limits_{t\to \infty} (\widetilde{\mathcal{Y}}(t)+\mathcal{S}(t) )=\frac{  \sigma^{2}_{v}
\|x^{*}\|^{2} +C^{2}_{v}  }{2\kappa^2},$ which leads to (\ref{ybounded}).

Now, we prove (\ref{unibouofz})-(\ref{l2qiroop}).
Let $Y_{1}(t)=\widetilde{\mathcal{Y}}(t)$, $Y_{2}(t)=\mathcal{S}(t)$, $Y_{3}(t)=B(t)$, $a_{1}(t)=2\beta_{3}(t)
\lambda_2(\mathbb{L}_{A})$, $a_{2}(t)= m_{1}(t)-a_{1}(t)$, $a_{3}(t)=m_{2}(t)$, $a_{4}(t)=m_{3}(t)$, $b_{1}(t)=2\kappa_{2}\beta_{1}(t)\beta_{2}(t)$ and $b_{2}(t)=2\sqrt{2}\kappa\beta_{1}(t)\\ \beta_{2}(t)$ in Lemma \ref{decotwotimescale}.
By (\ref{ybounded}), (\ref{ASSinequaliofwidy}), (\ref{deriavteofS}),  (\ref{boundofB}), Lemma \ref{decotwotimescale} and Assumption \ref{assumption3}, we have
\begin{align}
\lim_{t \to \infty}\mathcal{R}(t)&=0, \ \ \lim_{t \to \infty}\widetilde{\mathcal{R}}(t)=0\label{yptento0}
\end{align}
and (\ref{l2qiroop}).
Then, by the above equalities and  Lemma \ref{variancetenzero}, we have $\sup_{t\geqslant 0, p\in [0,1] }E\big[ \|z_{p}(t)-E [z_{p}(t) ]
 \|^{2}\big]<\infty,$ $\sup_{t\geqslant 0} \int_{[0,1]}\big\|
E\left[z_{p}(t)\right]-\int_{[0,1]}E\left[z_{q}(t)\right]dq
\big\|^{2}dp<\infty$, $\sup_{t\geqslant 0}  \big\|\int_{[0,1]}E\left[z_{q}(t)\right]dq-x^{*}\big\|^{2}<\infty$, $\sup_{t\geqslant 0, p\in [0,1] }  E\big[\|\widetilde{y}_{p}(t)  -E[\widetilde{y}_{p}(t)]\|^{2}\big]<\infty$ and $\sup_{t\geqslant 0}\int_{[0,1]} \| E[\widetilde{y}_{p}(t)]\|^{2} dp<\infty$.
This together with $C_{r}$ inequality gives
\begin{align}
&\sup\limits_{t\geqslant 0}\int_{[0,1]}E\big[ \|z_{p}(t) \|^{2}\big]dp\notag\\
\leqslant & 4\sup\limits_{t\geqslant 0, p\in [0,1] }E\big[ \|z_{p}(t)-E [z_{p}(t) ]
 \|^{2}\big]+4\|x^{*}\|^{2}
 \notag\\ & +4\sup\limits_{t\geqslant 0}\int_{[0,1]}\Big\|
E\left[z_{p}(t)\right]-\int_{[0,1]}E\left[z_{q}(t)\right]dq
\Big\|^{2}dp\notag\\ &
+4\sup\limits_{t\geqslant 0}\Big\|\int_{[0,1]}\hspace{-5pt} E\left[z_{q}(t)\right]dq-x^{*}\Big\|^{2}
<\infty \label{seconofzbound}
\end{align}
and
$
 \sup_{t\geqslant 0}\int_{[0,1]} \hspace{-3pt} E \big[\|\widetilde{y}_{p}(t)\|^{2}\big] dp
 \leqslant  \sup_{t\geqslant 0,p\in[0,1]}\hspace{-3pt}E \big[\|\widetilde{y}_{p}(t) -E[\widetilde{y}_{p}(t)]\|^{2}\big]
+\sup\limits_{t\geqslant 0}\int_{[0,1]}  \| E[\widetilde{y}_{p}(t)]\|^{2} dp
 <\infty.$
 Then, similar to the proof of (\ref{boudeforjoint}) in Theorem \ref{jointdecoupleconsensus} and by Assumption \ref{assumption0} (i)  and Assumption \ref{assumption3}, we have  (\ref{unibouofz}) and (\ref{unibouofyw}).
  By (\ref{unibouofz}), (\ref{unibouofyw}), (\ref{yptento0}) and Theorem \ref{jointdecoupleconsensus}, we have (\ref{zpconsensus1}) and (\ref{zpconsensus}).
$\hfill\blacksquare$

\vskip 1.5mm

\emph{Proof of Theorem \ref{therelast}:}
By Lemma \ref{variancetenzero},  Lemma \ref{sigleconseussss} and $C_{r}$ inequality, we have (\ref{tenoptimal}).
Similar to the proof of (\ref{meanyiszero}) in Lemma \ref{sigleconseussss} and by Assumption \ref{assumption0} (ii), we have $\int_{[0,1]}E [\widetilde{y}_{q}(t) ]dq=0$ and $\nabla_{x}\big(\int_{[0,1]} V(q,x^{*})dq\big)=0$.  This together with Assumption \ref{assumption0} (ii)-(iii) and $C_{r}$ inequality gives
\begin{align}
&E E\bigg[ \Big\|y_{p}(t) - \nabla_{x}\Big(\int_{[0,1]} V(q,x^{*})dq \Big) \Big\|^{2}\bigg]\notag\\
\leqslant & 3 E\big[\|\widetilde{y}_{p}(t)-E[\widetilde{y}_{p}(t)]\|^{2}\big] +3\beta^{2}_{2}(t)E\big[\|\nabla_{x}
 V(z_{p}(t),p)\|^{2}\big]  +3\left\|E[\widetilde{y}_{p}(t)]\right\|^{2}\notag\\
\leqslant&
3 E\big[\|\widetilde{y}_{p}(t)  -E[\widetilde{y}_{p}(t)]\|^{2}\big] +6C_{v}^{2} \beta^{2}_{2}(t)+3\left\|E[\widetilde{y}_{p}(t)] \right\|^{2}\notag\\
&
+6\sigma_{v}^{2} \beta^{2}_{2}(t) \sup_{p \in [0,1], \ t \geqslant 0}E\big[ \|z_{p}(t) \|^{2}\big].\label{tracklast}
\end{align}
 By  Assumption \ref{assumption3}, Lemma \ref{variancetenzero}, Lemma \ref{sigleconseussss} and (\ref{tracklast}),  we have (\ref{tenoptimay}).
 $\hfill\blacksquare$

 \setcounter{equation}{0}
\renewcommand{\theequation}{D.\arabic{equation}}
\setcounter{lemma}{0}
\renewcommand{\thelemma}{D.\arabic{lemma}}
\setcounter{theorem}{0}
\renewcommand{\thetheorem}{D.\arabic{theorem}}
\setcounter{assumption}{0}
\renewcommand{\theassumption}{D.\arabic{assumption}}
\section{Proof of Theorem  \ref{hheNwucha}}

Denote the undirected   graph as $G^N$ with adjacency matrix $(A^N(\frac{i}{N},\frac{j}{N}))_{1\leq i,j\leq N}$.
The continuous-time approximation of (\ref{EQinuous}) is given by
\begin{align}
 x_i^{N,h}(t)
=&x_{\frac{i}{N}}(0) -\int_{0}^{t} \alpha_2(s_h) \nabla_{x} V\left(\frac{i}{N},x_i^{N,h}(s_h)\right)ds\notag\\
&  +\int_{0}^{t}  \alpha_1(s_h)\frac{1}{N}\sum_{j=1}^{N}A^N \Big(\frac{i}{N},\frac{j}{N} \Big)\left(x_j^{N,h}(s_h)-x_i^{N,h}(s_h)\right)ds\notag\\
& -\int_{0}^{t} \alpha_2(s_h)\Sigma_1 d w_{\frac{i}{N}}(s),\ t\geq 0,\ i =1,2,...,N,\label{discrete00}
\end{align}
where  $s_h=\lfloor s/h\rfloor h$.


To show the error between the systems (\ref{Mckeanvlasov0}) and (\ref{EQinuous}), we introduce the following continuous-time system
\begin{align}\label{discretecontinuous}
   x_i^{N}(t)
=&x_{\frac{i}{N}}(0) -\int_{0}^{t} \alpha_2(s) \nabla_{x} V\left(\frac{i}{N},x_i^{N}(s)\right)ds +\int_{0}^{t}  \alpha_1(s)\frac{1}{N}\sum_{j=1}^{N}A^N \Big(\frac{i}{N},\frac{j}{N} \Big)\left(x_j^{N}(s)-x_i^{N}(s)\right)ds \notag\\
&  -\int_{0}^{t} \alpha_2(s)\Sigma_1d w_{\frac{i}{N}}(s),\ t\geq 0, \ i =1,2,...,N.
\end{align}

The following lemma shows that the states of (\ref{discretecontinuous}) are uniformly bounded with respect to the time  and the number of nodes.


\begin{lemma}\label{lemmmyjielis0000}
For the  system  (\ref{discretecontinuous}), if Assumptions \ref{assumption0},  \ref{assumptiondescent02} and \ref{assumptiondescent1} hold, then there exists $M>0$, such that
\begin{align}
\sup_{N\in \mathbb{N}, \ t \geq 0} \frac{1}{N}\sum_{i=1}^{N}E\left[\left\| x_i^{N}(t)\right\|^2\right] \leq M. \label{boxin}
\end{align}
\end{lemma}

\begin{proof}
By (\ref{discretecontinuous}) and It\^{o} formula, we have
\begin{align}
&d\left(\|x_i^{N}(t)\|^2\right)\notag\\
=&-2\alpha_2(t)
\left(x_i^{N}(t)\right)^\top
\nabla_x V\left(\frac{i}{N},x_i^{N}(t)\right)dt +2\alpha_1(t)
\left(x_i^{N}(t)\right)^\top
\left[
\frac{1}{N}\sum_{j=1}^{N}A^N(i,j)
\left(x_j^{N}(t)-x_i^{N}(t)\right)
\right]dt \notag\\
&+\alpha_2^2(t)\operatorname{Tr}\left(\Sigma_1\Sigma_1^\top\right)dt  -2\alpha_2(t)
\left(x_i^{N}(t)\right)^\top
\Sigma_1 d w_{\frac{i}{N}}(t).
\end{align}
  Taking expectations on the both sides of the above equality and by Assumption  \ref{assumption0}, we have
  \begin{align}
&d\left(\frac{1}{N}\sum_{i=1}^{N}E\left[\|x_i^{N}(t)\|^2\right]\right)\notag\\
=&-2\alpha_2(t)\frac{1}{N}\sum_{i=1}^{N}
E\left[\left(x_i^{N}(t)\right)^\top
\nabla_x V\left(\frac{i}{N},x_i^{N}(t)\right)\right]dt\notag\\
& +2\alpha_1(t)\frac{1}{N^2}\sum_{i,j=1}^{N}E\left[ A^N(i,j)\left(x_i^{N}(t)\right)^\top
\left(x_j^{N}(t)-x_i^{N}(t)\right)\right]
  dt+\alpha_2^2(t)\operatorname{Tr}\left(\Sigma_1\Sigma_1^\top\right) dt \notag\\
  \leq & -2\kappa_2 \alpha_2(t)\frac{1}{N}\sum_{i=1}^{N} E\left[\|x_i^{N}(t)\|^2\right]
  -2 \alpha_2(t)\frac{1}{N}\sum_{i=1}^{N} E\left[\left(x_i^{N}(t)\right)^\top
\nabla_x V\left(\frac{i}{N},0\right)\right]dt\notag\\
& +\alpha_2^2(t)\operatorname{Tr}\left(\Sigma_1\Sigma_1^\top\right) dt\notag\\
 \leq & -2\kappa_2 \alpha_2(t)\frac{1}{N}\sum_{i=1}^{N} E\left[\|x_i^{N}(t)\|^2\right]
  + 2 \sqrt{2}C_v \alpha_2(t)\left(\frac{1}{N}\sum_{i=1}^{N} E\left[\|x_i^{N}(t)\|^2\right]\right)^{\frac{1}{2}} \notag\\
& +\alpha_2^2(t)\operatorname{Tr}\left(\Sigma_1\Sigma_1^\top\right)dt,\notag
\end{align}
which together with Lemma \ref{seconesstmate}, Assumptions  \ref{assumptiondescent02}-\ref{assumptiondescent1} gives  (\ref{boxin}).
\end{proof}

The following lemma shows that the states of (\ref{EQinuous}) are uniformly bounded with respect to  the time, the step-size and the number of nodes.

\begin{lemma}\label{lemmmyjielis0}
For the  system  (\ref{EQinuous}), if Assumptions \ref{assumption0}, \ref{assumptiondescent02}, \ref{assumptiondescent1} and \ref{asssdd} hold, then  there exists $h_0>0$ and $ M(h_0)>0$,   such that
\begin{align}
\sup_{N\in \mathbb{N},\ k\in \mathbb{N},\ h\in(0,h_0)} \frac{1}{N}\sum_{i=1}^{N}E\left[\left\| x_i^{N,h}(kh)\right\|^2\right] \leq M(h_0).\label{lemmmyjielis}
\end{align}
\end{lemma}

\begin{proof}
By (\ref{EQinuous}), we have
\begin{align}
&\left\| x_i^{N,h}((k+1)h) \right\|^2-\left\|x_i^{N,h}(kh) \right\|^2\notag\\
=& 2 \left(x_i^{N,h}(kh) \right)^{\mathrm{T}} \left( x_i^{N,h}((k+1)h)-  x_i^{N,h}(kh) \right)+\left\| x_i^{N,h}((k+1)h)- x_i^{N,h}(kh)\right\|^2\notag\\
=& 2 (x_i^{N,h}(kh))^{\mathrm{T}} \bigg( -  h \alpha_2(kh) \nabla_{x} V\left(\frac{i}{N},x_i^{N,h}(kh)\right) +  \frac{h \alpha_1(kh)}{N}\sum_{j=1}^{N}A^N \big(\frac{i}{N},\frac{j}{N} \big)\left(x_j^{N,h}(kh )-x_i^{N,h}(kh)\right) \bigg)\notag\\
&+ \left\| -  h \alpha_2(kh) \nabla_{x} V\left(\frac{i}{N},x_i^{N,h}(kh)\right) +  \frac{h \alpha_1(kh)}{N}\sum_{j=1}^{N}A^N \Big(\frac{i}{N},\frac{j}{N} \Big)\left(x_j^{N,h}(kh )-x_i^{N,h}(kh)\right)\right\|^2\notag\\
&+\left\| - \alpha_2(kh)\Sigma_1  \Delta_kw_{\frac{i}{N}} \right\|^2- 2\alpha_2(kh) \left(x_i^{N,h}(kh) \right)^{\mathrm{T}} \left(\Sigma_1  \Delta_kw_{\frac{i}{N}} \right)\notag\\
&-2 \alpha_2(kh)\left(\Sigma_1  \Delta_kw_{\frac{i}{N}} \right)^{\mathrm{T}}\bigg(-  h \alpha_2(kh) \nabla_{x} V\left(\frac{i}{N},x_i^{N,h}(kh)\right)\notag\\
&+ h \alpha_1(kh) \frac{1}{N}\sum_{j=1}^{N}A^N \Big(\frac{i}{N},\frac{j}{N} \Big)\left(x_j^{N,h}(kh)-x_i^{N,h}(kh )\right)\bigg).\label{EQinuous1}
\end{align}
Denote $\beta^{N,h}(k)=\frac{1}{N}\sum_{i=1}^{N}E\left[\left\|  x_i^{N,h}(kh)\right\|^2\right]$. Then, taking expectations and summation on the both sides of (\ref{EQinuous1}) leads to
\begin{align}
& \beta^{N,h}(k+1) - \beta^{N,h}(k) \notag\\
\leq & \frac{-2 h \alpha_2(kh)}{N}\sum_{i=1}^{N} E\left[ \left(x_i^{N,h}(kh) \right)^{\mathrm{T}} \left(    \nabla_{x} V\left(\frac{i}{N},x_i^{N,h}(kh)\right)   \right)\right]\notag\\
& +\frac{2 h \alpha_1(kh) }{N} \sum_{i=1}^{N} E\left[ \left(x_i^{N,h}(kh) \right)^{\mathrm{T}} \left(  \frac{1}{N}\sum_{j=1}^{N}A^N \Big(\frac{i}{N},\frac{j}{N} \Big)\left(x_j^{N,h}(kh )-x_i^{N,h}(kh)\right) \right)\right]\notag\\
&+\frac{2h^2 \alpha_2^2(kh)}{N} \sum_{i=1}^{N}E\left[  \left\| \nabla_{x} V\left(\frac{i}{N},x_i^{N,h}(kh)\right)  \right\|^2\right]
\notag\\
& +\frac{2 h^2 \alpha_1^2(kh)}{N} \sum_{i=1}^{N}E\left[  \left\| \frac{1}{N}\sum_{j=1}^{N}A^N \Big(\frac{i}{N},\frac{j}{N} \Big)\left(x_j^{N,h}(kh )-x_i^{N,h}(kh )\right)  \right\|^2\right]+h \alpha_2^2(kh)\left\| \Sigma_1 \right\|^2.\label{EQinuous2}
\end{align}
By Assumption \ref{assumption0}, H\"{o}lder inequality and $C_r$ inequality, we have
\begin{align}
&\frac{-2 h \alpha_2(kh)}{N}\sum_{i=1}^{N} E\left[ \left(x_i^{N,h}(kh) \right)^{\mathrm{T}} \left(    \nabla_{x} V\left(\frac{i}{N},x_i^{N,h}(kh)\right)   \right)\right]\notag\\
\leq &- 2 \kappa_2  h \alpha_2(kh) \beta^{N,h}(k)+2 \sqrt{2} C_v  h \alpha_2(kh) \left(\beta^{N,h}(k)\right)^{\frac{1}{2}},\label{EQinuous3}
\end{align}
\begin{align}
 \frac{2h^2 \alpha_2^2(kh)}{N} \sum_{i=1}^{N}E\left[  \left\| \nabla_{x} V\left(\frac{i}{N},x_i^{N,h}(kh)\right)  \right\|^2\right]
\leq   4\sigma_v^2 h^2 \alpha_2^2(kh) \beta^{N,h}(k)+4C_v^2 h^2 \alpha_2^2(kh),\label{EQinuous4}
\end{align}
and
\begin{align}
 \frac{2h^2 \alpha_1^2(kh)}{N} \sum_{i=1}^{N}E\left[  \left\| \frac{1}{N}\sum_{j=1}^{N}A^N \Big(\frac{i}{N},\frac{j}{N} \Big)\left(x_j^{N,h}(kh )-x_i^{N,h}(kh )\right)  \right\|^2\right]
\leq    8 h^2 \alpha_1^2(kh) \beta^{N,h}(k).\label{EQinuous5}
\end{align}
By the symmetry of $A^{N}$, we know that
\begin{align}\label{EQinuous6}
  \frac{2 h \alpha_1(kh) }{N} \sum_{i=1}^{N} E\left[ \left(x_i^{N,h}(kh) \right)^{\mathrm{T}} \left(  \frac{1}{N}\sum_{j=1}^{N} A^N \Big(\frac{i}{N},\frac{j}{N} \Big)\left(x_j^{N,h}(kh )-x_i^{N,h}(kh )\right) \right)\right]\leq 0.
\end{align}
By (\ref{EQinuous1})-(\ref{EQinuous6}), we have
\begin{align}
& \beta^{N,h}(k+1)  -  \beta^{N,h}(k) \notag\\
\leq & \left(- 2 \kappa_2  h \alpha_2(kh)+4\sigma_v^2 h^2 \alpha_2^2(kh)+ 8 h^2 \alpha_1^2(kh) \right) \beta^{N,h}(k)+2 \sqrt{2} C_v  h \alpha_2(kh) \left(\beta^{N,h}(k)\right)^{\frac{1}{2}}\notag\\
&+h \alpha_2^2(kh)\left\| \Sigma_1 \right\|^2+4C_v^2 h^2 \alpha_2^2(kh).\label{EQinuous7}
\end{align}
By Assumptions  \ref{assumptiondescent1} and \ref{asssdd}, we know that there exist $C_{01},\ C_{02}>0$, such that $\alpha_2^2(t)\leq C_{01} \alpha_2(t)$, $\alpha_1^2(t)\leq C_{02}\alpha_2(t), \ \forall \ t\geq 0$. Besides, there exists $h_0$, such that $- 2 \kappa_2    +4\sigma_v^2 C_{01} h  + 8 h C_{02}\leq - \kappa_2,\ \forall \ h\in(0,h_0)$. Then, we have
\begin{align}
& \beta^{N,h}(k+1) -  \beta^{N,h}(k) \notag\\
\leq & -\kappa_2 h \alpha_2(kh)
\left[
\left(\beta^{N,h}(k)\right)^{\frac12}
-\frac{\sqrt{2}C_v}{\kappa_2}
\right]^2
+\frac{2C_v^2}{\kappa_2}h\alpha_2(kh) \notag\\
&\quad
+h\alpha_2^2(kh)\operatorname{Tr}\left(\Sigma_1\Sigma_1^\top\right)
+2C_v^2h^2\alpha_2^2(kh),\  \forall \ h\in(0,h_0).\label{EQinuous8}
\end{align}
Then, if $\left(\beta^{N,h}(k)\right)^{1/2}
>
\frac{\sqrt{2}C_v}{\kappa_2}
+
\left(
\frac{2C_v^2}{\kappa_2^2}
+
\frac{\alpha_2(kh)\operatorname{Tr}\left(\Sigma_1\Sigma_1^\top\right)}{\kappa_2}
+
\frac{2C_v^2h\alpha_2(kh)}{\kappa_2}
\right)^{1/2}$, then $ \beta^{N,h}(k+1) <\\   \beta^{N,h}(k)$.
Then, by Assumption  \ref{assumption0}, we have
\begin{align}
 \beta^{N,h}(k)\leq &  \beta^{N,h}(0)+
\frac{8C_v^2}{\kappa_2^2}
+
\frac{2 \sup_{t\geq 0}\alpha_2(t)\operatorname{Tr}\left(\Sigma_1\Sigma_1^\top\right) }{\kappa_2}
+
\frac{4C_v^2h\sup_{t\geq 0}\alpha_2(t)}{\kappa_2}\notag\\
\leq & \zeta_2+
\frac{8C_v^2}{\kappa_2^2}
+
\frac{2 \sup_{t\geq 0}\alpha_2(t)\operatorname{Tr}\left(\Sigma_1\Sigma_1^\top\right)}{\kappa_2}
+
\frac{4C_v^2h\sup_{t\geq 0}\alpha_2(t)}{\kappa_2},
\end{align}
which leads to (\ref{lemmmyjielis}).
\end{proof}

The following lemma estimates the within-step variation of the states of the  system (\ref{discrete00}).
\begin{lemma}\label{yxbuc}
For the  system  (\ref{discrete00}), if Assumptions \ref{assumption0}, \ref{assumptiondescent02} and \ref{assumptiondescent1} and \ref{asssdd} hold, then
\begin{align}\label{errorwitht}
&\frac{1}{N}\sum_{i=1}^{N} E\left[\left\|x_i^{N,h}(t)-x_i^{N,h}(t_h) \right\|^2\right]\notag\\
\leq & \sup_{t\geq0}\alpha_1^2(t) \Big(( 6\sigma_v^2  C_0^2 +12) h^2 M(h_0) + 6 C_v^2 h^2 C_0^2    +h  C_0^2 \operatorname{Tr}\left(\Sigma_1\Sigma_1^\top\right)\Big),\ t\geq 0, \ \forall \ h\in(0,h_0),
\end{align}
where $M$ is given by  Lemma \ref{lemmmyjielis0000} and $h_0$ and $ M(h_0)$ are given by Lemma \ref{lemmmyjielis0}.

\end{lemma}

\begin{proof}
By (\ref{discrete00}), Assumption \ref{assumption0} and $C_r$ inequality, we have
\begin{align}
&\frac{1}{N}\sum_{i=1}^{N} E\left[\left\|x_i^{N,h}(t)-x_i^{N,h}(t_h) \right\|^2\right]\notag\\
=& \frac{1}{N}\sum_{i=1}^{N} E\Bigg[\bigg\|-(t-t_h) \alpha_2(t_h) \nabla_{x} V\left(\frac{i}{N},x_i^{N,h}(t_h)\right) \notag\\
&+(t-t_h)  \alpha_1(t_h)\frac{1}{N}\sum_{j=1}^{N}A^N \Big(\frac{i}{N},\frac{j}{N} \Big)\left(x_j^{N,h}(t_h)-x_i^{N,h}(t_h)\right) \notag\\
&-  \alpha_2(t_h)\Sigma_1   (w_{\frac{i}{N}}(t) -w_{\frac{i}{N}}(t_h)  )\bigg\|^2\Bigg]\notag\\
\leq &  \frac{3(t-t_h)^2   \alpha_2^2(t_h)}{N}\sum_{i=1}^{N} E\left[\left\|  \nabla_{x} V\left(\frac{i}{N},x_i^{N,h}(t_h)\right) \right\|^2\right]\notag\\
&+ \frac{3(t-t_h)^2   \alpha_1^2(t_h)}{N}\sum_{i=1}^{N} E\left[\left\| \frac{1}{N}\sum_{j=1}^{N}A^N \Big(\frac{i}{N},\frac{j}{N} \Big)\left(x_j^{N,h}(t_h)-x_i^{N,h}(t_h)\right) \right\|^2\right]  +3(t-t_h)  \alpha_2^2(t_h)\|\Sigma_1\|^2\notag\\
\leq &  \frac{\left(6\sigma_v^2 h^2   \alpha_2^2(t_h)+ 12 h^2  \alpha_1^2(t_h)\right)}{N}\sum_{i=1}^{N} E\left[\left\|   x_i^{N,h}(t_h) \right\|^2\right]+ 6 C_v^2 h^2 \alpha_2^2(t_h)  +3h  \alpha_2^2(t_h)\|\Sigma_1\|^2,\notag
\end{align}
which together with  Assumption \ref{asssdd}  and Lemma \ref{lemmmyjielis0} gives (\ref{errorwitht}).
\end{proof}

The next lemma shows the error between the systems  (\ref{discrete00}) and (\ref{discretecontinuous}).

\begin{lemma}\label{lemmmydsitolianxu0}
For the  systems  (\ref{discrete00}) and  (\ref{discretecontinuous}), if Assumptions \ref{assumption0}, \ref{assumptiondescent02}  and  \ref{assumptiondescent1}   hold, then  there exists $T>0$, $C_1,\ C_2>0$, such that
\begin{align}
 \frac{1}{N} \sum_{i=1}^{N} E\left[\|x_i^{N,h}(t)- x_i^{N}(t)\|^2\right]
\leq e^{-\int_{h_0+T}^{t}2\kappa_2\alpha_2(s) ds} \left(M+M(h_0)\right)  + C_1 \sqrt h+C_2 h,\ \forall \ t\geq h_0+T,  \label{hliantidi}
\end{align}
where
$h_0$ is given by Lemma \ref{lemmmyjielis0}.
\end{lemma}

\begin{proof}
By (\ref{discrete00}) and (\ref{discretecontinuous}), we have
\begin{align}
& d( x_i^{N,h}(t)- x_i^{N}(t))\notag\\
=& -\left(\alpha_2(t_h) \nabla_{x} V\left(\frac{i}{N},x_i^{N,h}(t_h)\right)-\alpha_2(t) \nabla_{x} V\left(\frac{i}{N},x_i^{N}(t)\right)\right)dt\notag\\
& +\left( \alpha_1(t_h)\frac{1}{N}\sum_{j=1}^{N}A^N \Big(\frac{i}{N},\frac{j}{N} \Big)\left(x_j^{N,h}(t_h)-x_i^{N,h}(t_h)\right)-\alpha_1(t)\frac{1}{N}\sum_{j=1}^{N}A^N \Big(\frac{i}{N},\frac{j}{N} \Big)\left(x_j^{N}(t)-x_i^{N}(t)\right)\right)dt\notag\\
& -\left(\alpha_2(t_h)-\alpha_2(t)\right)\Sigma_1 d w_{\frac{i}{N}}(t). \notag
\end{align}
Then, by It\^{o} formula, we have
\begin{align}
&d( \|x_i^{N,h}(t)- x_i^{N}(t)\|^2)\notag\\
=&-2\left(x_i^{N,h}(t)-x_i^{N}(t)\right)^\top
\Bigg[
\alpha_2(t_h) \nabla_{x} V\left(\frac{i}{N},x_i^{N,h}(t_h)\right)
-\alpha_2(t) \nabla_{x} V\left(\frac{i}{N},x_i^{N}(t)\right)
\Bigg]dt \notag\\
&+2\left(x_i^{N,h}(t)-x_i^{N}(t)\right)^\top
\Bigg[
\alpha_1(t_h)\frac{1}{N}\sum_{j=1}^{N}A^N \Big(\frac{i}{N},\frac{j}{N} \Big)
\left(x_j^{N,h}(t_h)-x_i^{N,h}(t_h)\right) \notag\\
&
-\alpha_1(t)\frac{1}{N}\sum_{j=1}^{N}A^N \Big(\frac{i}{N},\frac{j}{N} \Big)
\left(x_j^{N}(t)-x_i^{N}(t)\right)
\Bigg]dt \notag\\
&+\left(\alpha_2(t_h)-\alpha_2(t)\right)^2
\operatorname{Tr}\left(\Sigma_1\Sigma_1^\top\right)dt -2\left(\alpha_2(t_h)-\alpha_2(t)\right)
\left(x_i^{N,h}(t)-x_i^{N}(t)\right)^\top
\Sigma_1 d w_{\frac{i}{N}}(t).\notag
\end{align}
Denote $B_{N,h}(t)=\frac{1}{N} \sum_{i=1}^{N} E\left[\|x_i^{N,h}(t)- x_i^{N}(t)\|^2\right]$. Taking expectations on the both sides of the above equality leads to
\begin{align}
&dB_{N,h}(t)\notag\\
=&-\frac{2}{N}\sum_{i=1}^{N}
\mathbb E\Bigg[
\left(x_i^{N,h}(t)-x_i^{N}(t)\right)^\top
\Bigg(
\alpha_2(t_h) \nabla_{x} V\left(\frac{i}{N},x_i^{N,h}(t_h)\right)
-\alpha_2(t) \nabla_{x} V\left(\frac{i}{N},x_i^{N}(t)\right)
\Bigg)
\Bigg]dt \notag\\
&+\frac{2}{N}\sum_{i=1}^{N}
\mathbb E\Bigg[
\left(x_i^{N,h}(t)-x_i^{N}(t)\right)^\top
\Bigg(
\alpha_1(t_h)\frac{1}{N}\sum_{j=1}^{N}A^N(i,j)
\left(x_j^{N,h}(t_h)-x_i^{N,h}(t_h)\right) \notag\\
&-\alpha_1(t)\frac{1}{N}\sum_{j=1}^{N}A^N(i,j)
\left(x_j^{N}(t)-x_i^{N}(t)\right)
\Bigg)
\Bigg]dt  +\left(\alpha_2(t_h)-\alpha_2(t)\right)^2
\operatorname{Tr}\left(\Sigma_1\Sigma_1^\top\right) dt.\label{lemmmydsitolianxu01}
\end{align}
 For the first term on the r.h.s. of (\ref{lemmmydsitolianxu01}), by  H\"{o}lder inequality, Lemmas \ref{lemmmyjielis0}-\ref{yxbuc} and Assumption \ref{assumption0}, we have
 \begin{align}
& -\frac{2}{N}\sum_{i=1}^{N}
\mathbb E\Bigg[
\left(x_i^{N,h}(t)-x_i^{N}(t)\right)^\top
\Bigg(
\alpha_2(t_h) \nabla_{x} V\left(\frac{i}{N},x_i^{N,h}(t_h)\right)
-\alpha_2(t) \nabla_{x} V\left(\frac{i}{N},x_i^{N}(t)\right)
\Bigg)
\Bigg]\notag\\
=& -\alpha_2(t)\frac{2}{N}\sum_{i=1}^{N}
\mathbb E\Bigg[
\left(x_i^{N,h}(t)-x_i^{N}(t)\right)^\top
\Bigg(
 \nabla_{x} V\left(\frac{i}{N},x_i^{N,h}(t)\right)
- \nabla_{x} V\left(\frac{i}{N},x_i^{N}(t)\right)
\Bigg)
\Bigg]\notag\\
&-(\alpha_2(t_h)-\alpha_2(t))\frac{2}{N}\sum_{i=1}^{N}
\mathbb E\Bigg[
\left(x_i^{N,h}(t)-x_i^{N}(t)\right)^\top
\Bigg(
 \nabla_{x} V\left(\frac{i}{N},x_i^{N,h}(t_h)\right)
\Bigg)
\Bigg]\notag\\
&-\alpha_2(t)\frac{2}{N}\sum_{i=1}^{N}
\mathbb E\Bigg[
\left(x_i^{N,h}(t)-x_i^{N}(t)\right)^\top
\Bigg(
 \nabla_{x} V\left(\frac{i}{N},x_i^{N,h}(t_h)\right)-
 \nabla_{x} V\left(\frac{i}{N},x_i^{N,h}(t)\right)
\Bigg)
\Bigg]\notag\\
\leq & -2 \kappa_2 \alpha_2(t) B_{N,h}(t)+ 2|\alpha_2(t_h)-\alpha_2(t)| \left(B_{N,h}(t)\right)^{\frac{1}{2}}\left(\frac{1}{N}\sum_{i=1}^{N} E\left[\left\|  \nabla_{x} V\left(\frac{i}{N},x_i^{N,h}(t_h)\right)\right\|^2\right]\right)^{\frac{1}{2}}\notag\\
& + \alpha_2(t)\left(B_{N,h}(t)\right)^{\frac{1}{2}}\left(\frac{1}{N}\sum_{i=1}^{N} E\left[\left\|
 \nabla_{x} V\left(\frac{i}{N},x_i^{N,h}(t_h)\right)-
 \nabla_{x} V\left(\frac{i}{N},x_i^{N,h}(t)\right)\right\|^2\right]\right)^{\frac{1}{2}}\notag\\
 \leq &  -2 \kappa_2 \alpha_2(t) B_{N,h}(t)+ 2\sqrt{2} |\alpha_2(t_h)-\alpha_2(t)| \left(B_{N,h}(t)\right)^{\frac{1}{2}}\left(\sigma_v^2M(h_0)+C_v^2\right)^{\frac{1}{2}}\notag\\
& + \alpha_2(t)\Big(\sup_{t\geq 0} \alpha_1(t)\Big)\kappa \left(B_{N,h}(t)\right)^{\frac{1}{2}}\left(( 6\sigma_v^2  C_0^2 +12) h^2 M(h_0) + 6 C_v^2 h^2 C_0^2    +h  C_0^2 \operatorname{Tr}\left(\Sigma_1\Sigma_1^\top\right) \right)^{\frac{1}{2}}.\label{lemmmydsitolianxu02}
 \end{align}
 For the second term on the r.h.s. of (\ref{lemmmydsitolianxu01}), by  H\"{o}lder inequality, Lemmas \ref{lemmmyjielis0}-\ref{yxbuc} and Assumption \ref{assumption0}, we have
\begin{align}
& \frac{2}{N}\sum_{i=1}^{N}
\mathbb E\Bigg[
\left(x_i^{N,h}(t)-x_i^{N}(t)\right)^\top
\Bigg(
\alpha_1(t_h)\frac{1}{N}\sum_{j=1}^{N}A^N(i,j)
\left(x_j^{N,h}(t_h)-x_i^{N,h}(t_h)\right) \notag\\ &
-\alpha_1(t)\frac{1}{N}\sum_{j=1}^{N}A^N(i,j)
\left(x_j^{N}(t)-x_i^{N}(t)\right)
\Bigg)
\Bigg] \notag\\
=& \frac{2(\alpha_1(t_h)-\alpha_1(t))}{N}\sum_{i=1}^{N}
\mathbb E\Bigg[
\big(x_i^{N,h}(t)-x_i^{N}(t)\big)^\top
\Bigg(
\frac{\sum_{j=1}^{N}A^N(i,j)}{N}
\left(x_j^{N,h}(t_h)-x_i^{N,h}(t_h)\right)
\Bigg)
\Bigg] \notag\\
&\hspace{-3pt} + \hspace{-3pt}  \frac{2 \alpha_1(t) \sum_{i=1}^{N}}{N}
\mathbb E\bigg[
\big(x_i^{N,h}(t)-x_i^{N}(t)\big)^\top
\bigg(
\frac{\sum_{j=1}^{N}A^N(i,j)}{N}
\left(x_j^{N,h}(t_h)-x_i^{N,h}(t_h)-(x_j^{N,h}(t)-x_i^{N,h}(t))\right)
\bigg)
\bigg] \notag\\
&+ \frac{2 \alpha_1(t) }{N}\sum_{i=1}^{N}
\mathbb E\Bigg[
\left(x_i^{N,h}(t)-x_i^{N}(t)\right)^\top
\Bigg(
\frac{1}{N}\sum_{j=1}^{N}A^N(i,j)
\left(x_j^{N,h}(t)-x_i^{N,h}(t)-(x_j^{N}(t)-x_i^{N}(t))\right)
\Bigg)
\Bigg] \notag\\
\leq&
2|\alpha_1(t_h)-\alpha_1(t)|
\left(B_{N,h}(t)\right)^{\frac12}
\left(
\frac{1}{N}\sum_{i=1}^{N}
\mathbb E\left[
\left\|
\frac{1}{N}\sum_{j=1}^{N}A^N(i,j)
\left(x_j^{N,h}(t_h)-x_i^{N,h}(t_h)\right)
\right\|^2
\right]
\right)^{\frac12} \notag\\
&\hspace{-3pt} +\hspace{-3pt} 
2\alpha_1(t)
\left(B_{N,h}(t)\right)^{\frac12}
\left(
\frac{1}{N}\sum_{i=1}^{N}
\mathbb E\left[
\left\|
\frac{1}{N}\sum_{j=1}^{N}A^N(i,j)
\Big[
x_j^{N,h}(t_h)-x_i^{N,h}(t_h)
-\left(x_j^{N,h}(t)-x_i^{N,h}(t)\right)
\Big]
\right\|^2
\right]
\right)^{\frac12} \notag\\
\leq&
4|\alpha_1(t_h)-\alpha_1(t)|
\left(B_{N,h}(t)\right)^{\frac12}
\left(M(h_0)\right)^{\frac12} \notag\\
&
+4\alpha_1(t)
\left(B_{N,h}(t)\right)^{\frac12}
\left(
\frac{1}{N}\sum_{i=1}^{N}
\mathbb E\left[
\left\|x_i^{N,h}(t)-x_i^{N,h}(t_h)\right\|^2
\right]
\right)^{\frac12}\notag\\
\leq &
4|\alpha_1(t_h)-\alpha_1(t)|
\left(B_{N,h}(t)\right)^{\frac12}
\left(M(h_0)\right)^{\frac12} \notag\\
&
+4\alpha_1(t)\alpha_1(t_h)
\left(B_{N,h}(t)\right)^{\frac12}
\Big(
(6\sigma_v^2 C_0^2+12)h^2M(h_0)
+6C_v^2h^2C_0^2
+hC_0^2\|\Sigma_1\|
\Big)^{\frac12}.\label{lemmmydsitolianxu03}
\end{align}
By \eqref{lemmmydsitolianxu01}, \eqref{lemmmydsitolianxu02} and \eqref{lemmmydsitolianxu03}, we have
\begin{align}
dB_{N,h}(t)
\leq &
\Big[
-2\kappa_2\alpha_2(t)B_{N,h}(t)
+\Gamma_h(t)\left(B_{N,h}(t)\right)^{\frac12}
+\left(\alpha_2(t_h)-\alpha_2(t)\right)^2\operatorname{Tr}\left(\Sigma_1\Sigma_1^\top\right)
\Big]dt,
\label{lemmmydsitolianxu04}
\end{align}
where
$
\Gamma_h(t)
 =
2\sqrt{2}|\alpha_2(t_h)-\alpha_2(t)|
\left(\sigma_v^2M(h_0)+C_v^2\right)^{\frac12}
+4|\alpha_1(t_h)-\alpha_1(t)|
\left(M(h_0)\right)^{\frac12} +\big(\alpha_2(t)\\ \big(\sup_{t\geq 0} \alpha_1(t)\big)\kappa
+4\alpha_1(t)\alpha_1(t_h)\big)
\big(
(6\sigma_v^2C_0^2+12)h^2M(h_0)
+6C_v^2h^2C_0^2
+hC_0^2\operatorname{Tr}\left(\Sigma_1\Sigma_1^\top\right)
\big)^{\frac12}.
$
By Assumption \ref{asssdd}, we know that there exists $T>0$, such that if $t-h_0>T$, then   $|\dot\alpha_1(t)|$ and $|\dot\alpha_2(t)|$ are  non-increasing.
 This together with mean value theorem gives
 \begin{align}
& dB_{N,h}(t)\notag\\
\leq &
\Big[
-2\kappa_2\alpha_2(t)B_{N,h}(t)
+\Gamma_{1,h_0}(t)\left(B_{N,h}(t)\right)^{\frac12}
+\left(\dot\alpha_2(t-h_0)\right)^2 h^2 \operatorname{Tr}\left(\Sigma_1\Sigma_1^\top\right)
\Big]dt,\ t>h_0+T,
\label{lemmmydsitolianxu05}
\end{align}
 where $
\Gamma_{1,h_0}(t)
 =
2\sqrt{2}h |\dot\alpha_2(t-h_0)|
(\sigma_v^2M(h_0)+C_v^2)^{\frac12}
+4h|\dot\alpha_1(t-h_{0})|
\left(M(h_0)\right)^{\frac12}
 +\big(\alpha_2(t)\\ \big(\sup_{t\geq 0} \alpha_1(t)\big)\kappa
+4\alpha_1(t)\alpha_1(t-h_0)\big)
\big(
(6\sigma_v^2C_0^2+12)h^2M(h_0)
+6C_v^2h^2C_0^2
+hC_0^2\operatorname{Tr}\left(\Sigma_1\Sigma_1^\top\right)
\big)^{\frac12}.
$
Notice that by Lemmas \ref{lemmmyjielis0000}-\ref{lemmmyjielis0}, we have $B_{N,h}(t)\leq M+M(h_0),$ which together with (\ref{lemmmydsitolianxu05}) gives
\begin{align}
dB_{N,h}(t)
\leq &
\Big[
-2\kappa_2\alpha_2(t)B_{N,h}(t)
+\Gamma_{1,h_0}(t)\left(M+M(h_0)\right)^{\frac12}
+\left(\dot\alpha_2(t-h_0)\right)^2 h^2 \operatorname{Tr}\left(\Sigma_1\Sigma_1^\top\right)
\Big]dt.\notag
\end{align}
Then, we have
\begin{align}
 B_{N,h}(t)
\leq & e^{-\int_{h_0+T}^{t}2\kappa_2\alpha_2(s) ds} B_{N,h}(h_0+T)\notag\\
& + \int_{h_0+T}^{t} e^{-\int_{s}^{t}2\kappa_2\alpha_2(s^{\prime}) ds^{\prime}}
\Big[  \Gamma_{1,h_0}(s)\left(M+M(h_0)\right)^{\frac12}
+\left(\dot\alpha_2(s-h_0)\right)^2 h^2 \operatorname{Tr}\left(\Sigma_1\Sigma_1^\top\right)
\Big]ds\notag\\
\leq & e^{-\int_{h_0+T}^{t}2\kappa_2\alpha_2(s) ds} \left(M+M(h_0)\right)\notag\\
& + \int_{h_0+T}^{t} e^{-\int_{s}^{t}2\kappa_2\alpha_2(s^{\prime}) ds^{\prime}}
\Big[  \Gamma_{1,h_0}(s)\left(M+M(h_0)\right)^{\frac12}
+\left(\dot\alpha_2(s-h_0)\right)^2 h^2 \operatorname{Tr}\left(\Sigma_1\Sigma_1^\top\right)
\Big]ds.\notag
\end{align}
This together with Assumption \ref{asssdd}  leads to (\ref{hliantidi}).
\end{proof}

The next lemma shows the consensus error of the system  (\ref{discretecontinuous}).

\begin{lemma}\label{Nconsensus}
For the  system  (\ref{discretecontinuous}), if Assumptions \ref{assumption0},  \ref{assumptiondescent02} and \ref{assumptiondescent1}   hold,  then
\begin{align}\label{Nconsensus0}
 & \frac{1}{N}\sum_{i=1}^{N}E\left[\left\|x_i^{N}(t)-\frac{1}{N}\sum_{j=1}^{N}x_j^{N}(t) \right\|^2\right]\notag\\
 \leq& e^{-2\lambda_2(\mathbb{L}_{A}) \int_0^t\alpha_1(s)ds}M +\frac{\Big|\lambda_2(\mathbb{L}_{A})  -
 \frac{\lambda_2(L^N)}{N} \Big|M}{4\lambda_2(\mathbb{L}_{A})  } \notag\\
& +
\int_0^t
e^{- 2\lambda_2(\mathbb{L}_{A})\int_s^t\alpha_1(r)dr}
\bigg[ \alpha_2(s)(M+ 2\sigma_v^2 M+ 2C_v^2 ) +2\alpha_2^2(s)\operatorname{Tr}\left(\Sigma_1\Sigma_1^\top\right)
\bigg]ds,\ t\geq 0.
\end{align}
\end{lemma}

\begin{proof}
 By (\ref{discretecontinuous}) and It\^{o} formula, we have
 \begin{align}
 &d\left\|x_i^{N}(t)-\frac{1}{N}\sum_{j=1}^{N}x_j^{N}(t)\right\|^2\notag\\
 =& -2\alpha_2(t)\left( x_i^{N}(t)-\frac{1}{N}\sum_{j=1}^{N}x_j^{N}(t)\right)\left(\nabla_{x} V\left(\frac{i}{N},x_i^{N}(t)\right)-\frac{1}{N}\sum_{j=1}^{N} \nabla_{x} V\left(\frac{j}{N},x_j^{N}(t)\right) \right)\notag\\
 &+ 2\alpha_1(t)\left( x_i^{N}(t)-\frac{1}{N}\sum_{j=1}^{N}x_j^{N}(t)\right)\left( \frac{1}{N}\sum_{j=1}^{N}A^N (i,j)\left(x_j^{N}(t)-x_i^{N}(t)\right) \right)\notag\\
 &+ \alpha_2^2(t)\left(2-\frac{2}{N}\right)\operatorname{Tr}\left(\Sigma_1\Sigma_1^\top\right) -  \alpha_2(t)\left( x_i^{N}(t)-\frac{1}{N}\sum_{j=1}^{N}x_j^{N}(t)\right)\Sigma_1 P_j d w(t),\notag
 \end{align}
 where $w^{N}(t)=(w_{\frac{1}{N}}(t),\cdots,w_{1}(t))$, $P_j=\operatorname{diag}(p_{j,1},p_{j,2},\ldots,p_{j,N}),$ $p_{j,i}=
\begin{cases}
1-\dfrac{1}{N}, & i=j,\\[4pt]
\dfrac{1}{N}, & i\neq j.
\end{cases}
 $
Denote $B_N(t)=\frac{1}{N}\sum_{i=1}^{N}E\big[\|x_i^{N}(t)-\frac{1}{N}\sum_{j=1}^{N}x_j^{N}(t) \|^2\big]$. By the above equality, we have
  \begin{align}
 &d B_N(t)\notag\\
 =& -2\alpha_2(t)\frac{1}{N}\sum_{i=1}^{N}E\left[ \left( x_i^{N}(t)-\frac{1}{N}\sum_{j=1}^{N}x_j^{N}(t)\right)\left(\nabla_{x} V\left(\frac{i}{N},x_i^{N}(t)\right)-\frac{1}{N}\sum_{j=1}^{N} \nabla_{x} V\left(\frac{j}{N},x_j^{N}(t)\right) \right)\right]\notag\\
 &+ 2\alpha_1(t)\frac{1}{N}\sum_{i=1}^{N}E\left[\left( x_i^{N}(t)\right)\left( \frac{1}{N}\sum_{j=1}^{N}A^N (i,j)\left(x_j^{N}(t)-x_i^{N}(t)\right) \right)\right].\notag\\
 \label{Nconsensus1}
 \end{align}
 For the first term on the r.h.s. of the above equality, by Assumption \ref{assumption0}, Lemma \ref{lemmmyjielis0000},  H\"{o}lder inequality, we have
 \begin{align}
 &-2\alpha_2(t)\frac{1}{N}\sum_{i=1}^{N}E\left[ \left( x_i^{N}(t)-\frac{1}{N}\sum_{j=1}^{N}x_j^{N}(t)\right)\left(\nabla_{x} V\left(\frac{i}{N},x_i^{N}(t)\right)-\frac{1}{N}\sum_{j=1}^{N} \nabla_{x} V\left(\frac{j}{N},x_j^{N}(t)\right) \right)\right]\notag\\
 \leq & -2\alpha_2(t)\frac{1}{N}\sum_{i=1}^{N}E\left[ \left( x_i^{N}(t)-\frac{1}{N}\sum_{j=1}^{N}x_j^{N}(t)\right)\left(\nabla_{x} V\left(\frac{i}{N},x_i^{N}(t)\right) \right)\right]\notag\\
 \leq & \alpha_2(t) B_N(t)+ \alpha_2(t)\frac{1}{N}\sum_{i=1}^{N} E\left[ \left\|\nabla_{x} V\left(\frac{i}{N},x_i^{N}(t)\right) \right\|^2\right]\notag\\
 \leq & \alpha_2(t) M+ 2\alpha_2(t)\sigma_v^2 M+ 2\alpha_2(t)C_v^2.\label{Nconsensus2}
 \end{align}
 For the second term on the r.h.s. of (\ref{Nconsensus1}), by the symmetry of $ A^N$, we have
 \begin{align}
 &2\alpha_1(t)\frac{1}{N}\sum_{i=1}^{N}E\left[\left( x_i^{N}(t)\right)\left( \frac{1}{N}\sum_{j=1}^{N}A^N (i,j)\left(x_j^{N}(t)-x_i^{N}(t)\right) \right)\right]\notag\\
 \leq & -2\alpha_1(t) E\left[
 \frac{\lambda_2(L^N)}{N} \frac{1}{N}\sum_{i=1}^{N}\Big\|x_i^{N}(t)-\frac{1}{N}\sum_{j=1}^{N}x_j^{N}(t) \Big\|^2\right]\notag\\
 =& -2\lambda_2(\mathbb{L}_{A}) \alpha_1(t)  B_N(t)
 +2 \alpha_1(t) \left(\lambda_2(\mathbb{L}_{A})  -
 \frac{\lambda_2(L^N)}{N} \right)   B_N(t).\label{Nconsensus3}
 \end{align}
 Then, by H\"{o}lder inequality and Lemma \ref{lemmmyjielis0000}, we have
 \begin{align}
  &2\alpha_1(t)\frac{1}{N}\sum_{i=1}^{N}E\left[\left( x_i^{N}(t)\right)\left( \frac{1}{N}\sum_{j=1}^{N}A^N (i,j)\left(x_j^{N}(t)-x_i^{N}(t)\right) \right)\right]\notag\\
 \leq &
 -2\lambda_2(\mathbb{L}_{A}) \alpha_1(t)  B_N(t)
 +8\alpha_1(t)\left|\lambda_2(\mathbb{L}_{A})  -
 \frac{\lambda_2(L^N)}{N} \right|M.\notag
 \end{align}
 This together with (\ref{Nconsensus1}) and (\ref{Nconsensus2}) gives
 \begin{align}
   d B_N(t)
  \leq &  -2\lambda_2(\mathbb{L}_{A}) \alpha_1(t)) B_N(t)+ \alpha_2(t)(M+ 2\sigma_v^2 M+ 2C_v^2 ) +2\alpha_2^2(t)\operatorname{Tr}\left(\Sigma_1\Sigma_1^\top\right) \notag\\
& +8\alpha_1(t)\left|\lambda_2(\mathbb{L}_{A})  -
 \frac{\lambda_2(L^N)}{N} \right|M.
 \end{align}
Then, we have
\begin{align}
B_N(t)
\leq&
e^{-2\lambda_2(\mathbb{L}_{A}) \int_T^t\alpha_1(s)ds}M  \notag\\
& +
\int_0^t
e^{- 2\lambda_2(\mathbb{L}_{A})\int_s^t\alpha_1(r)dr}
\bigg[ \alpha_2(s)(M+ 2\sigma_v^2 M+ 2C_v^2 ) +2\alpha_2^2(s)\operatorname{Tr}\left(\Sigma_1\Sigma_1^\top\right) \notag\\
&  +8\alpha_1(s)\Big|\lambda_2(\mathbb{L}_{A})  -
 \frac{\lambda_2(L^N)}{N} \Big|M
\bigg]ds \notag\\
=&e^{-2\lambda_2(\mathbb{L}_{A}) \int_0^t\alpha_1(s)ds}M +\frac{\Big|\lambda_2(\mathbb{L}_{A})  -
 \frac{\lambda_2(L^N)}{N} \Big|M}{4\lambda_2(\mathbb{L}_{A})  } \notag\\
& +
\int_0^t
e^{- 2\lambda_2(\mathbb{L}_{A})\int_s^t\alpha_1(r)dr}
\bigg[ \alpha_2(s)(M+ 2\sigma_v^2 M+ 2C_v^2 ) +2\alpha_2^2(s)\operatorname{Tr}\left(\Sigma_1\Sigma_1^\top\right)
\bigg]ds,\notag
\end{align}
which leads to (\ref{Nconsensus0}).
\end{proof}

The following lemma shows the error between  the averaged states of the system  (\ref{discretecontinuous}) and the global minimizer of $V$.

\begin{lemma}\label{Nconsensustomini}
For the  system  (\ref{discretecontinuous}), if Assumptions \ref{assumption0}, \ref{assumptiondescent02}, \ref{assumptiondescent1}  and \ref{VDLIP} hold, 
 then there exists $ C_4>0,$ such that
\begin{align}\label{Nconsensustomini0}
   & E\left[\left\|\frac{1}{N}\sum_{i=1}^{N}x_i^{N}(t)
  -x^*\right\|^2\right]\notag\\
  \leq &
\frac{C_4}{N}+ 2(\zeta_2+2\|x^*\|^2)
e^{-2\kappa_2 \int_{0}^{t}\alpha_2(s)ds}   \notag\\
&+
\int_{0}^{t}
e^{-2\kappa_2 \int_{s}^{t}\alpha_2(s^{\prime})ds^{\prime}}
2\alpha_2(s)(M+\|x^*\|^2)^{\frac12}
\bigg[
e^{-\lambda_2(\mathbb L_A)\int_0^s\alpha_1(\tau)d\tau}M^{\frac12}
\notag\\
&+
2(\lambda_2(\mathbb L_A))^{-\frac{1}{2}}
\left(\left|
\lambda_2(\mathbb L_A)
-
\dfrac{\lambda_2(L^N)}{N}
\right|M\right)^{\frac{1}{2}}
\notag\\
&
+
\bigg(
\int_0^s
e^{-2\lambda_2(\mathbb L_A)\int_\tau^s\alpha_1(r)dr}
\bigg[
\alpha_2(\tau)(M+2\sigma_v^2M+2C_v^2)
+
2\alpha_2^2(\tau)
\operatorname{Tr}\left(\Sigma_1\Sigma_1^\top\right)
\bigg]d\tau
\bigg)^{\frac12}
\bigg]ds.
\end{align}
\end{lemma}
\begin{proof}
By (\ref{discretecontinuous}), we have
\begin{align}
 d\left(\frac{1}{N}\sum_{i=1}^{N}x_i^{N}(t)\right)
=& -  \alpha_2(t)\frac{1}{N}\sum_{i=1}^{N} \nabla_{x} V\left(\frac{i}{N},x_i^{N}(t)\right)dt-  \alpha_2(t)\Sigma_1 \frac{1}{N}\sum_{i=1}^{N}d w_{\frac{i}{N}}(t).\notag
\end{align}
Denote $S_N(t)=E\left[\left\|\frac{1}{N}\sum_{i=1}^{N}x_i^{N}(t)-x^*\right\|^2\right]$. By It\^{o} formula and the above equality, we have
 \begin{align}
 &d S_N(t) \notag\\
=& -2\alpha_2(t)
E\left[
\left(
\frac{1}{N}\sum_{i=1}^{N}x_i^{N}(t)-x^*\right)^{T}
\left(\frac{1}{N}\sum_{i=1}^{N}
\nabla_x V\left(\frac{i}{N},x_i^N(t)\right)
\right)
\right]dt \notag\\
&\quad
+
\frac{\alpha_2^2(t)}{N}
\operatorname{Tr}(\Sigma_1^\top\Sigma_1)\,dt\notag\\
=& -2\alpha_2(t)
E\left[
\left(
\frac{\sum_{i=1}^{N}x_i^{N}(t)}{N}-x^*\right)^{T}
\left(\frac{\sum_{i=1}^{N}
\nabla_x V\left(\frac{i}{N},x_i^N(t)\right)}{N}-\frac{1}{N}\sum_{i=1}^{N}
\nabla_x V\left(\frac{i}{N},\frac{1}{N}\sum_{j=1}^{N}x_j^{N}(t)\right)
\right)
\right]dt \notag\\
& -2\alpha_2(t)
E\left[
\left(
\frac{1}{N}\sum_{i=1}^{N}x_i^{N}(t)-x^*\right)^{T}
\left(\frac{1}{N}\sum_{i=1}^{N}
\nabla_x V\left(\frac{i}{N},\frac{1}{N}\sum_{j=1}^{N}x_j^{N}(t)\right)-\frac{1}{N}\sum_{i=1}^{N}
\nabla_x V\left(\frac{i}{N},x^*\right)
\right)
\right]dt \notag\\
& -2\alpha_2(t)
E\left[
\left(
\frac{1}{N}\sum_{i=1}^{N}x_i^{N}(t)-x^*\right)^{T}
\left( \frac{1}{N}\sum_{i=1}^{N}
\nabla_x V\left(\frac{i}{N},x^*\right)-\int_{[0,1]}\nabla V(p,x^*)dp
\right)
\right]dt \notag\\
&\quad
+
\frac{\alpha_2^2(t)}{N}
\operatorname{Tr}(\Sigma_1^\top\Sigma_1)\,dt.\label{Nconsensustomini1}
 \end{align}
By Assumption \ref{assumption0}, H\"{o}lder inequality and Lemma \ref{lemmmyjielis0000}, we have
\begin{align}
  d S_N(t)
 \leq &  2\alpha_2(t) (M+\|x^*\|^2)^{\frac{1}{2}} \left( \frac{1}{N}\sum_{i=1}^{N}E\left[\left\|x_i^{N}(t)-\frac{1}{N}\sum_{j=1}^{N}x_j^{N}(t) \right\|^2\right]\right)^{\frac{1}{2}} dt \notag\\
& -2\kappa_2\alpha_2(t)   S_N(t) dt+ \frac{2L_0}{N} \alpha_2(t)(M+\|x^*\|^2)^{\frac{1}{2}} dt
+
\frac{\alpha_2^2(t)}{N}
\operatorname{Tr}(\Sigma_1^\top\Sigma_1)\,dt.\label{Nconsensustomini2}
 \end{align}
 Then,  by Assumption \ref{assumptiondescent02} and  Lemma \ref{Nconsensus}, we have
 \begin{align}
 S_N(t) \leq & e^{-2\kappa_2 \int_{0}^{t}\alpha_2(s)ds  }  S_N(0) \notag\\
 &+ \int_{0}^{t} e^{-2\kappa_2 \int_{s}^{t}\alpha_2(s^{\prime})ds^{\prime}  }   2\alpha_2(s) (M+\|x^*\|^2)^{\frac{1}{2}} \left( \frac{1}{N}\sum_{i=1}^{N}E\left[\left\|x_i^{N}(s)-\frac{1}{N}\sum_{j=1}^{N}x_j^{N}(s) \right\|^2\right]\right)^{\frac{1}{2}}ds\notag\\
 & + \int_{0}^{t} e^{-2\kappa_2 \int_{s}^{t}\alpha_2(s^{\prime})ds^{\prime}  }   \left( \frac{2L_0}{N} \alpha_2(s)(M+\|x^*\|^2)^{\frac{1}{2}}
+
\frac{\alpha_2^2(s)}{N}
\operatorname{Tr}(\Sigma_1^\top\Sigma_1)\right) ds\notag\\
\leq &2(\zeta_2+2\|x^*\|^2)
e^{-2\kappa_2 \int_{0}^{t}\alpha_2(s)ds}   \notag\\
&+
\int_{0}^{t}
e^{-2\kappa_2 \int_{s}^{t}\alpha_2(s^{\prime})ds^{\prime}}
2\alpha_2(s)(M+\|x^*\|^2)^{\frac12}
\bigg[
e^{-\lambda_2(\mathbb L_A)\int_0^s\alpha_1(\tau)d\tau}M^{\frac12}
\notag\\
&+
2(\lambda_2(\mathbb L_A))^{-\frac{1}{2}}
\left(\left|
\lambda_2(\mathbb L_A)
-
\dfrac{\lambda_2(L^N)}{N}
\right|M\right)^{\frac{1}{2}}
\notag\\
&
+
\bigg(
\int_0^s
e^{-2\lambda_2(\mathbb L_A)\int_\tau^s\alpha_1(r)dr}
\bigg[
\alpha_2(\tau)(M+2\sigma_v^2M+2C_v^2)
+
2\alpha_2^2(\tau)
\operatorname{Tr}\left(\Sigma_1\Sigma_1^\top\right)
\bigg]d\tau
\bigg)^{\frac12}
\bigg]ds
\notag\\
&+
\int_{0}^{t}
e^{-2\kappa_2 \int_{s}^{t}\alpha_2(s^{\prime})ds^{\prime}}
\left(
\frac{2L_0}{N}\alpha_2(s)(M+\|x^*\|^2)^{\frac12}
+
\frac{\alpha_2^2(s)}{N}
\operatorname{Tr}(\Sigma_1^\top\Sigma_1)
\right) ds.
 \end{align}
This together with Assumption \ref{assumptiondescent1} leads to (\ref{Nconsensustomini0}).
\end{proof}

Combining the above lemmas,  we can prove Theorem \ref{hheNwucha}.

\textbf{Proof of Theorem \ref{hheNwucha}:}
 By $C_r$ inequality, Assumptions  \ref{assumption0},  \ref{assumptiondescent02}, \ref{assumptiondescent1}, \ref{VDLIP} and \ref{asssdd}, we have
 \begin{align}
  & \frac{1}{N}\sum_{i=1}^{N}E\left[\left\| x_i^{N,h}(kh)
  -x^*\right\|^2\right]\notag\\
  \leq & \frac{3}{N}\sum_{i=1}^{N}E\left[\left\| x_i^{N,h}(kh)
  -x_i^{N}(kh)\right\|^2\right]+ \frac{3}{N}\sum_{i=1}^{N}E\left[\left\| x_i^{N}(kh)-\frac{1}{N}\sum_{j=1}^{N}x_j^{N}(kh)
  \right\|^2\right]\notag\\
  &+    3  E\left[\left\| \frac{1}{N}\sum_{j=1}^{N}x_j^{N}(kh)
  -x^*\right\|^2\right].\notag
 \end{align}
 This together with Lemmas \ref{lemmmydsitolianxu0}-\ref{Nconsensustomini} and  Assumption \ref{assumptiondescent1} gives (\ref{lisantox}). By Assumption    \ref{assumptiondescent1}, we have    $\lim\limits_{k\to\infty}\phi_1(kh)=0$.     By Assumption \ref{continuousgrapho} and Lemma \ref{lamda2sklian}, we know that $\lim\limits_{N\to\infty}\phi_3(N)=0.$
  $\hfill\blacksquare$

\vskip 2mm

In the following lemma, \(L^2([0,1]^2)\) is the space of square-integrable functions on
\([0,1]^2\), \(L^\infty([0,1])\) is the space of essentially bounded functions
on \([0,1]\), and \(\mathcal L(L^2)\) denotes the space of bounded linear
operators on \(L^2([0,1])\), equipped with the standard operator norm.

\begin{lemma}\label{lamda2sklian}
Assume that the following conditions hold.
\begin{enumerate}
\item The discontinuity set of  $A$ has two-dimensional Lebesgue measure zero.

\item There exist constants $m_0\in\mathbb N$ and $L_1>0$ such that, for every fixed
$p\in[0,1]$, the function
$
q\mapsto A(p,q)
$
has at most $m_0$ discontinuity points and is $L_1$-Lipschitz continuous on each
continuity interval.

\item The degree function
$
d_A(p)=\int_0^1 A(p,q)\,dq
$
is Lipschitz continuous on $[0,1]$.
\end{enumerate}

Then
$
\lim_{  N \to \infty}\frac{\lambda_2(L^N)}{N}
=
\lambda_2(\mathbb L_A).
$
\end{lemma}

\begin{proof}
We first prove that
$
\|\mathbb L_{A^N}-\mathbb L_A\|_{\mathcal L(L^2)}
\to0.
$
As the discontinuity set of $A$ has two-dimensional Lebesgue measure zero,
we have
$
A^N(p,q)\to A(p,q)
$
for almost every $(p,q)\in[0,1]^2$. Moreover, by $0\leq A\leq 1$,
$
|A^N(p,q)-A(p,q)|^2\leq 1.
$
By the dominated convergence theorem, we have
\begin{align}\label{Aslia}
 \|A^N-A\|_{L^2([0,1]^2)}\to0.
\end{align}

Next, define
$
d_N(p)=\int_0^1 A^N (p,q)\,dq$.
For $p\in \left(\frac{i-1}{N},\frac{i}{N}\right]$, we have
$
d_N(p)
=
\frac1N\sum_{j=1}^N
A\left(\frac{i}{N},\frac{j}{N}\right).
$
We will show that
$
\|d_N-d_A\|_{L^\infty([0,1])}\to0.
$
%
By assumption, $q\mapsto A(x,q)$ has at most $m_0$ discontinuity points.
The intervals containing these discontinuity points are called bad intervals.
There are at most $m_0$ such intervals. On each bad interval, by  $0\leq A\leq 1$,
we have
$
\big|
\frac1N A\left(x,\frac{j}{N}\right)
-
\int_{\left(\frac{j-1}{N},\frac{j}{N}\right]} A(x,q)\,dq
\big|
\leq
\frac{2m_0}{N}.
$
On the remaining intervals, $q\mapsto A(x,q)$ is $L_1$-Lipschitz continuous.
Therefore,
\begin{align}
&\left|
\frac1N W\left(x,\frac{j}{N}\right)
-
\int_{\left(\frac{j-1}{N},\frac{j}{N}\right]} W(x,q)\,dq
\right|\notag\\
 \leq&
\int_{\left(\frac{j-1}{N},\frac{j}{N}\right]}
\left|
W\left(x,\frac{j}{N}\right)-W(x,q)
\right|dq  \notag\\
 \leq&
\int_{\left(\frac{j-1}{N},\frac{j}{N}\right]}
L_1\left|\frac{j}{N}-q\right|dq
 \leq
\frac{L_1}{2N^2}.\notag
\end{align}
Thus,
$
\sup_{x\in[0,1]}
\left|
\frac1N\sum_{j=1}^N A\left(x,\frac{j}{N}\right)
-
\int_0^1 A(x,q)\,dq
\right|
\leq
\frac{2m_0}{N}+\frac{L_1}{2N}.
$
For $p\in \left(\frac{i-1}{N},\frac{i}{N}\right]$, we have
\begin{align}
|d_N(p)-d_A(p)|
\leq&
\left|
\frac1N\sum_{j=1}^N
A\left(\frac{i}{N},\frac{j}{N}\right)
-
d_A\left(\frac{i}{N}\right)
\right|
+
\left|
d_A\left(\frac{i}{N}\right)-d_A(p)
\right|.\notag
\end{align}
As $d_A$ is Lipschitz continuous, there exists $L_d>0$ such that
$
\left|
d_A\left(\frac{i}{N}\right)-d_A(p)
\right|
\leq
L_d\left|\frac{i}{N}-p\right|
\leq
\frac{L_d}{N}.
$
Therefore,
$
|d_N(p)-d_A(p)|
\leq
 \frac{2m_0}{N}+\frac{L_1}{2N}+\frac{L_d}{N}.
$
Taking the supremum over $p\in[0,1]$, we obtain
\begin{align}\label{dslia}
 \|d_N-d_W\|_{L^\infty([0,1])}
\leq
\frac{2m_0}{N}+\frac{L_1}{2N}+\frac{L_d}{N}
\to0.
\end{align}
Define
$
(T_Az)(p)=\int_0^1 A(p,q)z(q)\,dq,\
(T_{A^N}z)(p)=\int_0^1 A^N(p,q)z(q)\,dq .
$
For any $z\in L^2([0,1])$, by the Cauchy--Schwarz inequality, we have
\begin{align*}
\|(T_{A^N}-T_A)z\|_{L^2}^2
=&
\int_0^1
\left|
\int_0^1 (A^N(p,q)-A(p,q))z(q)\,dq
\right|^2 dp\\
\leq&
\int_0^1
\left(
\int_0^1 |A^N(p,q)-A(p,q)|^2dq
\right)
\left(
\int_0^1 |z(q)|^2dq
\right)dp\\
=&
\|A^N-A\|_{L^2([0,1]^2)}^2
\|z\|_{L^2}^2 .
\end{align*}
Thus,
$
\|T_{A^N}-T_A\|_{\mathcal L(L^2)}
\leq
\|A^N-A\|_{L^2([0,1]^2)}.
$
Denote
$
(M_{d_A}z)(p)=d_A(p)z(p),\
(M_{d_N}z)(p)=d_N(p)z(p),
$
then
$
\|M_{d_N}-M_{d_A}\|_{\mathcal L(L^2)}
=
\|d_N-d_A\|_{L^\infty([0,1])}
\to0 .
$
By $
\mathbb L_A=M_{d_A}-T_A,\
\mathbb L_{A^N}=M_{d_N}-T_{A^N},
$ (\ref{Aslia}) and (\ref{dslia}),
we get
\begin{align}
\|\mathbb L_{A^N}-\mathbb L_A\|_{\mathcal L(L^2)}
\leq&
\|M_{d_N}-M_{d_A}\|_{\mathcal L(L^2)}
+
\|T_{A^N}-T_A\|_{\mathcal L(L^2)}\notag\\
\leq&
\|d_N-d_A\|_{L^\infty([0,1])}
+
\|A^N-A\|_{L^2([0,1]^2)}
\to0 .
\end{align}

It remains to show that this operator convergence implies the convergence of
the normalized algebraic connectivity.

Define the step-function subspace
$
\mathcal S_N
=
\left\{
z\in L^2([0,1]):
z(p)=x_i,\ p\in \left(\frac{i-1}{N},\frac{i}{N}\right],\ i=1,\ldots,N
\right\}.
$
For any $x=(x_1,\ldots,x_N)^\top\in\mathbb R^{nN}$, define $z_N\in\mathcal S_N$ by
$
z_N(p)=x_i,\ p\in  \left(\frac{i-1}{N},\frac{i}{N}\right].
$
Then
$
\|z_N\|_{L^2}^2
=
\frac1N\sum_{i=1}^N |x_i|^2
$
and
$
\int_0^1 z_N(p)\,dp
=
\frac1N\sum_{i=1}^N x_i.
$
Thus $z_N\perp 1$ in $L^2([0,1])$ if and only if
$
x^\top \mathbf 1_{nN}=0.
$
For $p\in  \left(\frac{i-1}{N},\frac{i}{N}\right]$,
\begin{align*}
(\mathbb L_{A^N}z_N)(p)
=
\int_0^1 A^N(p,q)(z_N(p)-z_N(q))\,dq
=
\frac1N
\sum_{j=1}^N A\Big(\frac{i}{N},\frac{j}{N}\Big)(x_i-x_j)
=
\frac1N(L^Nx)_i.
\end{align*}
Therefore, on $\mathcal S_N$, the operator $\mathbb L_{A^N}$ corresponds to
the matrix $L^N/N$. Moreover,
\begin{align*}
\langle \mathbb L_{A^N}z_N,z_N\rangle_{L^2}
=
\sum_{i=1}^N
\int_{ \left(\frac{i-1}{N},\frac{i}{N}\right]}
\frac1N(L^Nx)_i x_i\,dp
=
\frac1{N^2}x^\top L^Nx.
\end{align*}
Hence
$
\frac{
\langle \mathbb L_{A^N}z_N,z_N\rangle_{L^2}
}{
\|z_N\|_{L^2}^2
}
=
\frac1N
\frac{x^\top L^Nx}{x^\top x}.
$
Taking the infimum over all nonzero \(z_N\in\mathcal S_N\) with zero mean,
equivalently over all nonzero \(x\in\mathbb R^{nN}\) satisfying
\(x^\top\mathbf 1_{nN}=0\), gives
$$
\lambda_{2,N}^{\rm step}
:=
\inf_{\substack{z\in\mathcal S_N,\ \int_0^1z(p)\,dp=0\\
\|z\|_{L^2}=1}}
\langle \mathbb L_{A^N}z,z\rangle_{L^2}
=
\frac{\lambda_2(L^N)}{N}.
$$

We now prove
$
\lambda_{2,N}^{\rm step}
\to
\lambda_2(\mathbb L_A).
$
For any \(z\in\mathcal S_N\) satisfying
$
\int_0^1z(p)\,dp=0,\
\|z\|_{L^2}=1,
$
we have
\begin{align*}
\langle \mathbb L_{A^N}z,z\rangle_{L^2}
&=
\langle \mathbb L_Az,z\rangle_{L^2}
+
\langle(\mathbb L_{A^N}-\mathbb L_A)z,z\rangle_{L^2}\\
&\geq
\lambda_2(\mathbb L_A)
-
\|\mathbb L_{A^N}-\mathbb L_A\|_{\mathcal L(L^2)} .
\end{align*}
Taking the infimum over all such $z\in\mathcal S_N$, we obtain
$
\lambda_{2,N}^{\rm step}
\geq
\lambda_2(\mathbb L_A)
-
\|\mathbb L_{A^N}-\mathbb L_A\|_{\mathcal L(L^2)} .
$
Therefore,
\begin{align}\label{LAMDAYULA}
\liminf_{N\to\infty}
\lambda_{2,N}^{\rm step}
\geq
\lambda_2(\mathbb L_A).
\end{align}
For any $\varepsilon>0$, by the definition of
$\lambda_2(\mathbb L_A)$, there exists $z\in L^2([0,1])$ such that
$
\int_0^1 z(p)\,dp=0,\
\|z\|_{L^2}=1,
$
and
$
\langle \mathbb L_Az,z\rangle_{L^2}
\leq
\lambda_2(\mathbb L_A)+\varepsilon .
$
Let \(P_Nz\in\mathcal S_N\) be the piecewise average of \(z\), defined by
\[
(P_Nz)(p)
=
N\int_{\left(\frac{i-1}{N},\frac{i}{N}\right]} z(q)\,dq,\ p\in \left(\frac{i-1}{N},\frac{i}{N}\right].
\]
Then
$
P_Nz\to z\ \text{in }L^2([0,1]),
$
and
$
\int_0^1 P_Nz(p)\,dp
=
\int_0^1 z(p)\,dp
=
0.
$
Since \(\|P_Nz\|_{L^2}\to \|z\|_{L^2}=1\), for all sufficiently large \(N\),
we  define
$
z_N=
\frac{P_Nz}{\|P_Nz\|_{L^2}} .
$
Then
$
z_N\in\mathcal S_N,\
\int_0^1 z_N(p)\,dp=0,\
\|z_N\|_{L^2}=1,
$
and
$
z_N\to z\ \text{in }L^2([0,1]).
$
Therefore,
$
\lambda_{2,N}^{\rm step}
\leq
\langle \mathbb L_{A^N}z_N,z_N\rangle_{L^2}.
$
Moreover,
\begin{align}
&
\left|
\langle \mathbb L_{A^N}z_N,z_N\rangle_{L^2}
-
\langle \mathbb L_Az,z\rangle_{L^2}
\right|\notag\\
\leq &
\left|
\langle(\mathbb L_{A^N}-\mathbb L_A)z_N,z_N\rangle_{L^2}
\right|
+
\left|
\langle \mathbb L_Az_N,z_N\rangle_{L^2}
-
\langle \mathbb L_Az,z\rangle_{L^2}
\right|.\notag
\end{align}
The first term satisfies
$
\left|
\langle(\mathbb L_{A^N}-\mathbb L_A)z_N,z_N\rangle_{L^2}
\right|
\leq
\|\mathbb L_{A^N}-\mathbb L_A\|_{\mathcal L(L^2)}
\to0.
$
For the  second term,
\begin{align*}
&
\left|
\langle \mathbb L_Az_N,z_N\rangle_{L^2}
-
\langle \mathbb L_Az,z\rangle_{L^2}
\right|\\
\leq &
\left|
\langle \mathbb L_A(z_N-z),z_N\rangle_{L^2}
\right|
+
\left|
\langle \mathbb L_Az,z_N-z\rangle_{L^2}
\right|\\
\leq &
\|\mathbb L_A\|_{\mathcal L(L^2)}
\|z_N-z\|_{L^2}\|z_N\|_{L^2}
+
\|\mathbb L_Az\|_{L^2}\|z_N-z\|_{L^2}
\to0.
\end{align*}
Thus,
$
\langle \mathbb L_{A^N}z_N,z_N\rangle_{L^2}
\to
\langle \mathbb L_Az,z\rangle_{L^2}.
$
Consequently,
$$
\limsup_{N\to\infty}
\lambda_{2,N}^{\rm step}
\leq
\langle \mathbb L_Az,z\rangle_{L^2}
\leq
\lambda_2(\mathbb L_A)+\varepsilon .
$$
Since $\varepsilon>0$ is arbitrary,
\begin{align} \label{LAMDAYULA1}
\limsup_{N\to\infty}
\lambda_{2,N}^{\rm step}
\leq
\lambda_2(\mathbb L_A).
\end{align}
Combining (\ref{LAMDAYULA}) and (\ref{LAMDAYULA1}) yields
$
\lambda_{2,N}^{\rm step}
\to
\lambda_2(\mathbb L_A).
$
Finally, as
$
\lambda_{2,N}^{\rm step}
=
\frac{\lambda_2(L^N)}{N},
$
we conclude that
$
\lim_{  N \to \infty}\frac{\lambda_2(L^N)}{N}
=
\lambda_2(\mathbb L_A).
$
\end{proof}

\begin{lemma}\label{lamda2sklian1}
Let
$
A(p,q)
=
\big(1-2|p-q|\big)\mathbf 1_{\{|p-q|\le 1/4\}},\ (p,q)\in[0,1]^2 .
$
Then \(A\) satisfies the assumptions in  Lemma \ref{lamda2sklian}.
\end{lemma}

\begin{proof}
The only discontinuities of \(A\) occur when the indicator function
\(\mathbf 1_{\{|p-q|\leq 1/4\}}\) jumps, namely on the set
$
\mathcal D
=
\big\{
(p,q)\in[0,1]^2: |p-q|=\frac14
\big\}.
$
This set is the union of at most two line segments in \([0,1]^2\), and hence
has two-dimensional Lebesgue measure zero. Therefore, the discontinuity set
of \(A\) has two-dimensional Lebesgue measure zero.

For every fixed \(p\in[0,1]\), consider the function
$
q\mapsto A(p,q).
$
Its possible discontinuity points are
$
q=p-\frac14,\
q=p+\frac14,
$
provided that these points belong to \([0,1]\). Hence, for every fixed
\(p\in[0,1]\), the function \(q\mapsto A(p,q)\) has at most two discontinuity
points. Therefore, one may take
$
m_0=2.
$

On each interval not containing the above discontinuity points, the function
\(q\mapsto A(p,q)\) is either identically zero or equal to
$
1-2|p-q|.
$
The function \(q\mapsto 1-2|p-q|\) is Lipschitz continuous with Lipschitz
constant \(2\). Hence \(q\mapsto A(p,q)\) is \(2\)-Lipschitz continuous on
each continuity interval. Thus one may take
$
L_1=2.
$

It remains to verify that the degree function is Lipschitz continuous.
For \(p\in[0,1]\),
$
d_A(p)
=
\int_0^1
\big(1-2|p-q|\big)\mathbf 1_{\{|p-q|\le 1/4\}}\,dq .
$
A direct computation gives
\[
d_A(p)=
\begin{cases}
p-p^2+\dfrac{3}{16},
& p\in[0,1/4],\\[6pt]
\dfrac38,
& p\in[1/4,3/4],\\[6pt]
(1-p)-(1-p)^2+\dfrac{3}{16},
& p\in[3/4,1].
\end{cases}
\]
This function is continuous on \([0,1]\).
Moreover, on the three open intervals \((0,1/4)\), \((1/4,3/4)\), and
\((3/4,1)\), its derivative is given by
\[
d_A'(p)=
\begin{cases}
1-2p,
& p\in(0,1/4),\\[4pt]
0,
& p\in(1/4,3/4),\\[4pt]
1-2p,
& p\in(3/4,1).
\end{cases}
\]
Thus
$
|d_A'(p)|\leq 1
$
for almost every \(p\in[0,1]\). Since \(d_A\) is continuous and piecewise
\(C^1\), it follows that, for any \(p_1,p_2\in[0,1]\),
$
|d_A(p_1)-d_A(p_2)|
\leq
|p_1-p_2|.
$
Therefore $d_A$ is 1-Lipschitz continuous on \([0,1]\).

Consequently, \(A\) satisfies all assumptions in Lemma \ref{lamda2sklian}.
\end{proof}
\end{document}